\documentclass[12pt]{article}
\usepackage{adjustbox,lipsum}
\usepackage{float}
\usepackage{amssymb}
\usepackage{amsthm}
\usepackage{amsmath}
\usepackage{latexsym}
\usepackage{epsfig}
\usepackage{graphicx}
\usepackage{wasysym}
\usepackage{threeparttable}
\usepackage{natbib}
\usepackage{color}
\usepackage{epstopdf}
\usepackage{caption}
\usepackage{subcaption}

\newcommand{\ind}{\stackrel{\mathrm{ind}}{\sim}}

\usepackage[breaklinks]{hyperref}

\usepackage{enumitem}

\def\boxit#1{\vbox{\hrule\hbox{\vrule\kern6pt
          \vbox{\kern6pt#1\kern6pt}\kern6pt\vrule}\hrule}}

\def\bse{\begin{eqnarray*}}
\def\ese{\end{eqnarray*}}
\def\be{\begin{eqnarray}}
\def\ee{\end{eqnarray}}
\def\bq{\begin{equation}}
\def\eq{\end{equation}}
\def\bse{\begin{eqnarray*}}
\def\ese{\end{eqnarray*}}

\usepackage[ruled]{algorithm2e}
\usepackage{mathptmx}      
\usepackage{bm}

 {\begin{list}{}%
         {\setlength{\leftmargin}{#1}}%
         \item[]%
 }
 {\end{list}}
\usepackage{setspace}

\def\bu{\textbf{u}}
\def\bs{\textbf{s}}

\def\bz{\textbf{Z}}

\def\by{\textbf{Y}}

\begin{document}
\part*{}
\thispagestyle{empty} \baselineskip=28pt

\begin{center}
{\LARGE{\bf Regionalization of Multiscale Spatial Processes using a Criterion for Spatial Aggregation Error}}
\end{center}

\baselineskip=12pt

\vskip 2mm
\begin{center}
Jonathan R. Bradley\footnote{(\baselineskip=10pt to whom correspondence should be addressed) Department of Statistics, University of Missouri, 146 Middlebush Hall, Columbia, MO 65211, bradleyjr@missouri.edu},
Christopher K. Wikle\footnote{\baselineskip=10pt  Department of Statistics, University of Missouri, 146 Middlebush Hall, Columbia, MO 65211-6100},
Scott H. Holan$^2$
\end{center}
%
%
%
%
\vskip 4mm

\begin{center}
\large{{\bf Abstract}}
\end{center}
The modifiable areal unit problem and the ecological fallacy are known problems that occur when modeling multiscale spatial processes. We investigate how these forms of spatial aggregation error can guide a regionalization over {a} spatial domain of interest. By ``regionalization'' we mean a specification of geographies that define the spatial support for areal data. This topic has been studied vigorously by geographers, but has been given less attention by spatial statisticians. Thus, we propose a criterion for spatial aggregation error (CAGE), which we minimize to obtain an optimal regionalization. To define CAGE we draw a connection between spatial aggregation error and a new multiscale representation of the Karhunen-Lo\'{e}ve (K-L) expansion. This relationship between CAGE and the multiscale K-L expansion leads to illuminating theoretical developments including: connections between spatial aggregation error, squared prediction error, spatial variance, and a novel extension of Obled-Creutin eigenfunctions. The effectiveness of our approach is demonstrated through an analysis of two datasets, one using the American Community Survey and one related to environmental ocean winds.
\baselineskip=12pt

%
%
%

\baselineskip=12pt
\par\vfill\noindent
\par\medskip\noindent
\clearpage\pagebreak\newpage \pagenumbering{arabic}
\baselineskip=24pt
\section{Introduction} \label{sec:intro} 

There has long been interest in non-statistical methods for specifying geographies to summarize spatial data (e.g., \citet{openshaw}, \citet{Murtagh}, \citet{Martin}, \citet{Guo}, and \citet{Logan}). In general, this is known as ``regionalization,'' and it is an important (and sometimes required) task for many applications. For example, the American Community Survey (ACS) is an ongoing survey administered by the US Census Bureau that produces estimates of important US demographic variables. The ACS provides public-use data referenced over areal units (e.g., median household income over US counties). Similar to the decennial census, many of these geographic regions are required (e.g., states, counties, etc.), however, other regions are consistently being evaluated and changed (e.g., combined statistical areas, metropolitan divisions, metropolitan statistical areas, etc.) in a sub-optimal manner based on population controls (e.g., \citet{censusdistricting}). This suggests that there is a clear need for regionalization methodology. Thus, we develop a principled statistical methodology for evaluating spatial aggregation error and optimal statistical regionalization.\\
\indent Regionalization is a topic that has been considered primarily by geographers. The current state-of-the-art is the deterministic ``max-$p$ algorithm'' \citep{Duque,Spielman1,Folch,Spielman2}. In general, the max-$p$ algorithm is a greedy search algorithm (using any desired criterion) that groups data defined on $n_{A}$ areal units into $p$ $(\le n_{A})$ contiguous regions. The max-$p$ algorithm offers a solution, but there are many known pitfalls to this approach. The most significant issue from the perspective of multiscale spatial inference is that the regions obtained by this approach are not protected from the \textit{ecological fallacy} \citep{Robinson}. Hence, proper inferential conclusions must be limited to a single (often difficult to interpret) spatial support. 
\\
\indent We interpret the ecological fallacy as a type of spatial aggregation error, which will be critical to our approach for regionalization.  In particular, the ecological fallacy refers to the situation where conclusions at the point-level spatial support differ from conclusions at an aggregate-level spatial support. Similarly, \textit{ecological inference} is explicitly defined as inference on individual behavior drawn from aggregate data (also sometimes referred to as downscaling). This topic has experienced growing interest within a variety of subject matter disciplines.  For example, see \citet{king} for the sociological data setting; \citet{darby}, and the references therein, for applications in epidemiology; and \citet{mearns}, and the references therein, for the climatology setting. Following the terminology of \citet{Kolaczyk}, a similar problem is known as \textit{image segmentation}, which involves optimally dividing an image into smaller regions (e.g., see \citet{Kolaczyk3}, \citet{Kolaczyk2}, and \citet{FerreiraHolan}). For reviews of ecological inference and image segmentation see \citet{wakefield}, \citet{Wallergotway}, and \citet{marco}.
\\
\indent The \textit{modifiable areal unit problem} ({MAUP}) is another type of spatial aggregation error. \citet{Wallergotway} consider the MAUP to be the geographic manifestation of the ecological fallacy. That is, the MAUP refers to situations where conclusions on one aggregate spatial support differ from conclusions on another distinct aggregate spatial support, whereas, the ecological fallacy concerns conflicting conclusions at point-level and aggregate-level supports. The MAUP has a rich history, originally considered by \citet{gehike}, and later by \citet{openshawTaylor}. Recently, the MAUP has become a topic covered in standard textbooks including \citet{cressie}, \citet{Wallergotway}, \citet{cressie-wikle-book}, and \citet{banerjee-etal-2015}, among others.\\
\indent The aforementioned forms of spatial aggregation error are closely related to the \textit{spatial change of support} (COS) problem, which refers to conducting statistical inference on a support that differs from the spatial support of the data (e.g., \citet{Wallergotway}, \citet{cressie-wikle-book}, and \citet{banerjee-etal-2015}). Methods for spatial COS allow one to choose any support on which to perform statistical inference.  However, different choices for the spatial support result in different magnitudes of spatial aggregation error. Nevertheless, the inherent flexibility to use any desired spatial support for inference has made spatial COS a popular area of research in both multiscale spatial analysis and other subject matter disciplines. For example, see \citet{WileBerliner} for the environmental data setting; \citet{mugglin99} for the public health setting; \citet{bradleywikleholan} for the survey data setting; and \citet{Wallergotway} and \citet{gelfandCOSreview} for a review. To capitalize on the flexibility of spatial COS methods, we adopt a multiscale spatial perspective to quantify spatial aggregation error and to develop a method for regionalization. \\
\indent The known presence of spatial aggregation error suggests an approach for an optimal regionalization. Specifically, our primary inferential question is the following: can we choose a spatial support that minimizes spatial aggregation error? To motivate this perspective, consider an example dataset obtained from the ACS. In Figures~1(a) and 1(b), we plot 5-year period estimates of median household income by county and state, respectively, for 2013. Upon comparison, Figures 1(a) and 1(b) show that the state-level ACS estimates suffer from noticeable spatial aggregation error. For example, Figure 1(b) suggests that households in Virginia have moderately high income, yet Figure 1(a) shows that only households in counties near Richmond have high income. Similarly, Figure 1(b) suggests that households in New York state have a moderately high income while Figure 1(a) shows that only households in counties near Manhattan have high income. These examples, and many others that are quite obvious upon study of these figures, provide evidence that states are not an appropriate (i.e., optimal) spatial support to summarize median household income, political reasons notwithstanding. \\
\indent In what follows, we formalize this intuition and develop a criterion to quantify spatial aggregation error and an associated method for regionalization.
      \begin{figure}[t!]
      \begin{center}
      \begin{tabular}{c}
        \includegraphics[width=16.5cm,height=8cm]{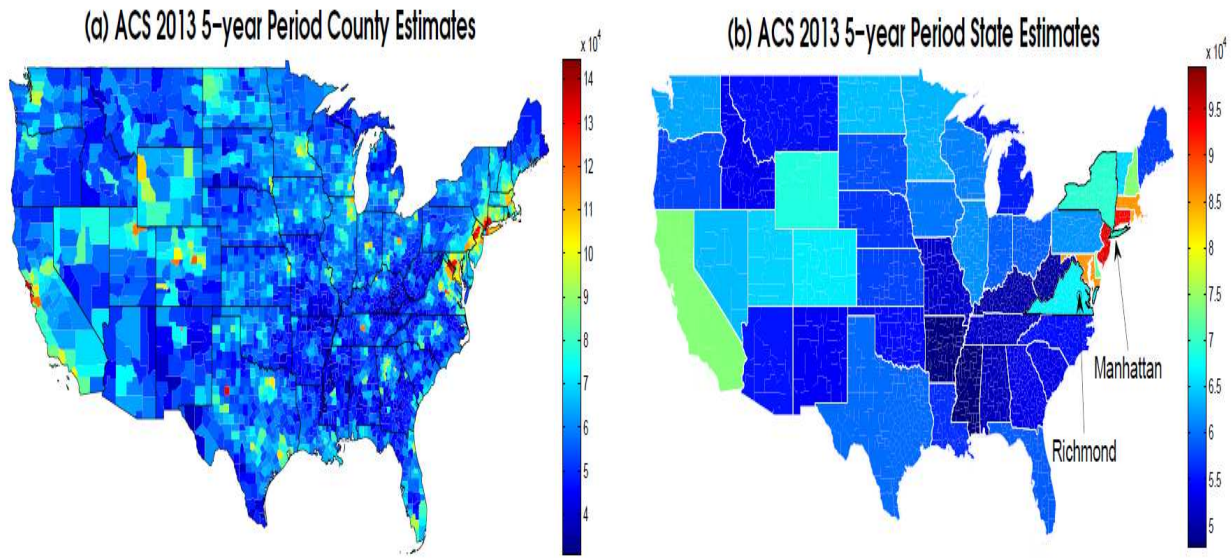}
      \end{tabular}
      \caption{ACS 5-year period estimates of median household income for 2013. In (a), we plot the ACS estimates by counties, and in (b) we plot the ACS estimates by state. We superimpose the state boundaries as a reference in both panels. Notice that the color-scales are different for each panel. In (b), the borders of the states are highlighted in white except for New York and Virginia, whose borders are highlighted in black. Also, Richmond Virginia and Manhattan are indicated with arrows in (b).}
      \end{center}
      \end{figure}
Our approach is to quantify spatial aggregation error using what we call the \textit{criterion for spatial aggregation error} ({CAGE}). Hence, an optimal spatial support is obtained by minimizing CAGE. The primary theoretical tool used to develop this criterion is the Karhunen-Lo\`{e}ve (K-L) expansion \citep{Karhunen,Loeve}, which is a well-known representation of a point-referenced process as the weighted sum of spatially varying eigenfunctions, where the weights are random. In more precise terms, we develop CAGE through a powerful technical result, which dictates that spatial aggregation error does not occur when the eigenfunctions of a spatial random process  are constant between spatial scales. Thus, CAGE is a measure of between spatial scale homogeneity of eigenfunctions within a novel multiscale representation of the K-L expansion.\\
\indent To date, there has been no such criterion that quantifies spatial aggregation error in this manner. The spatial statistics literature places an emphasis on prediction error (e.g., \citet{cressie}), and thus, such an aggregation-based approach for uncertainty quantification offers an exciting new perspective for spatial statistics. Therefore, to develop this perspective we provide technical results relating CAGE to prediction error and spatial variance.\\
\indent After having defined CAGE, we can choose a regionalization in a manner that mitigates spatial aggregation error. In particular, we propose an efficient search algorithm (with CAGE as the selection criterion) to specify a regionalization over the spatial domain of interest. This search algorithm involves two stages. In the first stage, a naive algorithm, say $k$-means (e.g., \citet{kmeans}) is used to determine a collection of spatial supports from which to select. Then, in the second stage CAGE is used to select a single spatial support from among the collection of spatial supports determined in the first stage of the search algorithm. This two-stage approach is extremely efficient because it uses an easy-to-compute deterministic algorithm to direct the path of spatial supports from which to choose. As such, it can be incorporated efficiently within a Bayesian framework using a Markov chain Monte Carlo (MCMC) implementation of a latent spatial model, which facilitates uncertainty quantification.
\\
\indent Finally, to apply our search algorithm in practice, we provide a specification for the multiscale eigenfunctions. Thus, we introduce a general class of eigenfunctions that leads to a consistent class of multiscale spatial processes. To do this, we utilize the often overlooked, but remarkable framework of \citet{obled-creutin}. \citet{obled-creutin} show that any class of geostatistical basis functions can be re-weighted so that they are eigenfunctions within a (single-scaled) K-L expansion. This notion of what we call \textit{generating basis functions} ({GBFs}), is central to our development of multiscale eigenfunctions. As interest in spatial and spatio-temporal processes has turned to ``big data'' problems with large numbers of prediction and/or data locations, the modeling focus has shifted to this basis function perspective incorporating complete, over-complete, and reduced-rank expansions \citep{bradleycs2014}. Thus, the use of GBFs greatly increases the generality and utility of our approach. Furthermore, the use of GBFs is a necessity for our approach to regionalization because they allow us to perform spatial COS without assuming some form of between scale homogeneity.\\
\indent The remainder of this paper is organized as follows. In Section~2, we introduce the multiscale K-L expansion and CAGE. Next, in Section~3 we describe how to use CAGE in practice, which includes details on truncating the multiscale K-L expansion and the introduction of the two stage regionalization algorithm. Section~4 provides derivations of a consistent class of multiscale eigenfunctions to use within the CAGE framework. Then, in Section~5 a demonstration is given using the motivating dataset of ACS 5-year period estimates of median household income from Figure~1. In addition to demonstrating the regionalization algorithm for ACS period estimates, this application also highlights an important use of optimal regionalization, namely, aggregation for the purpose of dimension reduction. Finally, Section~6 contains a concluding discussion. We provide additional Supplemental Materials including: the proofs of technical results, simulation studies, and an additional application using a dataset consisting of Mediterranean wind measurements (a subset of the data used in \citet{millif}). The Mediterranean wind example is used to illustrate that the two-stage regionalization algorithm is flexible enough to handle multiscale spatial data.
\section{Quantifying Aggregation Error}\label{Sec2}

Here, we provide requisite extensions of the K-L expansion to the multiscale setting (Section 2.1). These results are then used to formally define CAGE (Section 2.2).

\subsection{The Multiscale Karhunen-Lo\'{e}ve Expansion} \label{Sec21}
Consider a real-valued spatial process that is realized at (possibly) both point-level and aggregate-level spatial supports. That is, the values in the sets $\{Y_{s}(\bs): \bs \in D_{s}\}$ and $\{Y_{A}(A): A \in D_{A}\}$ can be realized, where $Y_{s}$ is a continuous spatial random process defined on $D_{s}$, $D_{s} \subset \mathbb{R}^{d}$, and $Y_{A}$ is a spatial random process defined on areal support $D_{A}$ with $D_{A}\equiv \{A_{i}: i = 1,...,n_{A}\}$ and $A_{i}\subset \mathbb{R}^{d}$. The set $A_{i}$ is an areal unit (e.g., a county, state, or census tract) and may be overlapping, contained in, or superimposed over another distinct areal unit $A_{j}\in D_{A}$ for $j\ne i$.\\
\indent The corresponding multiscale spatial process can be written as
\begin{eqnarray}\label{multiscale:process}
Y(\bu) =
\left\{
	\begin{array}{ll}
		Y_{s}(\bu) & \mbox{if } \bu \in D_{s} \\
		Y_{A}(\bu)  & \mbox{if }  \bu \in D_{A};\hspace{5pt}\bu \in D_{s}\cup D_{A}.
	\end{array}
\right.
\end{eqnarray}
We interpret $Y_{A}(\cdot)$ as being computed from the point-level process $\{Y_{s}(\cdot)\}$. In particular, as is standard in spatial statistics (e.g., \citet{cressie}, p. 284), we assume
\begin{eqnarray}\label{aggY}
Y_{A}(A) \equiv \frac{1}{|A|}\int_{A}Y_{s}(\bs) d\bs ;\hspace{5pt}A\in D_{A},
\end{eqnarray}
where $|A|$ represents the cardinality of the set $A$. Consequently, placing a statistical model on $Y_{s}$ implicitly places a statistical model on $Y_{A}$ and $Y$ through (\ref{multiscale:process}) and (\ref{aggY}). We explore this dependency between (\ref{multiscale:process}) and (\ref{aggY}) using the well-known K-L expansion (e.g., \citet{cressie-wikle-book}, p. 156),
\begin{eqnarray}\label{kl} 
Y_{s}(\bs) = \sum_{j=1}^{\infty} \phi_{j}(\bs) \alpha_{j}; \hspace{5pt} \bs \in D_{s},
\end{eqnarray}
\noindent where, without loss of generality, $\{Y_{s}(\cdot)\}$ is assumed to be mean-zero, the random variables in the set \(\{\alpha_{j}: j = 1, 2, ...\}\) are uncorrelated with associated variances \(\{\lambda_{j}: j = 1,2, ...\}\) (called eigenvalues), the orthonormal real-valued functions $ \{\phi_{j}(\bs) : j = 1, 2, ...\}$ (called eigenfunctions) have domain $D_{s}$, and satisfy a Fredholm integral equation for a given valid covariance function. (Note that the conditions needed for the K-L expansion are given in the statement of Proposition 1.) \\
\indent The use of the K-L expansion greatly increases the generality of our approach, since Mercer's theorem dictates that point-level covariance functions can be decomposed according to the K-L expansion \citep{mercer} under a very general set of assumptions \citep{genmerc}. This leads us to define a multiscale K-L expansion, which we formalize through Proposition 1 below.\\

\noindent
\textit{Proposition 1: Let $(\Omega,\mathcal{F},\mathcal{P})$ be a probability space, where $\Omega$ is a sample space, $\mathcal{F}$ is a sigma-algebra on $\Omega$, and $\mathcal{P}$ is a finite Borel measure. Let $Y_{s}(\bs)$ be defined by the mapping $Y_{s}: D_{s}\times \Omega \rightarrow \mathbb{R}$, such that $Y_{s}(\bs)$ is measurable for every $\bs \in D_{s}$, and $D_{s}\subset \mathbb{R}^{d}$ is a topological Hausdorff space. Assume that $C(\bs,\bu)\equiv\mathrm{cov} \left\lbrace Y_{s}(\bs), Y_{s}(\bu)\right\rbrace$ is a valid covariance function that exists for each $\bs,\bu \in D_{s}$. Let $L^{2}(\Omega)$ denote the Hilbert space of real-valued square integrable random variables. 
\begin{enumerate}[label=\roman*.]
\item Then, for each $A \subset D_{s}$ we have that 
\begin{eqnarray}\label{mkl}
Y_{A}(A) = \sum_{i = 1}^{\infty} \phi_{A,j}(A) \alpha_{j},
\end{eqnarray}
 in $L^{2}(\Omega)$, where for each positive integer $j$, $\phi_{A,j}(A)\equiv \int_{A}\phi_{j}(\bs)d\bs/|A|$, the random variables in the set \(\{\alpha_{j}: j = 1, 2, ...\}\) are uncorrelated with associated variances \(\{\lambda_{j}: j = 1,2, ...\}\) (called eigenvalues), the orthonormal real-valued functions $ \{\phi_{j}(\bs) : j = 1, 2, ...\}$ (called eigenfunctions) have domain $D_{s}$, and satisfy the Fredholm integral equation for $C(\bs,\bu)$. 
\item Then for any $A \subset D_{s}$ and $B \subset D_{s}$ we have that
\begin{eqnarray}\label{mercer2}
\mathrm{cov}\left\lbrace Y_{A}(A), Y_{A}(B)\right\rbrace = \underset{n \rightarrow \infty}{\mathrm{lim}}\hspace{10pt}\sum_{i = 1}^{n}\phi_{A,i}(A)\phi_{A,i}(B)\lambda_{i}.
\end{eqnarray}
\end{enumerate}
}

\noindent
The proof of this proposition can be found in the Supplemental Materials.\\

\noindent
\textbf{Remark 1:} We call the expression in (\ref{mkl}) the multiscale K-L expansion since Proposition $1.i$ extends the K-L expansion in (\ref{kl}) to a similar infinite-dimensional process that is a function of \textit{any} $A \subset D_{s}$. Similarly, the expression in (\ref{mercer2}) can be seen as an extension of Mercer's theorem to the multiscale spatial setting.\\

\noindent
\textbf{Remark 2:} In practice, the latent multiscale spatial process of interest $Y$ is not observed perfectly. Instead, we observe the $n$-dimensional data vector given by $\bz \equiv (Z(\bu): \bu \in D_{s}^{O}\cup D_{A}^{O})^{\prime}$, where the observed locations are denoted by $D_{s}^{O} \equiv \{\bs_{i}^{O}: i = 1,...,n_{s}^{O}\} \subset D_{s}$ and $D_{A}^{O} \equiv \{A_{j}: j = 1,...,n_{A}^{O}\} \subset D_{A}$, and $n = n_{s}^{O} + n_{A}^{O}$. We assume that the stochastic processes $Z: D_{s}\times \Omega \rightarrow \mathbb{R}$ and $Y$ are generated based on the generic probability space $(\Omega,\mathcal{F},\mathcal{P})$ such that the conditional probability density function of $Y(\bu)\vert \bz$ exists for each $\bu \in D_{s}\cup D_{A}$.\\

\noindent
\textbf{Remark 3:} For purposes of implementation it is helpful to define a set $D_{B}\equiv \{B_{j}: j = 1,...,n_{B}\}$ with $B_{j}\cap B_{\ell} = \emptyset$ for $j\ne \ell$ and $B_{j} \subset D_{s}$ for each $j$. Here, $D_{B}$ represents the finest resolution spatial support on which one is willing to perform inference. Then, after observing data $Z(\cdot)$, statistical inference is performed using sample draws from the distribution of $\by_{B}\vert \bz$, where the $n_{B}$-dimensional process vector is given by $\by_{B} \equiv (Y_{A}(B): B \in D_{B})^{\prime}$. 

\subsection{The Criterion for Spatial Aggregation Error (CAGE)}\label{Sec22}

There is an implicit conceptual challenge involved with quantifying spatial aggregation error. As Gotway and Waller (2011) discuss, the consequences of spatial aggregation error extend beyond between-scale differences of the values of a single statistic (e.g., correlation coefficient, mean, etc.). Thus, we say that spatial aggregation error occurs when there are between-scale differences for \textit{any} generic statistic. The multiscale K-L expansion in (\ref{mkl}) provides insight on a formalization of this concept, which we state in Proposition 2.\\

\noindent
\textit{Proposition 2: Assume that the conditions of Proposition 1 hold. Let $f$ be a measurable real-valued function with domain $\mathbb{R}^{n_{A}}$ that is discontinuous only on a set with measure zero. Let $\lambda_{k}$ be strictly greater than zero for each $k = 1, 2,...$. Define a generic point-level support $\{\textbf{x}_{j}: j = 1,...,n_{A}\}$, such that $\textbf{x}_{j} \in B_{j} \subset A_{j} \in D_{A}$ for $j = 1,...,n_{A}$, $\by_{s}^{(A)} \equiv  \left(Y_{s}(\textbf{x}_{j}): j = 1,...,n_{A}\right)^{\prime}$, $\by_{B}^{(A)} \equiv  \left(Y_{A}(B_{j}): j = 1,...,n_{A}\right)^{\prime}$, and $\by_{A}\equiv (Y_{A}(A): A \in D_{A})^{\prime}$. Then the following statements hold for $Y(\cdot)$ in (\ref{multiscale:process}):
\begin{enumerate}[label=\roman*.]
\item ${\phi}_{k}(\textbf{x}_{j})= {\phi}_{A,k}(A_{j})$ for $j = 1,...,n_{A}$ and every positive integer $k$, if and only if $f(\by_{s}^{(A)})$ = $f(\by_{A})$ almost surely.
\item ${\phi}_{k}(B_{j})= {\phi}_{k}(A_{j})$ for $j = 1,...,n_{A}$ and every positive integer $k$, if and only if $f(\by_{B}^{(A)})$ = $f(\by_{A})$ almost surely.
\item If ${\phi}_{k}(\textbf{x}_{j})= {\phi}_{k}(A_{j})$ for every positive integer $k$, and every $\textbf{x}_{j} \in B_{j}$ and $j$, then $f(\by_{B}^{(A)})$ = $f(\by_{A})$ almost surely.
\end{enumerate}
}

\noindent
\textbf{Remark 4:} Proposition 2 provides a condition so that there is no ecological fallacy between $\by_{s}^{(A)}$ and $\by_{A}$, and no MAUP between $\by_{B}^{(A)}$ and $\by_{A}$. By ``no ecological fallacy'' and ``no MAUP,'' we mean that for any real-valued, measurable, (almost) continuous statistic $f$, $f(\by_{s}^{(A)})=f(\by_{A})$ and $f(\by_{B}^{(A)})=f(\by_{A})$ almost surely. This ensures that conclusions using the summary statistic $f$ stay the same regardless of the scale of $Y$. In general terms, Propositions~$2.i$ and $2.ii$ show that ``no spatial aggregation error'' is equivalent to between-scale homogeneity of eigenfunctions within a multiscale K-L expansion. Furthermore, Propositions~$2.i$ and $2.iii$ provide a relationship between the ecological fallacy and the MAUP; namely, if there is uniformly no ecological fallacy for any of the sets in $\{B_{j}\}$ (i.e., $\bm{\phi}_{s}(\textbf{x}_{j})= \bm{\phi}(A_{j})$ for every $\textbf{x}_{j} \in B_{j}$ and $j$), then there is no MAUP.\\

Proposition 2 guarantees that spatial aggregation error does not occur when the point-level eigenfunctions are constant over each region in $D_{A}$. This leads naturally to a criterion that measures departures from the absence of spatial aggregation error. Specifically, we define CAGE as follows:
\begin{eqnarray}\label{gammafunc}
\mathrm{CAGE}(A) =  E\left[\int_{A}\frac{\sum_{j = 1}^{\infty} \left\lbrace\phi_{j}(\bs) - \phi_{A,j}(A)\right\rbrace^{2} \lambda_{j}}{|A|} d\bs \vert \bz\right],
\end{eqnarray}
\noindent
where $A$ is a generic areal unit (i.e., $A \subset D_{s}$), and the expectation is taken with respect to the conditional distribution given the data. The logic behind (\ref{gammafunc}) is straightforward: if CAGE$(A)$ is equal to zero there is no loss of information when aggregating $D_{s}$ to $D_{A}$, and if $\mathrm{CAGE}(A)$ is close to (far from) zero then we lose a small (large) amount of point-level information when aggregating to $A$. Hence, maps of $\{\mathrm{CAGE}(A_{i}): i = 1,...,n_{A}\}$ can be used to assess whether statistical inference on $Y_{A}$ is reasonable relative to the point level process. \\
\indent In some settings the latent process cannot realistically be defined at the point level. For example, the median (over counties) household income in Figure 1 cannot be interpreted on $D_{s}$ (see \citet{banerjee-etal-2015} for a discussion and more examples). Hence, for these settings the multiscale K-L expansion is used for spatial change of support, and the lowest spatial resolution on which $Y$ is defined is $D_{B}$. We use the following discretized CAGE (abbreviated as ``DCAGE'') in these settings:
\begin{eqnarray}\label{gammafunc2}
\mathrm{DCAGE}(C) \equiv  E\left[\underset{h\in H}{\sum}\frac{\sum_{j = 1}^{\infty} \left\lbrace\phi_{A,j}(B_{j}) - \phi_{A,j}(C)\right\rbrace^{2} \lambda_{j}}{|C|} \vert \bz\right],
\end{eqnarray}
\noindent
where $C = \cup_{h \in H} B_{h}$, $H \subset \{1,...,n_{B}\}$, and $B_{h} \in D_{B}$ for each $h\in H$. Proposition $2.ii$ implies the following logic for (\ref{gammafunc2}): if DCAGE$(C)$ is equal to zero there is no loss of information when aggregating $D_{B}$ to higher spatial resolutions, and if D$\mathrm{CAGE}(C)$ is close to (far from) zero then we lose a small (large) amount of lower resolution information when aggregating $D_{B}$ to higher spatial resolutions (see Remark 3).\\
\indent To date there has been no attempt to quantify the magnitude of spatial aggregation error using criteria like (\ref{gammafunc}) and (\ref{gammafunc2}). In the geostatistical setting, emphasis is usually placed on minimizing the squared prediction error \citep{cressie}. From this point-of-view, it is worthwhile to note that there are connections between the squared prediction error, spatial variance, and CAGE in (\ref{gammafunc}), which we formally state in Proposition 3 below.\\

\noindent
\textit{Proposition 3: Assume that the conditions of Proposition 1 hold. Also, assume that the stochastic process $Z: D_{s}\times \Omega \rightarrow \mathbb{R}$ is generated based on a generic probability space $(\Omega,\mathcal{F},\mathcal{P})$ such that the conditional probability density function of $Y(\bu)\vert \bz$ exists for each $\bu \in D_{s}\cup D_{A}$, where $Z$ is defined in Remark 2. Then, CAGE in (\ref{gammafunc}) has the following alternative expressions:
\begin{eqnarray}
\label{intuitive1}
CAGE(A) \hspace{-8pt}&=&\hspace{-15pt} \hspace{5pt}E\left[ \int_{A}\frac{\left\lbrace Y_{s}(\bs) - Y_{A}(A)\right\rbrace^{2}}{|A|} d\bs \vert \bz\right]\\
\label{intuitive2}
CAGE(A) \hspace{-8pt}&=&\hspace{-15pt} \hspace{5pt}E\left[\int_{A}\frac{\mathrm{var}\left\lbrace Y_{s}(\bs)\right\rbrace}{|A|} d\bs - \mathrm{var}\left\lbrace Y_{A}(A)\right\rbrace \vert \bz\right]\\
\label{intuitive3}
\hspace{-400pt}CAGE(A) \hspace{-10pt}&=&\hspace{-15pt}\hspace{5pt}E\left[ \int_{A}\frac{\left\lbrace Y_{s}(\bs) -\widehat{Y}_{A}(A)\right\rbrace^{2}}{|A|} d\bs \vert \bz\right] - E\left[ \left\lbrace\widehat{Y}_{A}(A) - Y_{A}(A)\right\rbrace^{2} \vert \bz\right]\hspace{-5pt}, 
\end{eqnarray}
\noindent
where $A$ is a generic areal unit (i.e., $A \subset D_{s}$), and $\widehat{Y}_{A}(A) \equiv E(Y_{A}(A)|\bz)$.}\\

\noindent
\textbf{Remark 5:} Each expression in Proposition 3 provides interesting motivation for CAGE. For example, (\ref{gammafunc}) was motivated by Proposition 2 (i.e., by measuring the departure from the absence of spatial aggregation error), however, one could argue to use (\ref{intuitive1}) from a practical perspective. That is, intuition suggests that it is reasonable to make finer scale inference using the aggregate process if $Y_{s}(\bs)$ is consistently ``close'' to ${Y}_{A}(A)$. However, it is important to note that our use of the K-L expansion is important because it allows us to perform spatial change of support to obtain $Y_{A}$ without assumptions of {between-scale} homogeneity. Additionally, the expression in (\ref{intuitive2}) is especially interesting from a historical perspective, since many of the early references on spatial aggregation error focused on second order statistics \citep{Robinson}. Here, we see that {between-scale} differences of variances have a connection (through Propositions~1, 2, and 3) to between-scale differences of any statistic.\\

\noindent
\textbf{Remark 6:} The ``ANOVA-type'' decomposition in (\ref{intuitive3}) offers a different perspective in which to interpret (\ref{gammafunc}). The first term on the right-hand-side of (\ref{intuitive3}) (from left to right) represents a within-areal unit prediction error.  Specifically, the first term represents the prediction error between the point-level process $Y_{s}$ and the aggregate-level estimator $\widehat{Y}_{A}$. The second term in (\ref{intuitive3}) shows that a minimax-type approach is used for between areal unit error. That is, we minimize the squared prediction error to obtain $\widehat{Y}_{A}$, but penalize for choosing $A$ so that $Y_{A}$ is close to $\widehat{Y}_{A}$. One could conceive of a version of Proposition 3 that provides similar identities for the DCAGE in (\ref{gammafunc2}). In Supplemental Materials, we provide the statement and proof of this technical result. 

\section{Statistical Methodology for Regionalization}\label{Sec3}
In practice, higher order components, of the infinite sum in (\ref{kl}), correspond to a decreasing percentage of variation. Thus, it is standard practice to truncate the K-L expansion, and assume that the residual is negligible (e.g., see \citet{obled-creutin}, and \citet{cressie-wikle-book} p. 267). In this section, we extend the results from Section 2 to accommodate this common assumption. In particular, for our applications we truncate the multiscale K-L expansion (Section 3.1), which leads to another version of CAGE (Section 3.2). With these details in place, we can describe how to use CAGE for regionalization (Section 3.3). 

\subsection{The Truncated Multiscale Karhunen-Lo\'{e}ve Expansion} \label{Sec31}
A common simplification of the K-L expansion is to truncate the infinite sum in (\ref{kl}) and assume that
\begin{eqnarray}\label{approxZ}
Y_{s}(\bs; \hspace{1pt} \bm{\phi}_{s}) = \sum_{j = 1}^{r}  \phi_{s,j}(\bs) \alpha_{j} \equiv \bm{\phi}_{s}(\bs)^{\prime}\bm{\alpha};\hspace{5pt} \bs \in D_{s},
\end{eqnarray}
\noindent where $\emph{r}$ is a fixed and ``known'' integer, the $r$-dimensional vector of eigenfunctions is given by $\bm{\phi}_{s}(\cdot)\equiv (\phi_{s,1}(\cdot),...,\phi_{s,r}(\cdot))^{\prime}$, and the associated $r$-dimensional random vector is $\bm{\alpha}\equiv (\alpha_{1},...,\alpha_{r})^{\prime}$.  It is important to note that $Y_{s}(\bs; \hspace{1pt} \bm{\phi}_{s}) \neq Y_{s}(\bs)$ in general due to the truncation in (\ref{approxZ}).  
%
\\
\indent Now, (\ref{aggY}) and (\ref{approxZ}) provide an immediate expression for $Y_{A}$, namely,
\begin{eqnarray}\label{approxZA}
Y_{A}(A; \hspace{1pt} \bm{\phi}_{s}) = \sum_{j = 1}^{r}  \frac{1}{|A|}\left\lbrace \int_{A} \phi_{s,j}(\bs)d\bs \right\rbrace \hspace{1pt}\alpha_{j} \equiv \bm{\phi}(A; \hspace{1pt} \bm{\phi}_{s})^{\prime}\bm{\alpha};\hspace{5pt} A \in D_{A},
\end{eqnarray}
\noindent
where $\bm{\phi}(A; \hspace{1pt} \bm{\phi}_{s})\equiv \left(\frac{1}{|A|}\int_{A} \phi_{s,j}(\bs)d\bs: j = 1,...,r\right)^{\prime}$. Then, (\ref{multiscale:process}), (\ref{approxZ}), and (\ref{approxZA}) imply the following expression for the truncated K-L expansion of the multiscale spatial process $Y$,
\begin{eqnarray}\label{multiscale:process:kl}
Y(\bu; \hspace{1pt} \bm{\phi}_{s}) =
\left\{
	\begin{array}{ll}
		\bm{\phi}_{s}(\bu)^{\prime}\bm{\alpha} & \mbox{if } \bu \in D_{s} \\
		\bm{\phi}(\bu; \hspace{1pt}\bm{\phi}_{s})^{\prime}\bm{\alpha}  & \mbox{if }  \bu \in D_{A};\hspace{5pt}\bs \in D_{s}\cup D_{A},
	\end{array}
\right.
\end{eqnarray}
\noindent where it is important to note that the $r$-dimensional random vector $\bm{\alpha}$ is the same for both supports. Validity of the implied covariance function for $Y$ follows immediately from the quadratic form (see Supplemental Materials for more details).\\
\indent The distributional assumptions governing Propositions 1$\--$3 were very general (see Remark 2). For the truncated multiscale K-L expansion we incorporate additional distributional assumptions. In particular, we assume the following:
\begin{eqnarray}\label{additive}
Z(\bu)\vert Y(\cdot), \bm{\theta}_{D}\ind \mathrm{Normal}\left\lbrace Y(\bu), \sigma_{Z}^{2}(\bu)\right\rbrace;\hspace{5pt} \bu \in D_{s}\cup D_{A},
\end{eqnarray}
\noindent where $\sigma_{Z}^{2}(\bu)>0$, and
\begin{eqnarray}\label{nu}
Y(\bu) = \mu + Y(\bu; \hspace{1pt} \bm{\phi}_{s}) + \delta(\bu; \hspace{1pt}\bm{\xi});\hspace{5pt} \bu \in D_{s} \cup D_{A},
\end{eqnarray}
\noindent is the unknown process of interest. In principal, one could easily adopt the generalized linear mixed effects model framework and replace the normal distribution in (\ref{additive}) with the appropriate probability density function from the exponential class of distributions. For example, if $Z(\cdot)$ is count-valued than one might let $Z(\bu)\vert Y(\bu), \bm{\theta}_{D}$ be distributed as Poisson with the log link. \\
\indent The unknown real value $\mu$ is interpreted as a constant ``trend term.'' Additionally, in (\ref{nu}) we assume that $\bm{\alpha}$ is an $r$-dimensional random vector with mean zero and covariance matrix $\bm{\Lambda} \equiv \mathrm{diag}(\lambda_{1},...,\lambda_{r})$. The specification of $\bm{\phi}_{s}$, the distribution of $\bm{\alpha}$, and associated prior distributions for $\bm{\phi}_{s}$ and $\bm{\Lambda}$, are stated in Section 5. It is important to note that it is typically straightforward to take an empirical Bayesian approach by directly estimating $\bm{\phi}_{s}$ and $\bm{\Lambda}$ instead of placing prior distributions on these unknown quantities. \\
\indent The $\delta$ process represents ``fine-scale variability.'' We adopt the models for $\delta$ used in \citet{WileBerliner} and \citet{bradleywikleholan}. That is, let $\bm{\xi}\equiv (\xi_{j}: j = 1,...,n_{B})^{\prime}$ consist of i.i.d. random variables with mean zero and variance $\sigma_{\xi}^{2}$, and let
\begin{eqnarray}
\delta(\bs;\bm{\xi})= \xi_{j},\label{finescale}
\end{eqnarray}
\nonumber
for any $\bs \in D_{s}$ such that $\bs$ is in the $j$-th areal unit in $D_{B}$. Thus, $\delta(B_{j}; \bm{\xi}) = (1/|B_{j}|)\int_{B_{j}}\delta(\bs;\bm{\xi})d\bs = \xi_{j}$ for $B_{j} \in D_{B}$. In general, (\ref{finescale}) implies that the fine-scale variability term is constant within each of the $j = 1,...,n_{B}$ areal units in $D_{B}$ (with the respective value $\xi_{j}$). The specification of the distribution of $\bm{\xi}$ and a prior for $\sigma_{\xi}^{2}$ shall also be given in Section 5.
\subsection{CAGE for the Truncated Karhunen-Lo\'{e}ve Expansion} \label{Sec32}It is not immediate that Proposition 2 (which motivated CAGE) holds for the process $Y$ in (\ref{nu}). Thus, we provide an extension of Proposition 2 that develops the spatial aggregation error properties of $Y$ in (\ref{nu}). We formally state this result in Proposition 4.\\

\noindent
\textit{Proposition 4: Let $f$ be any real-valued function with domain $\mathbb{R}^{n_{A}}$, and $\lambda_{k}$ be strictly greater than zero for each $k = 1,...,r$. Recall that a regionalization of $D_{B}$ is given by $D_{C} = \{C_{\ell}: \ell = 1,...,n_{C}\}$ with $C_{j}\cap C_{\ell} = \emptyset$ for $j\ne \ell$, $C_{\ell} = \cup_{h \in H} B_{h}$, $H \subset \{1,...,n_{B}\}$, and $B_{h} \in D_{B}$ for $\ell = 1,...,n_{C}\le n_{B}$. Define a generic point-level support $\{\textbf{x}_{j}: j = 1,...,n_{C}\}$, such that $\textbf{x}_{j} \in B_{j} \in D_{B}$, where $B_{j}\subset C_{j}$ and $j = 1,...,n_{C}$. Let $\textbf{Y}_{s}^{(C)} \equiv  \left(Y_{s}(\textbf{x}_{j}): j = 1,...,n_{C}\right)^{\prime}$, $\textbf{Y}_{B}^{(C)} \equiv  \left(Y_{A}(B_{j}): j = 1,...,n_{C}\right)^{\prime}$, and $\textbf{Y}_{C}\equiv (Y_{A}(C): A \in D_{C})^{\prime}$. Then the following statements hold for $Y$ in (\ref{nu}):
\begin{enumerate}[label=\roman*.]
\item $\bm{\phi}_{s}(\textbf{x}_{j})= \bm{\phi}(C_{j}; \hspace{1pt}\bm{\phi}_{s})$ for $j = 1,...,n_{C}$, if and only if $f(\textbf{Y}_{s}^{(C)})$ = $f(\textbf{Y}_{C})$ almost surely.
\item $\bm{\phi}(B_{j}; \hspace{1pt}\bm{\phi}_{s})= \bm{\phi}(C_{j}; \hspace{1pt}\bm{\phi}_{s})$ for $j = 1,...,n_{A}$, if and only if $f(\textbf{Y}_{B}^{(C)})$ = $f(\textbf{Y}_{C})$ almost surely.
\item If $\bm{\phi}_{s}(\textbf{x}_{j})= \bm{\phi}(C_{j}; \hspace{1pt}\bm{\phi}_{s})$ for every $\textbf{x}_{j} \in B_{j}$ and $j$, then $f(\textbf{Y}_{B}^{(C)})$ = $f(\textbf{Y}_{C})$ almost surely.\\
\end{enumerate}
}

\noindent
\textbf{Remark 7:} For the process $Y$ in (\ref{nu}) to have no spatial aggregation error on $D_{C}$ we (again) require between scale homogeneity of the eigenfunctions. There are two key differences between Propositions~2 and 4. The first difference is that Proposition 4 can be seen as an extension of Proposition 2 from the multiscale K-L expansion in (\ref{kl}) to the truncated process $Y$ in (\ref{nu}). The second difference is that Proposition 4 can be seen as a discretized version of Proposition 2. That is, Proposition 2 allows $B_{j}$ to be any subset of $A_{j}$, and Proposition 4 requires $B_{j}$ to be defined on the (discrete) areal support $D_{B}$.\\

\noindent
\textbf{Remark 8:} The choice to set $r<\infty$ is intimately related to the concept of spatial aggregation error. It is well known that predictors based on spatial basis functions with $r$-large display more fine-level details than predictors based on spatial basis functions with $r$-small \citep{steinr,bradleycompare}. Thus, if $r$ is chosen to be ``too small'' then predictions of $Y_{s}$ will have less variability over $D_{s}$ (i.e., be more constant), and consequently the differences between $Y_{s}$ and $Y_{A}$ (or CAGE; see Proposition $3.i$) will be smaller than they should be. We strongly recommend performing an in-depth sensitivity analysis to choose $r$ when using CAGE. To investigate the consequences of choosing $r$ ``too small'' we provide a small sensitivity study in the Supplemental Materials. Additionally, in the Supplemental Materials we provide a sensitivity analysis for the choice of $r$ for the application in Section 5.\\

Similar to Proposition 2, we have that Proposition 4 guarantees that spatial aggregation error does not occur for the spatial process in (\ref{nu}) when \textit{a finite number} of point-level eigenfunctions are constant over each region in $D_{A}$. This leads naturally to a definition of CAGE for the spatial process in (\ref{nu}):
\begin{eqnarray}
\label{Cage_trunc}
\hspace{-5pt}\mathrm{CAGE}(A)\hspace{-8pt} &\equiv& \hspace{-8pt} E\left[\int_{A}\frac{\left\lbrace\bm{\phi}_{s}(\bs) - \bm{\phi}(A; \hspace{1pt}\bm{\phi}_{s})\right\rbrace^{\prime}\bm{\Lambda}\left\lbrace\bm{\phi}_{s}(\bs) - \bm{\phi}(A; \hspace{1pt}\bm{\phi}_{s})\right\rbrace}{|A|} d\bs \vert \bz\right]\\
\hspace{-100pt}\mathrm{DCAGE}(C)\hspace{-8pt}&\equiv& \hspace{-8pt} E\left[\underset{h\in H}{\sum}\frac{\left\lbrace\bm{\phi}(B_{h};\hspace{1pt}\bm{\phi}_{s}) - \bm{\phi}(C;\hspace{1pt}\bm{\phi}_{s})\right\rbrace^{\prime}\bm{\Lambda}\left\lbrace\bm{\phi}(B_{h};\hspace{1pt}\bm{\phi}_{s}) - \bm{\phi}(C;\hspace{1pt}\bm{\phi}_{s})\right\rbrace}{|C|} \vert \bz\right]\hspace{-4pt},\label{Cage_trunc2}
\end{eqnarray}
\noindent
where $A$ is a generic areal unit (i.e., $A \subset D_{s}$), $\bm{\Lambda} \equiv \mathrm{diag}(\lambda_{i}: i = 1,...,r)$, $C = \cup_{h \in H} B_{h}$, $H \subset \{1,...,n_{B}\}$, $B_{h} \in D_{B}$ for each $h\in H$, and the expectation is taken with respect to the posterior distribution derived from (\ref{additive}) and (\ref{nu}). Notice that (\ref{Cage_trunc}) and (\ref{Cage_trunc2}) are the truncated versions of CAGE and DCAGE in (\ref{gammafunc}) and (\ref{gammafunc2}), respectively. In a similar manner a truncated version of Proposition 3 exists. We state and prove this result in Supplemental Materials.

%

\subsection{A Two-Stage Regionalization Algorithm}\label{Sec33}
The $\mathrm{CAGE}(A)$ measure allows us to evaluate whether or not the generic areal unit $A$ has poor spatial aggregation properties. However, it is not immediately clear how it can be used to specify an optimal spatial support. We now describe the use of CAGE to explicitly obtain an optimal regionalization. Recall that $D_{B}$ is the finest level aggregate support on which we wish to predict. In general, our approach is to consider many different regionalizations (combinations) of elements of $D_{B}$ and select from among them the support that produces the smallest average CAGE. By ``regionalizations of $D_{B}$'' we mean a generic set $D_{C} \equiv \{C_{\ell}: \ell = 1,...,n_{\ell}\}$, where $C_{j}\cap C_{\ell} = \emptyset$ for $j\ne \ell$ and for each $\ell$, $C_{\ell} = \cup_{h \in H} B_{h}$, $H \subset \{1,...,n_{B}\}$, and $B_{h} \in D_{B}$. 

A greedy search algorithm that seeks the minimum of the average CAGE (i.e., $\sum_{\ell = 1}^{n_{\ell}}\mathrm{CAGE}(C_{\ell})/n_{\ell}$) poses a considerable computational challenge (see \citet{Spielman1} for related discussion). To address this computational issue we use a two stage search algorithm. In the first stage, a naive clustering algorithm is applied to each of the $M$ samples of $\by_{B}$ from $[\by_{B}\vert \bz]$, denoted $\by_{B}^{[m]},$ for $m = 1,...,M$. For example, we could apply a $k$-means algorithm to $\by_{B}^{[m]}$ to define a set $D_{C}^{(k)}(\by_{B}^{[m]})\equiv \{C_{\ell}^{[m]}: \ell = 1,...,k\}$, where $C_{\ell}^{[m]}$ is the $\ell$-th cluster returned by the $k$-means algorithm. The superscript ``$(k)$'' denotes the number of areal units in $D_{C}^{(k)}$, and we keep track of the dependence of the $m$-th replicate $\by_{B}^{[m]}$. In this article, we consider using the $k$-means algorithm. We set the input of the $k$-means algorithm to be the centroids of the areal units in $D_{B}$ and $\by_{B}^{[m]}$. In the Supplemental Materials we also consider {\it structural hierarchical clustering} (\textit{SHC}) \citep{shm} in place of $k$-means. The choice of clustering algorithm depends on the application. In settings where computation is of particular interest $k$-means is preferable over structural hierarchical clustering. However, structural hierarchical clustering allows one to incorporate neighborhood information to obtain contiguous areal units, which is a preferred regionalization in some applications.

The first stage of our algorithm defines a collection of ``candidate'' spatial supports
\begin{eqnarray}
\label{candidates}
\mathcal{C} = \{D_{C}^{(k)}(\by_{B}^{[m]}): k = g_{L},...,g_{U}; m = 1,...,M\}. 
\end{eqnarray}
\noindent
Here, $g_{L}$ ($g_{U}$) represents the smallest (largest) number of areal units one is willing to consider, and both $g_{L}$ and $g_{U}$ must be pre-specified. Notice that there are a total of $M \times (g_{U} - g_{L}+1)$ spatial supports in $\mathcal{C}$, which is considerably fewer than the total number of possible candidate spatial supports to chose from.

In the second stage of the search algorithm we find the best (i.e., smallest average CAGE) subset of $\mathcal{C}$. To do this, we compute
\begin{eqnarray}
\label{select}
D_{C}^{op} = \underset{D_{C}^{(k)}(\by_{B}^{[m]})\in \mathcal{C}}{\mathrm{arg}\hspace{5pt}\mathrm{min}}\left[\frac{1}{k}\sum_{\ell = 1}^{k}\mathrm{CAGE}(C_{\ell}^{[m]})\right],
\end{eqnarray}
\noindent
where $D_{C}^{op}\equiv \{C_{j}^{op}: j = 1,...,n_{C}^{op}\}$ and $C_{k}^{op}\subset \mathbb{R}^{d}$ for $k = 1,...,n_{C}^{op}$. It should be noted that $D_{C}^{op}$, by definition, is optimal since it is obtained by minimizing error. However, one might obtain a smaller value for the average CAGE by optimizing over a different set than $\mathcal{C}$. Furthermore, one has to determine for their application whether or not it is appropriate to use CAGE or DCAGE in (\ref{select}); that is, in the case where the process is not interpretable on $D_{s}$ then one should replace CAGE in (\ref{select}) with DCAGE. A step-by-step presentation of the regionalization procedure is provided in the Supplemental Materials.

\section{A Class of Multiscale Eigenfunctions} \label{ocbf} 

Propositions 2 and 4 show that between scale differences in the eigenfunctions indicate that spatial aggregation error is present. Thus, the importance of the eigenfunctions for quantifying spatial aggregation error suggests that it should be parameterized. This will allow us to estimate eigenfunctions, and hence, CAGE can be informed by the data. Below, we discuss the construction of what we call Obled-Creutin (O-C) eigenfunctions as a weighted combination of generic GBFs. We then discuss the properties of these basis functions. 

\subsection{Obled-Creutin Eigenfunctions} \label{Sec41}It has become common to express spatial random processes in terms of a basis expansion on random effects. As such, there are many possible choices for basis functions \citep{wikleHandbook,bradleycompare}. The insight provided by \citet{obled-creutin} is that one can use {\it any} of these classes of point-level spatial basis functions to build an eigenfunction. We define an Obled-Creutin (O-C) eigenfunction as any real-valued function on $D_{s}$ that takes the following form:
\begin{eqnarray}
\label{lowrankphi}
\phi_{k}^{\mathrm{OC}}(\bs; \hspace{1pt} \textbf{F}) &\equiv& \sum_{i = 1}^{r}\psi_{i}(\bs) F_{ik}; \hspace{5pt} \bs \in D_{s}, k = 1,...,r,
\end{eqnarray}
\noindent
where $\textbf{F}$ is an $r\times r$ matrix with $(i,k)$-th element given by the real value weight $F_{ik}$, and the $r$-dimensional vector $\bm{\psi}(\cdot)\equiv \left\lbrace\psi_{1}(\cdot),...,\psi_{r}(\cdot)\right\rbrace^{\prime}$, with $\psi_{i}(\cdot): D_{s} \rightarrow \mathbb{R}$ for $i = 1,...,r$, corresponds to the aforementioned GBF basis vectors. One can organize the O-C eigenfunctions into the $r$-dimensional vector, $\bm{\phi}_{s}^{\mathrm{OC}}(\cdot; \hspace{1pt} \textbf{F})\equiv (\phi_{1}^{\mathrm{OC}}(\cdot; \hspace{1pt} \textbf{F}),...,\phi_{r}^{\mathrm{OC}}(\cdot; \hspace{1pt} \textbf{F}))^{\prime}$, which we call an Obled-Creutin (O-C) vector. 

It is not necessarily true that $Y(\cdot; \hspace{1pt}\bm{\phi}_{s}^{\mathrm{OC}})$ in (\ref{approxZ}) leads to a multiscale truncated K-L expansion. In Proposition 5 below, we specify the condition such that $Y(\cdot; \hspace{1pt}\bm{\phi}_{s}^{\mathrm{OC}})$ admits a multiscale truncated K-L expansion.\\

\noindent
\textit{Proposition 5: Let $Y\left\lbrace \cdot; \hspace{1pt}\bm{\phi}_{s}^{\mathrm{OC}}(\cdot; \hspace{1pt} \textbf{F})\right\rbrace$ be the multiscale spatial process defined in (\ref{multiscale:process:kl}), where $\lambda_{j}\ge 0$ and $>0$ for at least one $j = 1,...,r$. Here, $\psi_{1}(\cdot),...,\psi_{r}(\cdot)$ are $r$ real-valued functions with domain $D_{s}$. Additionally, let $\textbf{F}$ be an invertible $r\times r$ real$\--$valued matrix. If $\textbf{F}^{\hspace{1pt}\prime}\textbf{W}\textbf{F} = \textbf{I}$ then $Y\left\lbrace \cdot; \hspace{1pt}\bm{\phi}_{s}^{\mathrm{OC}}(\cdot; \hspace{1pt} \textbf{F})\right\rbrace$ admits a multiscale truncated K-L expansion, where $\textbf{I}$ is an $r\times r$ identity matrix and we define the $(i,j)$-th element of the $r\times r$ matrix $\textbf{W}$ as $W_{ij}\equiv \int_{D_{s}}\psi_{i}(\bs)\psi_{j}(\bs)d\bs$.}\\
 
 \noindent
 \textbf{Remark 9:} Proposition 5 is crucial for implementing the two-stage regionalization algorithm. That is, with a given GBF (i.e., radial basis functions, Fourier basis functions, wavelets, etc.) one can construct eigenfunctions, which can then be used within the two-stage regionalization algorithm from Section 3.3. There are many choices of GBFs available in the literature (e.g., \citet{bradleycs2014}), and in Section 5 we use the local bisquare functions from \citet{johan}. In the Supplemental Materials, we also consider using Wendland basis functions \citep{wendland}.

 \subsection{Specification of the O-C Weight Matrix, \textbf{F}} \label{Sec42} We capitalize on the fact that the $r\times r$ matrix $\textbf{F}$ is unknown. Estimating $\textbf{F}$ will allow the data to inform the value of CAGE. However, Proposition 5 suggests that one needs to specify $\textbf{F}$ with care; specifically, we require $\textbf{F}^{\prime}\textbf{W}\textbf{F} = \textbf{I}$ to ensure that $Y_{s}(\cdot \hspace{5pt}; \hspace{1pt} \bm{\phi}_{s}^{\mathrm{OC}})$ is a multiscale truncated K-L expansion. We achieve this by introducing a novel class of $\textbf{F}$ matrices. This contribution is formally stated in Proposition~6.\\

\noindent
\textit{Proposition 6: For a given $r$-dimensional vector of basis functions $\bm{\psi}$ let $\textbf{W}$ be positive definite. Let $\textbf{G}$ be an $r\times r$ real-valued orthogonal matrix. Then,
\begin{eqnarray}\label{solution}
\textbf{F}(\textbf{G}) \equiv \textbf{P}_{\mathrm{W}}\bm{\Lambda}_{\mathrm{W}}^{-1/2}\textbf{G},
\end{eqnarray}
\noindent
satisfies $\textbf{F}(\textbf{G})^{\prime}\textbf{W}\textbf{F}(\textbf{G}) =\textbf{I}$, where $\textbf{P}_{\mathrm{W}}\bm{\Lambda}_{\mathrm{W}}^{-1/2}$ is the Cholesky square root of the matrix $\textbf{W}^{-1}$.}

\noindent
\textbf{Remark 10:} For a given set of spatial basis functions $\{\psi_{i}\}$ we suggest verifying that $\textbf{W}$ is positive definite. Then from (\ref{multiscale:process:kl}), (\ref{lowrankphi}), and (\ref{solution}) one can write $Y_{s}$ as
\begin{eqnarray}\label{writtenout}
Y_{s}\left[\cdot; \hspace{1pt} \bm{\phi}_{s}^{\mathrm{OC}}\left\lbrace\cdot; \hspace{1pt}\textbf{F}(\textbf{G})\right\rbrace\right] = \bm{\phi}_{s}^{\mathrm{OC}}\left\lbrace\cdot; \hspace{1pt} \textbf{F}(\textbf{G})\right\rbrace^{\prime} \bm{\alpha} = \bm{\psi}(\cdot)' \textbf{F}(\textbf{G}) \bm{\alpha} = \bm{\psi}(\cdot)'
 \textbf{P}_{\mathrm{W}}\bm{\Lambda}_{\mathrm{W}}^{-1/2}\textbf{G} \bm{\alpha},
\end{eqnarray}
\noindent
where $\bm{\alpha}$ has mean-zero and $r\times r$ covariance matrix $\bm{\Lambda}$. If a closed form expression for $\textbf{W}$ is not available then numerical integration or direct Monte Carlo sampling can easily be applied to approximate $\textbf{W}$. In the case of the latter, one can randomly generate $n_w$ points $\{\bs_{k}: k = 1,...,n_w\} \subset D_{s}$ using a uniform distribution on $D_{s}$, and approximate $W_{im}$ with $(1/{n_w}) \sum_{k = 1}^{n_w}|D_{s}|\psi_{i}(\bs_{k})\psi_{m}(\bs_{k})$ \\ 
\indent In our Bayesian implementation given in Section 5, we use the following equivalent reparameterized expression of $Y_{s}\left[\cdot; \hspace{1pt} \bm{\phi}_{s}^{\mathrm{OC}}\left\lbrace\cdot; \hspace{1pt}\textbf{F}(\textbf{G})\right\rbrace\right]$ derived from the representation of $Y_{s}$ in (\ref{writtenout}):
\begin{eqnarray}\label{alternativerep}
Y_{s}\left[\cdot; \hspace{1pt} \bm{\phi}_{s}^{\mathrm{OC}}\left\lbrace\cdot; \hspace{1pt}\textbf{F}(\textbf{G})\right\rbrace\right] =\bm{\psi}^{*}(\bu)^{\prime}\bm{\eta}; \hspace{5pt} \bu \in D_{s}\cup D_{A},
\end{eqnarray}
where $\bm{\psi}^{*}(\bs)^{\prime} \equiv \bm{\psi}(\bs)^{\prime}\textbf{P}_{\mathrm{W}}\bm{\Lambda}_{\mathrm{W}}^{-1/2}$ for $\bs \in D_{s}$, $\bm{\psi}^{*}(A)^{\prime} \equiv \frac{1}{|A|}\int_{A}\bm{\psi}(\bs)^{\prime}d\bs\hspace{5pt}\textbf{P}_{\mathrm{W}}\bm{\Lambda}_{\mathrm{W}}^{-1/2}$ for $A \in D_{A}$, and $\bm{\eta}$ $(\equiv \textbf{G}\bm{\alpha})$ has mean zero and $r\times r$ covariance matrix $\textbf{Q} \equiv \textbf{G}\bm{\Lambda}\textbf{G}^{\prime}$. Additionally, we assume that $\textbf{Q}$ consists of random parameters that can be sampled. (For each application we undergo independent sensitivity analyses to select a prior distribution. For details behind the prior specification, and for related empirical results, see Supplemental Materials.) Then, it is straightforward to obtain samples of $\textbf{Q}$ and $\bm{\eta}$, respectively, via a MCMC algorithm. Note that if a closed form expression for $\frac{1}{|A|}\int_{A}\bm{\psi}(\bs)^{\prime}d\bs$ is not available then numerical integration or direct Monte Carlo sampling can easily be applied to obtain an approximation. In the case of the latter, one can randomly generate $n_w$ points $\{\bs_{k}: k = 1,...,n_w\} \subset A\subset D_{s}$ using a uniform distribution on $A$, and approximate $\frac{1}{|A|}\int_{A}\bm{\psi}(\bs)^{\prime}d\bs$ with $(1/{n_w}) \sum_{k = 1}^{n_w}\bm{\psi}(\bs_{k})^{\prime}$. {In general, we have found that the value of $n_{w}$ needs to be large for these approximations to be reasonable (in Section 5 we set $n_{w} = 20,000$).}\\
\indent Additionally, one can obtain samples of the eigenfunction $\bm{\phi}_{s}^{\mathrm{OC}}\left\lbrace\cdot; \hspace{1pt}\textbf{F}(\textbf{G}^{[m]})\right\rbrace$ to use within the expression of CAGE in (\ref{gammafunc}). That is, denote the $m$-th replicate of $\textbf{Q}$ with $\textbf{Q}^{[m]}$, and let the corresponding spectral decomposition be written as $\textbf{Q}^{[m]}=\textbf{G}^{[m]}\bm{\Lambda}_{\mathrm{Q}}^{[m]}\textbf{G}^{[m]\prime}$. Then, the corresponding $m$-th replicate of $\bm{\phi}_{s}^{\mathrm{OC}}\left\lbrace\cdot; \hspace{1pt}\textbf{F}(\textbf{G}^{[m]})\right\rbrace$ is given by
\begin{eqnarray}\label{OCclass}
\bm{\phi}_{s}^{\mathrm{OC}}\left\lbrace\cdot; \hspace{1pt}\textbf{F}(\textbf{G}^{[m]})\right\rbrace = \bm{\psi}^{*}(\cdot)^{\prime}\textbf{G}^{[m]}; \hspace{5pt} m = 1,...,M.
\end{eqnarray}
\noindent
We shall henceforth use the representation of $Y_{s}\left[\cdot; \hspace{1pt} \bm{\phi}_{s}^{\mathrm{OC}}\left\lbrace\cdot; \hspace{1pt}\textbf{F}(\textbf{G})\right\rbrace\right] $ in (\ref{alternativerep}), and the O-C eigenfunction $\bm{\phi}_{s}^{\mathrm{OC}}\left\lbrace\cdot; \hspace{1pt}\textbf{F}(\textbf{G}^{[m]})\right\rbrace$ in (\ref{OCclass}).

\section{Application: Median Household Income from the American Community Survey} \label{Sec5}

We revisit the ACS 5-year period estimates of median household income for 2013 presented in Figure~1. This data can be downloaded at \url{http://factfinder2.census.gov/}. This is an important example because there has been a growing interest in regionalizing data from ACS \citep{Spielman1,Spielman2}. \\
\indent For this example, $D_{s}^{O} = \emptyset$, and $D_{A}^{O} = D_{A}$ consists of the $n$ = 3,109 counties in the continental US. Since US counties are the finest spatial resolution of the dataset in Figure 1, we set $D_{B} = D_{A}$. Let $[Z(\cdot)\vert Y(\cdot)]$ be a normal probability density function with mean $Y(\cdot)$ and known variance $\sigma_{Z}(\cdot)>0$, which are computed from margin of error estimates that are publicly available. Here, $Z(\cdot)$ is the log median household income, and we let $Y(\cdot)$ be distributed according to (\ref{nu}). \\
\indent Both $\bm{\alpha}$ and $\bm{\xi}$ are assumed to be Gaussian, and we perform regionalization using $\bm{\phi}_{s}^{\mathrm{OC}}(\cdot; \hspace{1pt} \bm{\psi})$,  where $\bm{\psi}(\cdot)\equiv (\psi_{j}(\cdot): j = 1,...,75)^{\prime}$ is a $75$-dimensional vector consists of local bisquare functions \citep{johan}:
\begin{eqnarray}\label{cht3.bi}
\psi_{j}(\bs) \equiv
\left\lbrace
	\begin{array}{ll}
		\{1 - (||\bs - \textbf{c}_{j}||/w)^{2}\}^{2}  & \mbox{if } ||\bs - \textbf{c}_{j}|| \le w \\
		0  & \mathrm{otherwise};\hspace{5pt} \bs \in D_{s},
	\end{array}
\right.
\end{eqnarray}
\noindent
 with $j = 1,...,75$ equally spaced knots $\textbf{c}_{j}$, and where $w$ is 1.5 times the smallest distance between two different knots. The placement of knots was achieved using a space filling design \citep{NychkaSpaceFill}. We performed empirical studies that explore the relationship between $r$ and $n_{C}^{op}$ (see discussion in Remark 8). These investigations suggest that $r = 75$ is appropriate for this example. (From our experience, our method is rather robust to the placement and number of knots, and the empirical results guiding this experience are provided in the Supplemental Materials.) 
        \begin{figure}[t!]
         \begin{center}
         \begin{tabular}{cc}
            \includegraphics[width=7.5cm,height=5.8cm]{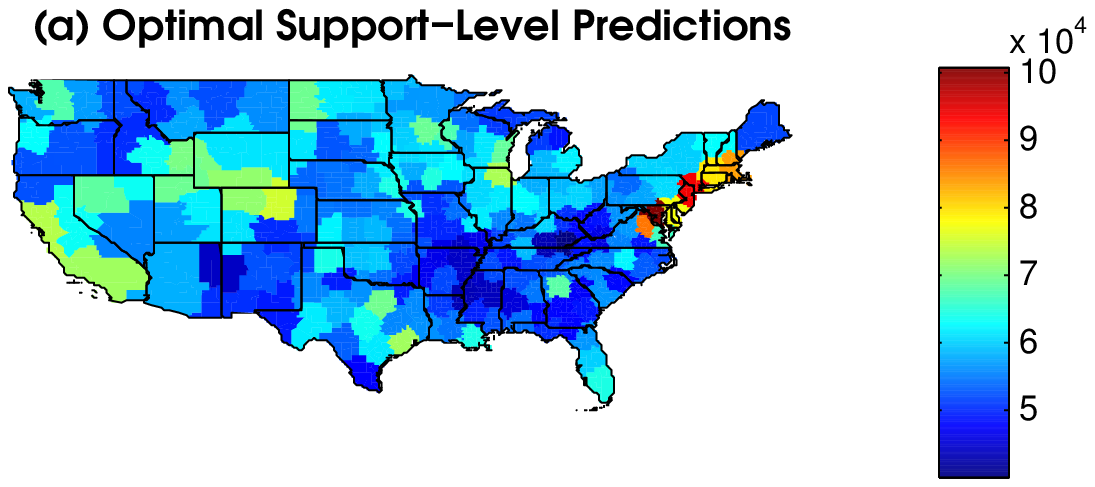}&
            \includegraphics[width=7.5cm,height=5.8cm]{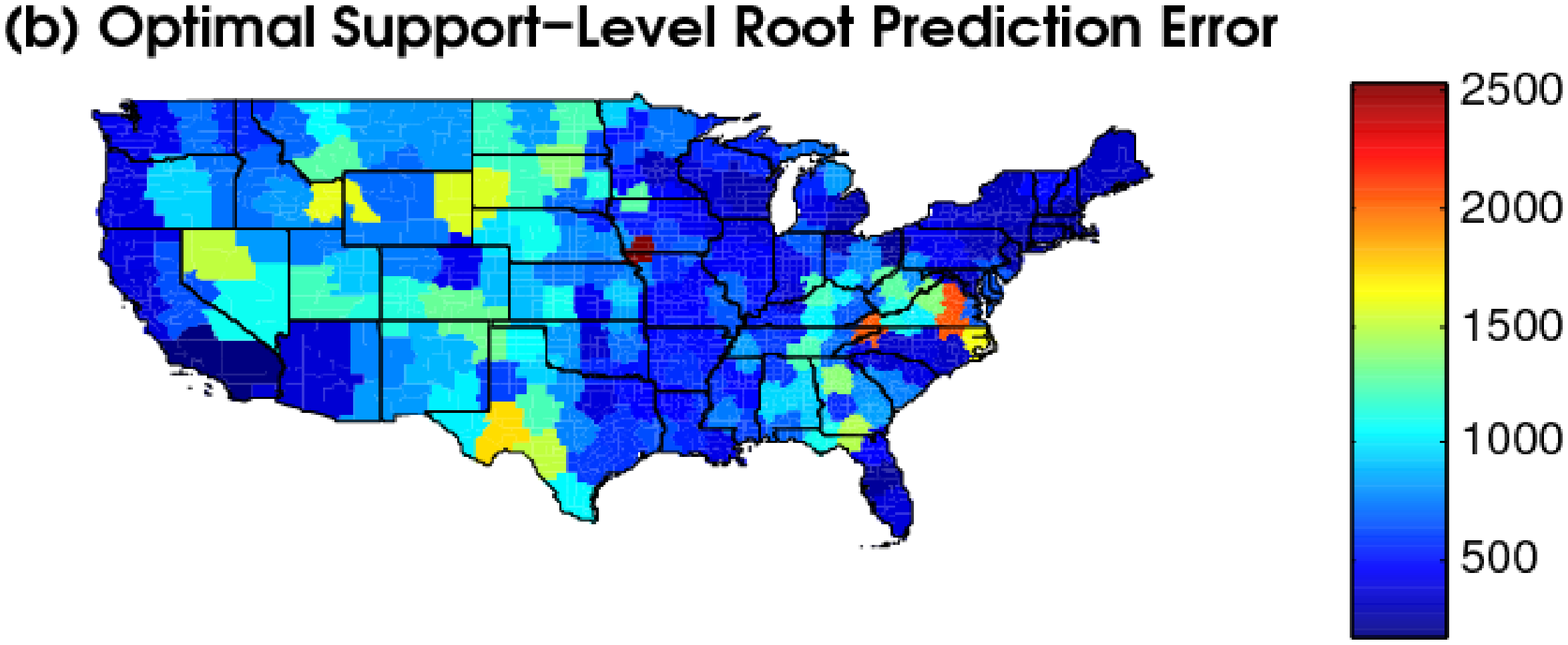}\\
            \includegraphics[width=7.5cm,height=5.8cm]{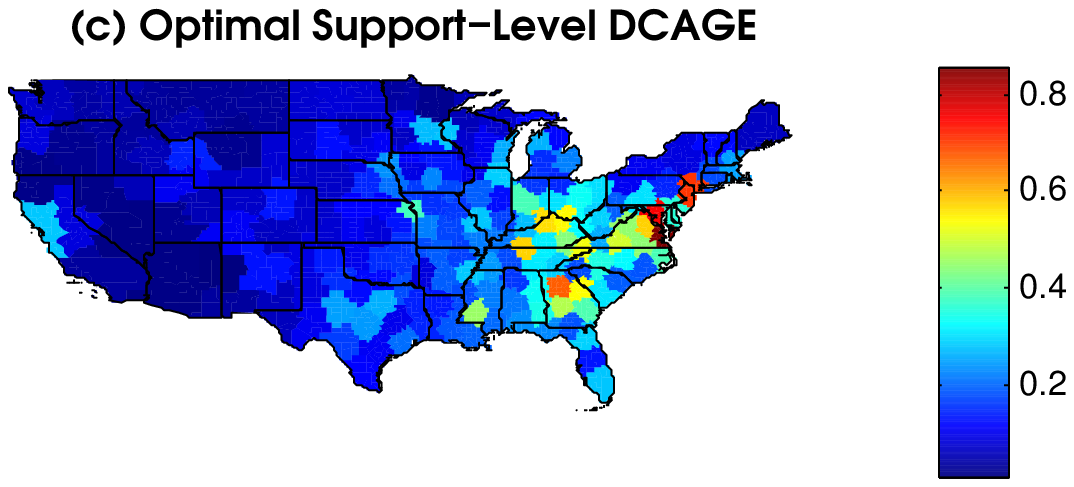}&
            \includegraphics[width=7.5cm,height=5.8cm]{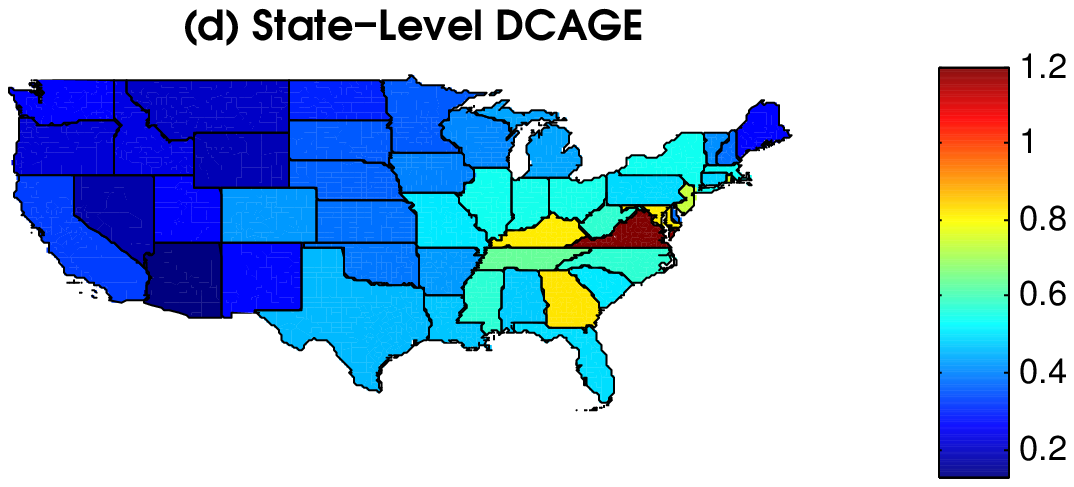}
         \end{tabular}
         \caption{\baselineskip=10pt In (a), we present maps (for the contiguous US) of predicted median household income ({US dollars}) defined on the optimal spatial support (i.e., $D_{C}^{op}$) consisting of 185 areal units. Recall, we consider areal units 175 through 195, and the value chosen using DCAGE is 185. We superimpose the state boundaries as a reference to compare to Figure 1(b). In (b) and (c), we present maps of the posterior standard deviations and DCAGE. In (d), we plot DCAGE by states.}\vspace{-20pt}
         \end{center}
         \end{figure}
 We considered many different choices of prior distributions for the $r\times r$ covariance matrix $\textbf{Q}$, and through independent sensitivity studies we found that the so-called MI prior \citep{bradleymst} appeared to be the most appropriate choice for this example (see the Supplemental Materials for more details). The $k$-means algorithm is used to define $\mathcal{C}$ in (\ref{candidates}), and we let $g_{L} = 175$ and $g_{U} = 195$. Since the latent field is not interpretable on $D_{s}$, we use DCAGE within the expression of $D_{C}^{op}$ in (\ref{select}). The variances of $\{\epsilon(A_{i}): i = 1,...,n\}$ are estimated \textit{a priori} by ACS, and hence, are assumed known.\\
\indent In Figure 2(a) {and 2(b)}, we present the predictions {and corresponding prediction error} of median household income on the optimal spatial support $D_{C}^{op}$ (and add state boundaries as a reference). {In Figure 2(b), the predictions appear fairly precise with largest prediction error occurring in regions near Virginia, which have posterior standard deviation around $2,500$  (which is roughly 5$\%$ of the mean median household income).} The problems with spatial aggregation error indicated by Figures 1(a) and 1(b) described in the Introduction are no longer present in $D_{C}^{op}$, which consists of 185 areal units. For example, counties near Richmond constitute a distinct region. Also, the state of New York is divided into multiple distinct regions: areas near and in Manhattan, western New York, and upstate New York are all separated. {However, it is worth noting that in Figure 2(c) the square root DCAGE values are comparatively larger around the state of Virginia}.\\
\indent The DCAGE can also be used for uncertainty quantification. That is, state-level representatives may not be interested in the optimal regionalization produced by the two stage search algorithm, and instead, be interested in the median income over states. The DCAGE can be used to identify which states have poor spatial aggregation error properties. In Figure 2(d), we plot DCAGE over states (i.e., treat states as fixed areal units), which has an average DCAGE of 0.24. This value is larger than the average DCAGE corresponding to the optimal solution, which is 0.19. Notice that the DCAGE corresponding to Virginia (and states near Virginia) are relatively high, while other states in the Midwest and West coast have comparatively smaller values of DCAGE. This would suggest that one should be concerned about assuming that statistics over Virginia can interpreted at lower spatial resolutions.

\section{Discussion} \label{Sec6}The ecological fallacy and MAUP have become popular pedagogical tools for discussion in geography and spatial statistics \citep{Robinson,openshawTaylor,cressie,cressie-wikle-book,banerjee-etal-2015}. However, very little has been done to characterize and mitigate these forms of spatial aggregation error from a statistical perspective. Thus, we provide a measure to formally characterize such error and a principled way to obtain an optimal (in terms of spatial aggregation error) regionalization defined over the generic continuous domain $D_{s} \subset \mathbb{R}^{d}$. Regionalization has traditionally been solved using techniques outside the realm of statistics \citep{Duque,Spielman1,Folch,Spielman2}, and our work offers a new perspective that respects the uncertainty of spatial random processes. Consequently, our methodology can significantly impact federal statistics, survey methodology, geography, spatial statistics, and remote sensing/data acquisition settings.\\
\indent The heart of our methodology lies in the criterion for spatial aggregation error (CAGE), which we minimize to obtain our optimal regionalization. The methodological development of CAGE is intricate and involves a novel multiscale Karhunen-Lo\`{e}ve (K-L) expansion. The introduction of a multiscale K-L expansion provides an approach to spatial COS that is not based on assumptions of between scale homogeneity. Furthermore, the multiscale K-L expansion leads to a powerful technical result that shows that any statistic does not suffer from spatial aggregation error as long as the multiscale eigenfunctions are homogeneous across scales. Thus, CAGE represents a measure of between scale homogeneity of eigenfunctions within a multiscale K-L expansion. There are many additional motivating features of CAGE, including connections to prediction error and across scale homogeneity of variances.\\
\indent To apply CAGE we need a parameterization of the multiscale eigenfunctions. This allows the eigenfunctions to be estimated, and hence, the CAGE can be informed by the data. Thus, we provide a new class of Obled-Creutin (O-C) eigenfunctions motivated by the seminal paper of \citet{obled-creutin}. The proposed class of O-C eigenfunctions has broad applicability in the sense that any class of generating basis functions (GBF) can be used to build eigenfunctions.\\
\indent Finally, CAGE is used within an efficient two-stage regionalization algorithm. In
the first stage of the algorithm (for a given number of areal units) a deterministic
clustering algorithm is applied to each of the M samples from the posterior distribution of the latent process. This
defines M spatial supports to select from. Then, in the second stage, the spatial
support with the smallest (average) CAGE is chosen. This approach is extremely
efficient, and accounts for the variability of the data by performing the search algorithm within the latent process space.
%
\\
\indent An illustration of our algorithm was given using American Community Survey (ACS) 5-year period estimates of median household income. Comparisons of the optimal spatial support to the state-level ACS estimates indicate that the optimal regionalization preserves the county-level spatial information. Additionally, the size of this dataset is 3,109, and notably, the optimal spatial support consists of just 185 areal units. The dramatic decrease of the dimensionality of the problem has important implications for modeling very large spatial datasets.\\
\indent The application of CAGE to reduce the dimensionality of spatial data is just one of many exciting avenues for future research. For example, the introduction of spatially varying covariates into the statistical model will undoubtedly effect the spatial aggregation error properties. Also, as previously mentioned, model selection considerations, such as the number of basis functions and class of basis functions, may effect the conclusions of the two-stage regionalization algorithm. The truncation of the multiscale K-L expansion is especially important from the point of view of regionalization, since fewer basis functions lead to less variable predictions of the latent process, which then leads to fewer areal units produced by the regionalization algorithm. Another interesting idea for future research would be to construct a prior distribution for the regionalization by using the values of the CAGE to define prior weights.\\
\indent There are minor modifications to CAGE and the two-stage regionalization algorithm that would be reasonable to consider. For example, Proposition~2 shows that spatial aggregation error does not occur when point-level eigenfunctions are constant over each region in the aggregate-level spatial support. Thus, we use the squared distance between point-level and aggregate-level eigenfunctions to measure departures from the absence of spatial aggregation error. However, other distances besides the squared distance might be used. This is similar to considering other forms of prediction error besides squared error. Also, there are a number of alternative search algorithms that one might consider. For example, one could use CAGE within a forward selection algorithm, or perhaps, one might use \citet{Spielman2}'s ACS regionalization (AReg) algorithm within the first stage of the two-stage algorithm. It would be difficult to incorporate AReg into the two-stage algorithm practically, since it is not computationally efficient for high-dimensional spatial datasets. The specifications we use are computationally efficient and are shown to give favorable results.

\section*{Acknowledgments} \label{ack}This research was partially supported by the U.S. National Science Foundation (NSF) and the U.S. Census Bureau under NSF grant SES-1132031, funded through the NSF-Census Research Network (NCRN) program. In addition, C.K. Wikle acknowledges the support of NSF grant DMS-1049093 and Office of Naval Research (ONR) grant ONR-N00014-10-0518.

\newpage
\part*{}
\thispagestyle{empty} \baselineskip=28pt

\begin{center}
{\LARGE{\bf Supplemental Materials: Regionalization of Multiscale Spatial Processes using a Criterion for Spatial Aggregation Error}}
\end{center}

\baselineskip=12pt

\vskip 2mm
\begin{center}
Jonathan R. Bradley\footnote{(\baselineskip=10pt to whom correspondence should be addressed) Department of Statistics, University of Missouri, 146 Middlebush Hall, Columbia, MO 65211, bradleyjr@missouri.edu},
Christopher K. Wikle\footnote{\baselineskip=10pt  Department of Statistics, University of Missouri, 146 Middlebush Hall, Columbia, MO 65211-6100},
Scott H. Holan$^2$
\end{center}
%
%
%
%
\vskip 4mm

\baselineskip=12pt

%
%
%

\baselineskip=12pt
\par\vfill\noindent
{\bf Keywords:} American Community Survey; Empirical orthogonal functions; MAUP; Reduced rank; Spatial basis functions; Survey data
\par\medskip\noindent
\clearpage\pagebreak\newpage \pagenumbering{arabic}
\baselineskip=24pt

\section*{I Introduction} In this supplement to ``Regionalization of Multiscale Spatial Processes using a Criterion for Spatial Aggregation Error,'' by J.R. Bradley, C.K. Wikle, and S.H. Holan, we give additional insight to CAGE and the two-stage regionalization algorithm outside of what was presented in the main text. In particular, we have applied the algorithm to another dataset, performed many different sensitivity analyses, and provided additional material that is meant to aid readers interested in implementing our procedure.

This supplement is organized as follows. In Section~II, we provide guidance on the implementation of our algorithm including: a summary of the statistical model used in Section~5, details on prior distribution considerations, a step-by-step outline of estimation and the two-stage regionalization procedure, and additional discussion on model and regionalization specifications. Note, we use Roman numerals for section titles in this Supplement to distinguish from section titles in the main text. In Section~III, we provide sensitivity analyses including: a comparison to a current state-of-art method for regionalization within the geography literature from \citet{speilman}, a sensitivity analysis to the choice of $D_{A}$, and a simulation study investigating the choice of the rank of the truncated multiscale K-L expansion. Next, in Section~IV we provide a demonstration of the two-stage regionalization algorithm to a dataset consisting of Mediterranean wind measurements (a subset of the data used in \citet{millif}), which is used illustrate that the two-stage regionalization algorithm is flexible enough to handle multiscale spatial data. Finally, in Section~V we provide the proofs to the technical results from the main-text.

\section*{II Additional Details for Implementation} 
\renewcommand{\theequation}{2.\arabic{equation}}
\setcounter{equation}{0}
Here, we give guidance on the implementation of our algorithm including: a summary of the statistical model used in Section~5 (Section~II.i), details on prior distribution considerations (Section~II.ii), a step-by-step outline of the estimation and the two-stage regionalization procedure (Section~II.iii), and additional discussions on model and regionalization specification (Section~II.iv).

\subsection*{II.i Outline of the Statistical Model} The statistical model introduced in Section~3.1 is summarized in Algorithm~1 below. We choose to describe this Bayesian hierarchical model using the data, process, and parameter model terminology from \citet{berlinhier}.\\

\begin{algorithm}[H]
\caption{Outline of the statistical model introduced in Section~3.1}
\begin{align}
\nonumber
&\mathrm{Data\hspace{5pt}Model:}\hspace{5pt}Z(\bu)\vert \mu, \bm{\eta},\textbf{Q},\bm{\xi}\ind \mathrm{Normal}\left\lbrace\mu + \bm{\psi}^{*}(\bu)^{\prime}\bm{\eta} + \delta(\bu;\hspace{1pt}\bm{\xi}), \sigma_{Z}^{2}(\bu)\right\rbrace;\\
\nonumber
&\mathrm{Process\hspace{5pt}Model\hspace{5pt}1:}\hspace{5pt} \bm{\eta}\vert \textbf{Q}\sim \mathrm{Gaussian}\left(\bm{0}, \textbf{Q}\right);\\
\nonumber
&\mathrm{Process\hspace{5pt}Model\hspace{5pt}2:}\hspace{5pt} \bm{\xi}\vert \sigma_{\xi}^{2} \sim \mathrm{Gaussian}\left(\bm{0}, \sigma_{\xi}^{2}\textbf{I}_{n_{B}}\right);\\
\nonumber
&\mathrm{Parameter\hspace{5pt}Model\hspace{5pt}1:}\hspace{5pt} \mu \sim \mathrm{Normal}\left(0, \sigma_{\mu}^{2}\right);\\
\nonumber
&\mathrm{Parameter\hspace{5pt}Model\hspace{5pt}2:}\hspace{5pt} \sigma_{\xi}^{2} \sim \mathrm{IG}\left(\alpha_{\xi}, \beta_{\xi}\right);\\
\nonumber
&\mathrm{Parameter\hspace{5pt}Model\hspace{5pt}3:}\hspace{5pt}  \textbf{Q}\sim [\textbf{Q}];\hspace{5pt} \bu \in D_{s}\cup D_{A}.
\end{align}
\end{algorithm} 
\vspace{10pt}
\noindent
Here, the $n_{B}$-dimensional random vector $\bm{\xi}\equiv \left(\xi_{1},..., \xi_{n_{B}}\right)^{\prime}$, $\sigma_{\mu}^{2}>0$, $\alpha_{\xi}>0$, $\beta_{\xi}>0$, and we let $[\textbf{Q}]$ denote a probability density function for the unknown $r\times r$ covariance matrix $\textbf{Q}$. We consider many different choices for $[\textbf{Q}]$, and provide these details in Section~II.ii. The value of $\sigma_{\mu}^{2}$ is chosen to be large so that the prior distribution on $\mu$ is interpreted to be vague, and similarly, we set $\alpha_{\xi}=\beta_{\xi} = 1$ so that the prior distribution on $\sigma_{\xi}^{2}$ is flat.

\subsection*{II.ii Prior Distributions to Consider} 
\renewcommand{\theequation}{2.\arabic{equation}}
\setcounter{equation}{0}
As \citet{ruehyper} discuss, the prior distribution (and the associated hyperparameters) on the $r\times r$ covariance matrix $\textbf{Q}$ affects posterior inference. As such, we consider {several} different choices for priors on covariance matrices. In particular, we consider three different prior distributions. The first prior distribution we consider is the conjugate inverse Wishart distribution. This is a fairly common choice because it allows for direct sampling of the full-conditional distribution corresponding to $\textbf{Q}$, however, in high-dimensions this prior is known to perform poorly \citep{hodges}. 

The second prior distribution we consider is from \citet{bradleyMSME} and \citet{bradleymst}, where it is assumed that
\begin{equation}\label{targ}
\textbf{Q} = \frac{1}{\sigma^{2}}\left[\textbf{R}_{B}^{-1}\mathcal{A}^{+}\left\lbrace\textbf{Q}_{B}^{\prime}\left(\textbf{I} - \textbf{A}\right)\textbf{Q}_{B}\right\rbrace \textbf{R}_{B}^{-1}\right]^{-1},
\end{equation}
\noindent
where $\mathcal{A}^{+}(\textbf{M})$ is the best positive approximate \citep{Higham} of a square real-valued matrix $\textbf{M}$, $\sigma^{2}>0$ is unknown, the $n_{B}\times r$ matrix $\bm{\Psi}_{B}\equiv \left(\bm{\psi}(B)^{\prime}: B\in D_{B}\right)^{\prime}$, $\bm{\Psi}_{B} = \textbf{Q}_{B}\textbf{R}_{B}$ is the QR decomposition, and $\textbf{A}$ is the $n_{B}\times n_{B}$ adjacency matrix corresponding to $D_{B}$. Notice that (\ref{targ}) incorporates spatial information, but is not spatially referenced. That is, this prior for $\textbf{Q}$ is motivated by specifying $\mathrm{cov}(\bm{\Psi}_{B}\bm{\eta})$ so that it is ``close'' to the covariance from an ICAR model on $D_{B}$, where $\bm{\Psi}_{B}$ is spatially referenced but $\bm{\eta}$ is not. An inverse gamma prior is placed on $\sigma^{2}$ where the hyperparameters are chosen based on the suggestions in Section~3.2 of \citet{ruehyper}. Following \citet{bradleyMSME} and \citet{bradleymst}, we refer to this prior specification as the ``MI'' prior distribution due to a connection to the Moran's I statistic.

The third prior distribution we consider is the Givens angle prior \citep{bergerga,bradleywikleholan}, where the spectral decomposition is written as $\textbf{Q} = \textbf{P}_{\mathrm{Q}}\bm{\Lambda}_{Q}\textbf{P}_{\mathrm{Q}}$, and the $r\times r$ diagonal matrix $\bm{\Lambda}_{Q}$ has diagonal entries set equal to the eigenvalues of (\ref{targ}). The parameter $\sigma^{2}$ is assumed to follow a flat inverse gamma distribution (i.e., with shape and scale set equal to 1). The $r\times r$ orthogonal matrix $\textbf{P}_{\mathrm{Q}}$ is decomposed into a Givens rotator product,
\begin{align*}
\textbf{P}_{\mathrm{Q}} \equiv (\bm{O}_{1,2} \times \bm{O}_{1,3}\times\cdots \times \bm{O}_{1,r})\times (\bm{O}_{2,3} \times \cdots \times \bm{O}_{2,r})\times\cdots\times \bm{O}_{r-1,r},
\end{align*} 
\noindent
where $\bm{O}_{i,j}$ is a $r\times r$ identity matrix with the $(i,i)$-th and $(j,j)$-th element replaced by $\mathrm{cos}(\theta_{i,j})$ and the $(i,j)$-th ($(j,i)$-th) element replaced by $-\mathrm{sin}(\theta_{i,j})$ ($\mathrm{sin}(\theta_{i,j})$). Here, $\theta_{i,j} \in [-\pi/2, \pi/2]$ is unknown, and let the shifted and rescaled $\theta_{i,j}$ be denoted as $\zeta_{i,j} \equiv 1/2 + \theta_{i,j}/\pi$. Then, it is assumed that
\begin{equation}\label{logit}
\mathrm{logit}(\zeta_{i,j}) = a + b \times g_{i,j}(\textbf{P}_{\mathrm{Q}});\hspace{4pt}i<j = 1,...,r,
\end{equation}
\noindent
where $\mathrm{logit}(\zeta_{i,j}) \equiv \mathrm{log}\{\zeta_{i,j}/(1-\zeta_{i,j})\}$, $a,b \in \mathbb{R}$, and $g_{i,j}(\textbf{P}_{\mathrm{Q}})$ represents the $(i,j)$-th Givens angle of $\textbf{P}_{\mathrm{Q}}$. Finally, a vague Gaussian prior is placed on $(a,b)^{\prime}$ (i.e., Gaussian with mean zero and variance 1000). For all of our analyses we considered all three prior distributions. These sensitivity analyses suggested that the MI prior lead to the best predictive performance for the application in Section~5, and the inverse Wishart prior led to the best predictive performance in Section~V.



\subsection*{II.iii Outline: Estimation and Implementation of Regionalization}In this section we give a brief outline of the two-stage regionalization algorithm. It should be acknowledged that, for any given application, minor modifications to these steps may be needed.
\begin{enumerate}
\item Define the spatial support $D_{B}$, which represents the finest resolution one is willing to predict on. If $D_{s}^{O} = \emptyset$ we suggest setting $D_{B} = D_{A}$, which is the finest resolution information that is available. When $D_{s}^{O} \ne \emptyset$ then one has the freedom to choose any spatial support for $D_{B}$, however, one should be mindful of the size and spatial coverage of the locations within $D_{s}^{O}$. Thus, for illustration, when $D_{s}^{O} \ne \emptyset$ we suggest setting $D_{B}$ to a fine resolution grid.
\item Obtain $M$ MCMC replicates of $\by_{B} \equiv (Y_{A}(B): B \in D_{B})^{\prime}$, using the statistical model in Algorithm~1. Specifically, let $\bm{\eta}^{[m]}$ represent the $m$-th replicate of $\bm{\eta}$ and $\bm{\xi}^{[m]}$ represent the $m$-th replicate of $\bm{\xi}$. Then, the $m$-th replicate of $\by_{B}$ can be computed as
\begin{equation*}
\by_{B}^{[m]} = \bm{\Psi}_{B}\bm{\eta}^{[m]}+\bm{\xi}^{[m]}; \hspace{5pt}m = 1,...,M,
\end{equation*}
where the $n_{B}\times r$ matrix $\bm{\Psi}_{B} \equiv \left(\bm{\psi}^{*}(\bu)^{\prime}: \bu \in D_{B}\right)^{\prime}$. The Bayesian procedure can easily implemented using a Metropolis with in Gibbs sampling algorithm.
\item Use a naive clustering algorithm to obtain $\mathcal{C}$ in (19). We consider two clustering algorithms to define $\mathcal{C}$, namely, the $k$-means algorithm, and structural hierarchical clustering. In general, the $k$-means algorithm takes on as it's argument an $n_{B}\times f$ real-valued matrix $\textbf{J}$, and returns a clustering of the rows of $\textbf{J}$. Let $\textbf{L}$ be a $n_{B}\times d$ matrix with the $j$-th row equaling the centroid of the $j$-th areal unit in $D_{B}$. Then, we let $f = d+1$ and set $\textbf{J} = [\textbf{L},\hspace{5pt}\by_{B}^{[m]}]$. The structural hierarchical clustering approach takes on two arguments $\textbf{J}= [\textbf{L},\hspace{5pt}\by_{B}^{[m]}]$ and the adjacency matrix corresponding to $D_{B}$. 
\item Choose the spatial support from $\mathcal{C}$ that minimizes CAGE. That is, compute $D_{C}^{op}$ according to (20). If $Y$ can not be interpreted on $D_{s}$ substitute CAGE with DCAGE.
\item Produce maps of the values in the sets $\{\widehat{Y}_{A}(C^{op}): C^{op} \in D_{C}^{op}\}$, $\{\mathrm{var}(Y_{A}(C^{op}|\bz)): C^{op} \in D_{C}^{op}\}$, and $\{\mathrm{CAGE}(C^{op}): C^{op} \in D_{C}^{op}\}$ (or $\{\mathrm{DCAGE}(C^{op}): C^{op} \in D_{C}^{op}\}$ when appropriate). This allows one to visualize the process and its corresponding prediction and spatial aggregation errors.
\end{enumerate}

\subsection*{II.iv Model and Regionalization Algorithm Specifications} To implement the two-stage regionalization algorithm, we need to specify: the number and placement of knots that define the $r$-dimensional GBF $\bm{\psi}$, and the lower and upper bounds on the number of areal units used within the two-stage regionalization algorithm (i.e., $g_{L}$ and $g_{U}$). We now provide discussion on to make these choices in practice.\\

\noindent
\textbf{Specification of Knots:} The choice of knots and $r$ is important for preserving the appropriate fine-scale features of $Y_{s}$. If the fine-scale features of $Y_{s}$ are ignored then the two-stage regionalization algorithm may produce too coarse of a regionalization (see simulation study in Section~IV.iii). However, the number of areal units produced by the two-stage regionalization algorithm appears to be robust to $r$ ``too large.'' Recall the number of areal units in $D_{C}^{op}$ is denoted with $n_{C}^{op}$. This interaction between the number of optimal areal units and $r$ suggests an approach for selecting the rank $r$, which we outline into the following steps:
\renewcommand{\labelenumi}{(\arabic{enumi}) }
\begin{enumerate}
\item Consider a fixed range of values for $r$ (i.e., $r = r_{L},..., r_{U}$).
\item For each $r = r_{L},...,r_{U}$, use the algorithm outlined in II.iv to find an optimal regionalization and $n_{C}^{op}$. There will be a different value of  $n_{C}^{op}$ for each each $r = r_{L},...,r_{U}$.
\item Plot $r$ versus $n_{C}^{op}$.
\item Choose the value of $r$ to be the point in which $n_{C}^{op}$ does not change dramatically as $r$ increases. 
\end{enumerate}
We follow the suggestion of \citet[][chap. 13, pp. 255-260]{Ruppert} and apply a space filling design algorithm to a set of randomly selected points $\{\textbf{c}_{j}: j = 1,...,r^{*}\}$, where we set $r^{*} = 600>r$. The space-filling design can be determined using the FUNFITS function in R \citep{NychkaFUNFITS}. Then, we choose $r$ according to steps 1$\--$3 above. For the applications in Section~5 and Section~V we found that, respectively, {$r = 75$ and  $r = 200$} are appropriate.\\

\noindent
\textbf{Specification of $g_{L}$ and $g_{U}$:} The widest range of values that we can consider for regionalization is $g_{L} = 2$ and $g_{U} = n_{B}-1$. To specify less extreme choices for $g_{L} = 2$ and $g_{U} = n_{B}-1$ we consider running a simplified version of the two-stage regionalization algorithm, and use the results of the ``simplified two-stage regionalization algorithm'' to inform a tighter range between $g_{L}$ and $g_{U}$. In particular, we first run the two-stage regionalization algorithm (outlined in Section~II.iii) with $M = 1$, $g_{L} = 2$, $g_{U} = n-1$, and use the $k$-means algorithm. Then, we choose $g_{L}$ and $g_{U}$ to be a tight range centered around $n_{C}^{op}$ found using this simplified two-stage regionalization algorithm.

\section*{III Simulations, Sensitivity Analyses, Comparisons, and Technical Clarifications}
\renewcommand{\theequation}{3.\arabic{equation}}
\setcounter{equation}{0}
 Here, we provide many different side-studies including: a simulation study to compare the two-stage regionalization algorithm to a current state-of-the-art alternative in the geography literature, \citet{Spielman2}'s ACS regionalization (AReg) algorithm (Section~III.i); a small sensitivity analysis on the choice of $D_{A}$ (Section~III.ii); and a simulation study investigating the choice of the rank of the spatial basis function expansion (Section~III.iii).

\subsection*{III.i Simulation Study: A Comparison to \citet{speilman}} 
In this section, we establish that our approach performs regionalization extremely well relative to the AReg algorithm available in the geography literature. To do this, we generate synthetic data based on a subset of the ACS 5-year period (from 2009 to 2013) estimates of the percentage of households below the poverty threshold. We generate the spatial field,
\begin{equation}\label{dataACS}
Z(A) = Y_{A}(A) + \epsilon(A);\hspace{5pt}A \in D_{A},
\end{equation}
\noindent
where $D_{A}$ is the set of 351 census tracts surrounding the city of Austin (TX). Let $\{Z(A)\}$ represent the perturbed version of the logit transformed percent below the poverty level ACS survey estimate (denoted by $\{Y_{A}(A)\}$). (Notice that we use the symmeterizing logit transformation, where, for a given percentage $p$, logit($p$) = $p$/(1-$p$).) The set $\{\epsilon(A): A \in D_{A}\}$ consists of independent normal random variables with mean-zero and known variance. The published variances for percent below the poverty level are transformed to the logit scale using the delta method \citep{delta}, and used as the known variances of $\{\epsilon(A)\}$. In practice, the ACS  estimates (i.e., $\{Y_{A}\}$ for this example) are publicly available and are, hence, observed. Nevertheless, for the purposes of this simulation study we will act as if the ACS estimates are an unobserved spatial field to be estimated from $Z$.

To obtain $D_{C}^{op}$, we model this data using the mixed effects model in Algorithm~1, where $\bm{\psi}(\cdot)\equiv (\psi_{j}(\cdot): j = 1,...,42)^{\prime}$ is a $42$-dimensional vector consists of local bisquare functions \citep{johan}:
\begin{equation}\label{cht3.bi}
\psi_{j}(\bs) \equiv
\left\{
	\begin{array}{ll}
		\{1 - (||\bs - \textbf{c}_{j}||/w)^{2}\}^{2}  & \mbox{if } ||\bs - \textbf{c}_{j}|| \le w \\
		0  & \mathrm{otherwise};\hspace{5pt} \bs \in D_{s},
	\end{array}
\right.
\end{equation}
\noindent
 with $j = 1,...,42$ equally spaced knots $\textbf{c}_{j}$, and $w$ is 1.5 times the smallest distance between two different knots. Note, that we are not restricted to using local bisquare functions, since our modeling framework is general enough to allow for any desired GBF. For computational convenience, we use the $k$-means algorithm to define $\mathcal{C}$ in (19), and let $g_{L} = 2$ and $g_{U} = 100$. The latent process in (\ref{dataACS}) is not defined on $D_{s}$, and thus, we shall use DCAGE within the expression of $D_{C}^{op}$ in (20). Additionally, we denote the output of AReg with $D_{A}^{\mathrm{AReg}} \equiv \{A_{k}^{\mathrm{AReg}}: k = 1,...,n_{A}^{\mathrm{AReg}}\}$, and compute it using software made available at \url{https://github.com/geoss/ACS_Regionalization/blob/master/README.md}.\\
\indent The goal of this simulation study is to compare the error properties of $D_{C}^{op}$, and $D_{A}^{\mathrm{AReg}}$. This is done using the following metrics:
\begin{align}\label{metrics}
\nonumber
\mathrm{ReMSPE}(Z_{A}) &\equiv \frac{\sum_{j = 1}^{n_{A}^{\mathrm{AReg}}} \frac{1}{|A_{j}^{\mathrm{AReg}}|}\left\lbrace Y_{A}(A_{j}^{\mathrm{AReg}}) - \widehat{Y}_{A}(A_{j}^{\mathrm{AReg}})\right\rbrace^{2}}{\sum_{j = 1}^{n_{C}^{op}} \frac{1}{|C_{j}^{op}|}\left\lbrace Y_{A}(C_{j}^{op}) - \widehat{Y}_{A}(C_{j}^{op})\right\rbrace^{2}}\\
\nonumber
\mathrm{ReCAGE}(Z_{A}) &\equiv \frac{\sum_{j = 1}^{351} \sum_{k = 1}^{n_{A}^{\mathrm{AReg}}}I(A_{j} \subset A_{k}^{\mathrm{AReg}})\left[\frac{\left\lbrace Y_{A}(A_{j}) - Y_{A}(A_{k}^{\mathrm{AReg}})\right\rbrace^{2}}{|A_{k}^{\mathrm{AReg}}|}\right]}{ \sum_{j = 1}^{351} \sum_{k = 1}^{n_{C}^{op}}I(A_{j} \subset C_{k}^{op})\left[\frac{\left\lbrace Y_{A}(A_{j}) - Y_{A}(C_{k}^{op})\right\rbrace^{2}}{|C_{k}^{op}|}\right]},
\end{align}
\noindent
where $I(\cdot)$ is the indicator function. Here, ReMSPE stands for ``relative mean squared prediction error'' and ReCAGE stands for ``relative spatial aggregation error,'' respectively. Values of ReMSPE that are larger (smaller) than 1.0 indicate that prediction on $D_{C}^{op}$ has smaller (larger) MSPE than when predicting on $D_{A}^{\mathrm{AReg}}$. Thus, values of ReMSPE that are larger (smaller) than 1.0 indicate that the two-stage algorithm (AReg) leads to better (worse) predictive performance. Likewise, values of ReCAGE that are larger than 1.0 indicate that the two-stage algorithm is preferable in terms of spatial aggregation error.

        \begin{figure}[H]
        \begin{center}
        \begin{tabular}{cc}
           \includegraphics[width=6.5cm,height=6cm]{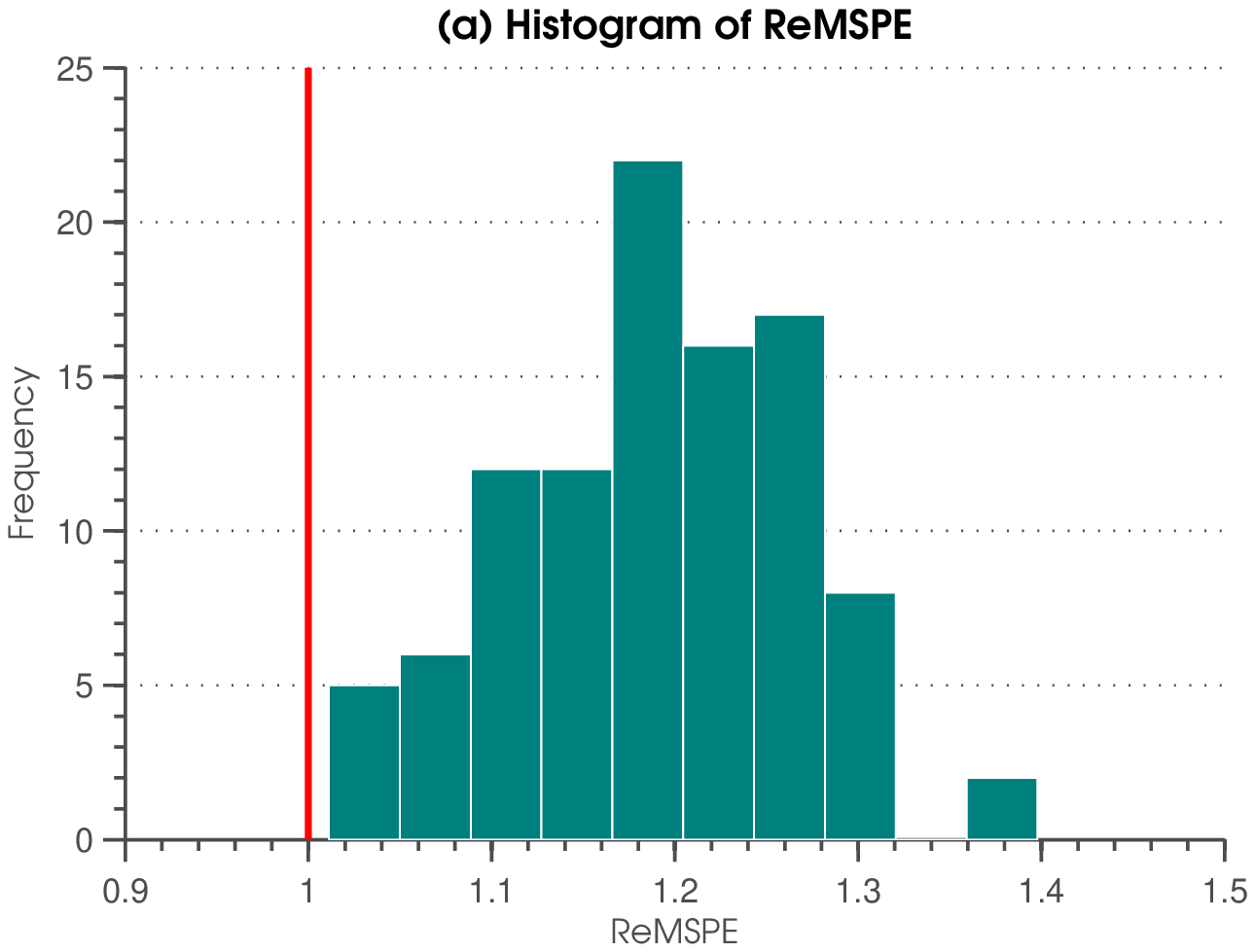}&
           \includegraphics[width=6.5cm,height=6cm]{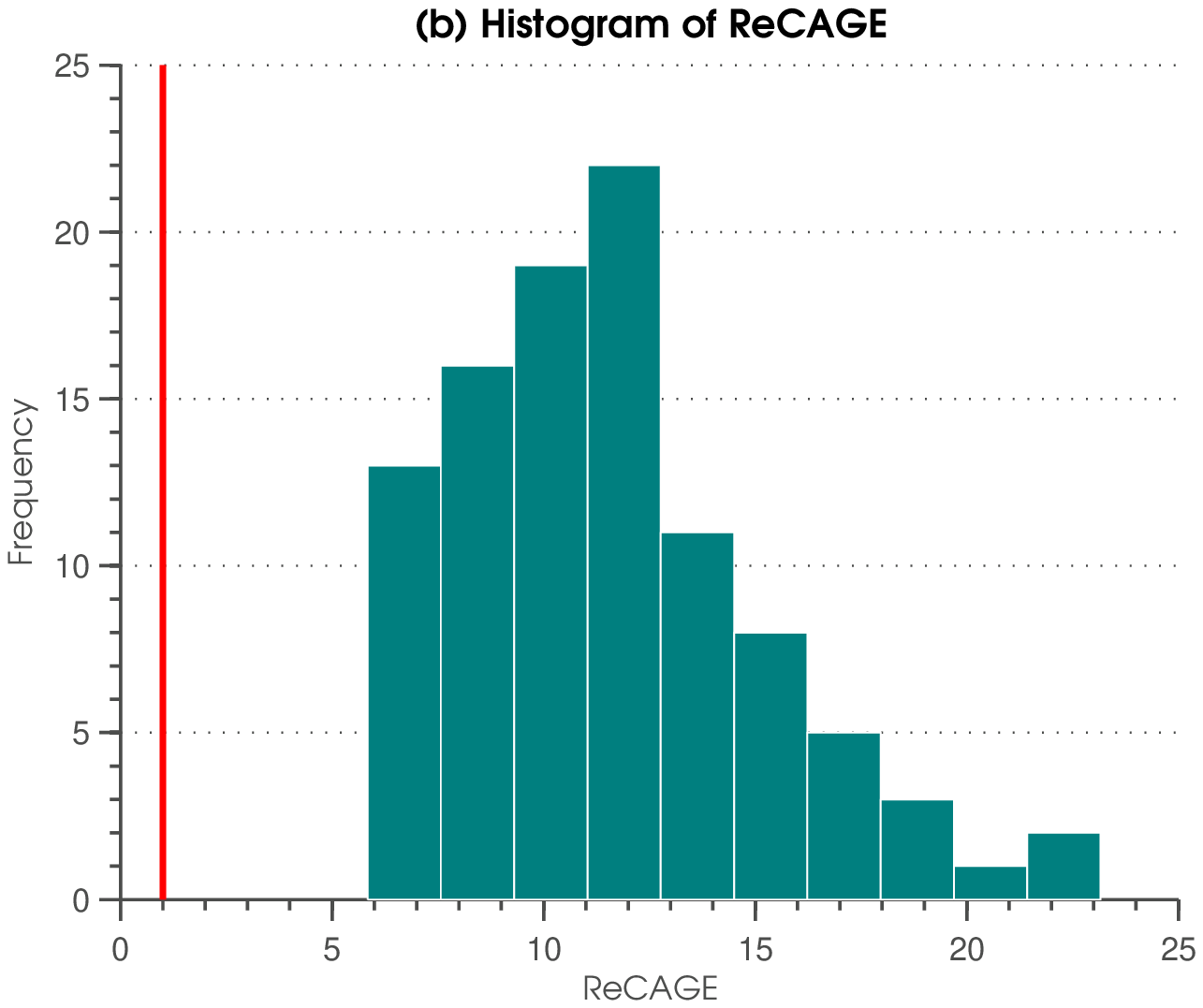}
        \end{tabular}
        \caption{\baselineskip=10pt In (a) and (b), we present histograms of ReMSPE and ReCAGE from taken over the 100 replicates of $Z$ defined in (21). The red line indicates the value of 1 in both panels. A value of ReMSPE and ReCAGE greater than 1.0 indicates that the two-stage regionalization algorithm is preferable over AReg.}
        \end{center}
        \end{figure}

We simulate 100 replicates of $Z$ in (\ref{dataACS}), and compute ReMSPE and ReCAGE for each of the 100 replicates. For both metrics our proposed algorithm consistently outperforms AReg. In fact, in each of the 100 replications of $Z$ we obtain an ReMSPE $>$ 1.0, and a ReCAGE $>$ 1.0, where ReMSPE ranges from 1.0112 to 1.3979 and ReCAGE ranges from 5.8408 to 23.1620, respectively (see Figure~1 for a histogram over the 100 replications of $Z$). It is somewhat expected that ReCAGE suggests that the two-stage regionalization algorithm is preferable over AReg because from Proposition 3, CAGE is directly related to the squared difference between the lower spatial resolution process and the aggregate-level estimator. However, it is rather interesting that ReMSPE suggests that the two stage algorithm is also preferrable in terms of squared prediction error, since AReg is motivated by reducing sampling error. This may be due to the fact that AReg does not take into account survey error (i.e., $\{\epsilon(A)\}$), while the two-stage regionalization algorithm accounts for this error by performing its search in latent space.

\subsection*{III.ii Sensitivity to $D_{A}$} Notice that the two-stage search algorithm takes on $D_{s}$ and $D_{A}$ (the spatial domains of interest) as an input. Thus, one might be interested in the sensitivity of our approach to the spatial domain of interest. For example, in Figure~2(a) we plot the optimal areal units (i.e., $D_{C}^{op}$), found in Section~5, over California, Oregon, Nevada, and Arizona. Now, suppose we let $D_{A}$ consist of the 126 counties in California, Oregon, Nevada, and Arizona, and we re-run the two stage search algorithm on this restricted domain (i.e., $D_{A}$ no longer consists of all counties in the mainland of US, but consists only of counties in California, Oregon, Nevada, and Arizona). The $D_{C}^{op}$ found under this restriction is given in Figure~2(b).

 There are 12 areal units in $D_{C}^{op}$ without restricting $D_{A}$, and 11 when one restricts $D_{A}$. Upon comparison of Figures~2(a) to 2(b) we see that the general pattern of the two-stage search algorithm is robust to this change in $D_{A}$, however, the final answer does change. We note that since the initialization of the $k$-means algorithm is random, the candidate set of areal units are not necessarily the same each time one runs the two-stage search algorithm.

        \begin{figure}
        \begin{center}
        \begin{tabular}{cc}
           \includegraphics[width=7cm,height=5cm]{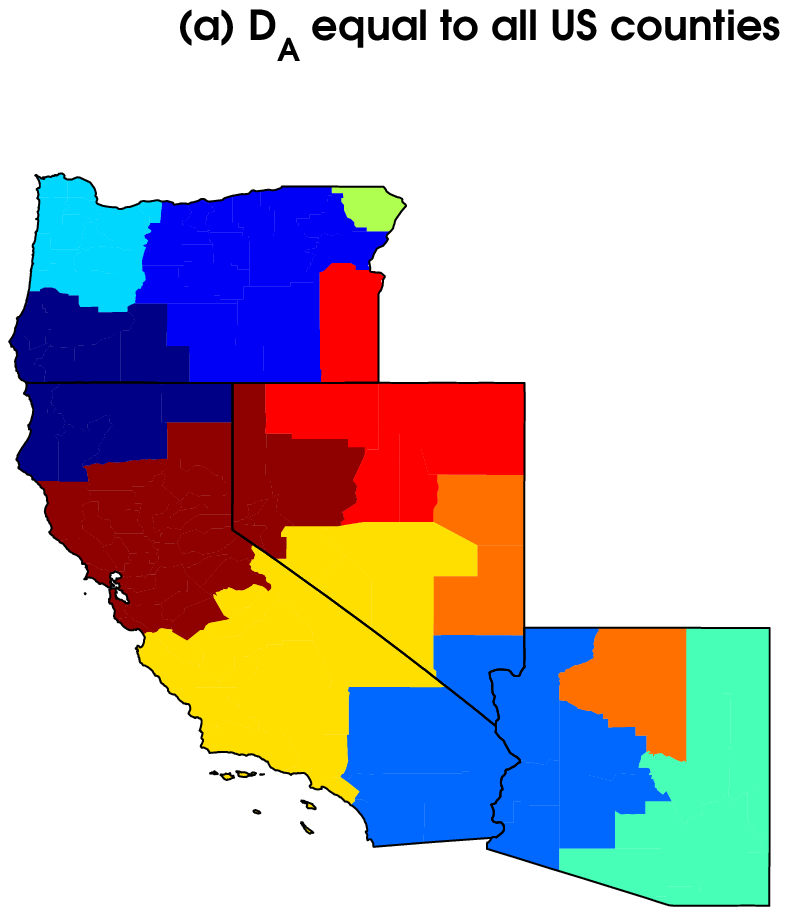}&
           \includegraphics[width=7cm,height=5cm]{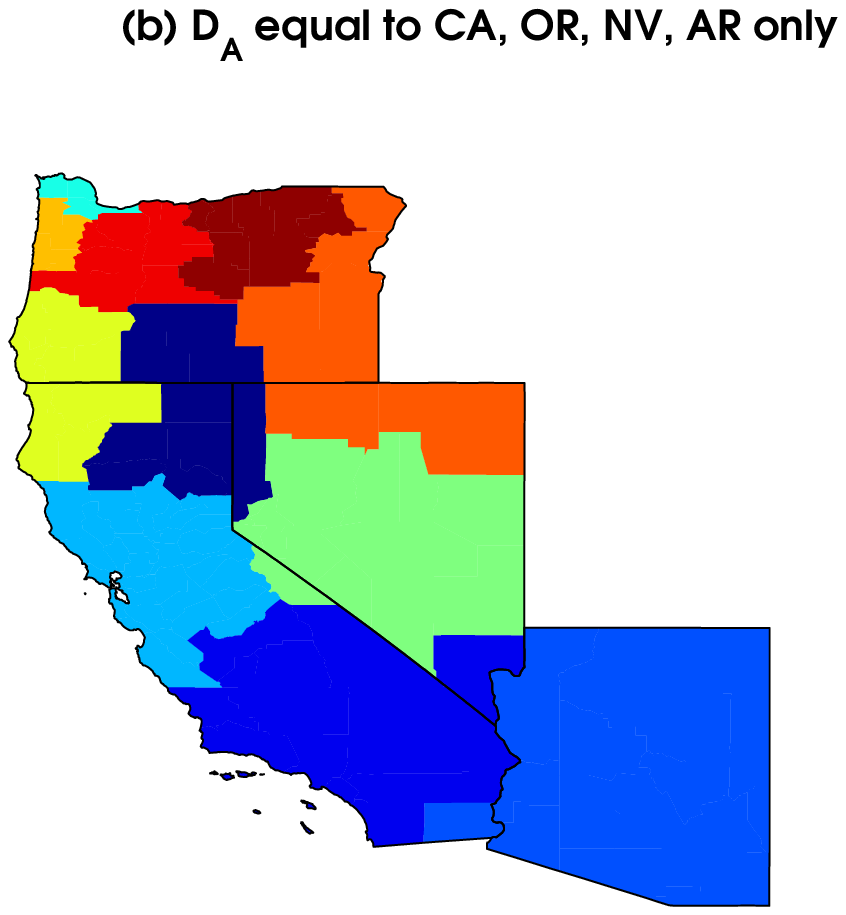}
        \end{tabular}
        \caption{\baselineskip=10pt In (a), we plot the optimal areal units (i.e., $D_{C}^{op}$), found in Section~5, over the state of California. In (b), we plot the $D_{C}^{op}$ found by restricting $D_{A}$ to consist only of counties in California, Oregon, Nevada, and Arizona. Each distinct color identifies a different areal unit, and the relative difference between each color is arbitrary. The state boundaries are superimposed as a reference.}
        \end{center}
        \end{figure}

\subsection*{III.iii Simulation Study: Selection of the Rank of the Truncated Multiscale K-L Expansion} In this section, we use simulation to investigate the impacts of misspecifying the rank of the truncated multiscale K-L expansion. In particular, we choose a simulation model with $r = 100$ random effects, and we perform regionalization with $r$ misspecified and $r$ correctly specified. The regionalization with $r$ correctly specified is treated as the ``correct'' regionalization, which we compare to.

Let the latent process of interest $Y_{s}$ be generated as follows:
\begin{eqnarray}\label{simmodel}
& Y_{s}(\bs) = \mu + Y_{s}(\bs; \bm{\phi}_{s}^{\mathrm{OC}}) + \delta(\bs;\bm{\xi}); \hspace{2pt} \bs \in D_{s},
\end{eqnarray}
\noindent
where $D_{s} \equiv \{\bs = (s_{1},s_{2})^{\prime}: s_{1}, s_{2} = [0.05, 0.1, 0.15,...,1]\times [0.05, 0.1, 0.15,...,1]\}$, recall $Y_{s}(\bs; \bm{\phi}_{s}^{\mathrm{OC}})$ is defined in (11), and let $\bm{\phi}_{s}^{\mathrm{OC}}$ be based generated from 100 equally spaced (over $D_{s}$) local bisquare basis functions. The corresponding dataset is generated as follows:
\begin{eqnarray}\label{datamodel}
\nonumber
Z_{s}(\bs) & = Y_{s}(\bs) + \epsilon_{s}(\bs); \bs \in D_{s}^{O} \subset D_{s}\\
Z_{A}(\bs) & = Y_{A}(A) + \epsilon_{A}(A); A \in D_{A},
\end{eqnarray}
      \begin{figure}[t!]
      \begin{center}
      \begin{tabular}{c}
        \includegraphics[width=15.5cm,height=9cm]{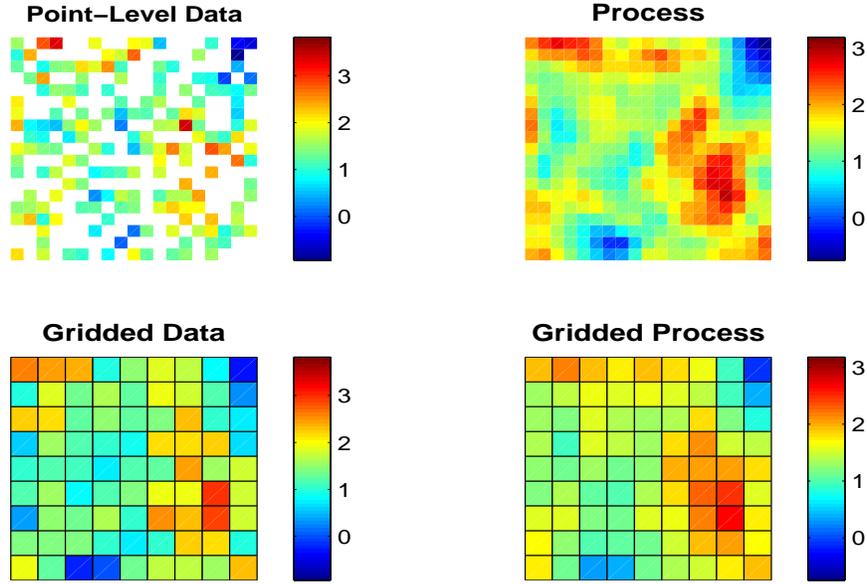}
      \end{tabular}
      \caption{Example simulated data and process. These maps are produced using (\ref{simmodel}) and (\ref{datamodel}). The top left panel contains simulated data on $D_{s}$ (with $50\%$ of the field being covered). The top right panel contains the simulated process on $D_{s}$. The bottom left panel contains the aggregate data process (i.e., $Z_{A}$), which has complete spatial coverage over $D_{A}$. The bottom right panel displays $Y_{A}$.}
      \end{center}
      \end{figure}
where we randomly select 50$\%$ of the observations from $D_{s}$ to define $D_{s}^{O}$, and $D_{A}$ consists of the $10\times 10$ grid cells that cover $[0,1]\times [0,1]$. We let $\epsilon_{s}(\cdot)$ be a mean zero white-noise process with variance $\sigma_{\epsilon}^{2} = 0.1820$ (so that the signal-to-noise ratio (=5) is large). Likewise, $\{\epsilon_{A}(A): A \in D_{A}\}$ consists of i.i.d. independent mean zero random variables with variance 0.1820, and is independent of the spatial random process $\epsilon_{s}(\cdot)$. An example of the data and the process is given in Figure~3.\\
\indent Consider performing regionalization using the outline in Section~II.iii, to the data in Figure~3 with $r = 9, 100,$ and $256$. For illustration let $D_{B} = D_{A}$, and set $g_{L}=2$ and $g_{U} = 99$ (the largest possible range). Here, $r = 9$ represents the case where $r$ is too small, $r = 100$ represents the case where $r$ is correct, and $r = 256$ represents the case when $r$ is too large. When $r$ is too small we obtain fewer areal units (6 areal units) than when $r$ is correct (13 areal units); however, the optimal regionalization algorithm is robust to the case where $r$ is too large, which produced 15 areal units. This conforms to intuition as it is well known that predictors based on spatial basis functions with $r$-large display more fine-level details than predictors based on spatial basis functions with $r$-small \citep{bradley2011,steinr,bradleycompare}. Thus, one would expect that if $r$ is chosen to be ``too small'' then predictions of $Y_{s}$ will have less variability over $D_{s}$ (i.e., be more constant), and consequently lead to coarser regionalizations.\\
\indent These conclusions are similar over multiple replications; in Figure~4 we provide histograms of $n_{C}^{op}$ obtained from the two-stage regionalization algorithm over 50 independent replications of $\{Z_{s}\}$ and $\{Z_{A}\}$. Notice, however, that the variability associated with $r$ too large is much higher than when $r$ is too small and when $r$ is correct. The $p$-value of a sign test comparing $n_{C}^{op}$ when $r = 9$ ($r = 256$), to $n_{C}^{op}$ when $r=100$ is 0.0494 (0.5716), which suggests that when $r$ is too small (too large) we obtain coarser (similar) results than when $r$ is correct. \\
\indent The fact that there is no significant change in the number of areal units when $r$ is too large also conforms to intuition; since there are enough spatial random effects to capture fine-scale behavior, and the remaining random effects are negligible. This interaction between the number of optimal areal units and $r$ suggest an approach for choosing $r$ (i.e., Steps 1$\--$3 in Section~II.iv). For the ACS application in Section~5, we consider $r = 25, 50, 75, 100, 125,$ and $150$. Likewise for the Mediterranean wind example we consider $r = 25, 50, 75, 100, 125,$ and $150$. {In Figure~5, we plot $n_{C}^{op}$ versus $r$ (i.e., Step 3 from Section~II.iv). Here, we see that for the applications in Section~5 and Section~V we found that, respectively, $r = 75$ and  $r = 200$ are appropriate.}
      \begin{figure}[t!]
      \begin{center}
      \begin{tabular}{c}
        \includegraphics[width=10.5cm,height=12cm]{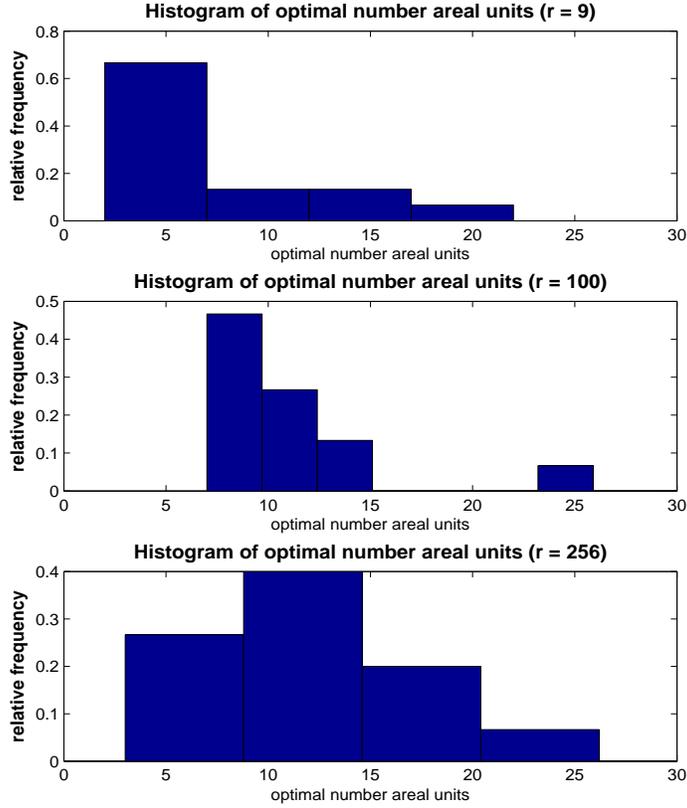}
      \end{tabular}
      \caption{Histograms of $n_{C}^{op}$ over 50 independent replications of $\{Z_{s}\}$ and $\{Z_{A}\}$. The value of $r$ used to fit Algorithm~1 is indicated in the title of the panel.}
      \end{center}
      \end{figure}
      
            \begin{figure}[t!]
            \begin{center}
            \begin{tabular}{cc}
              \includegraphics[width=6.5cm,height=5cm]{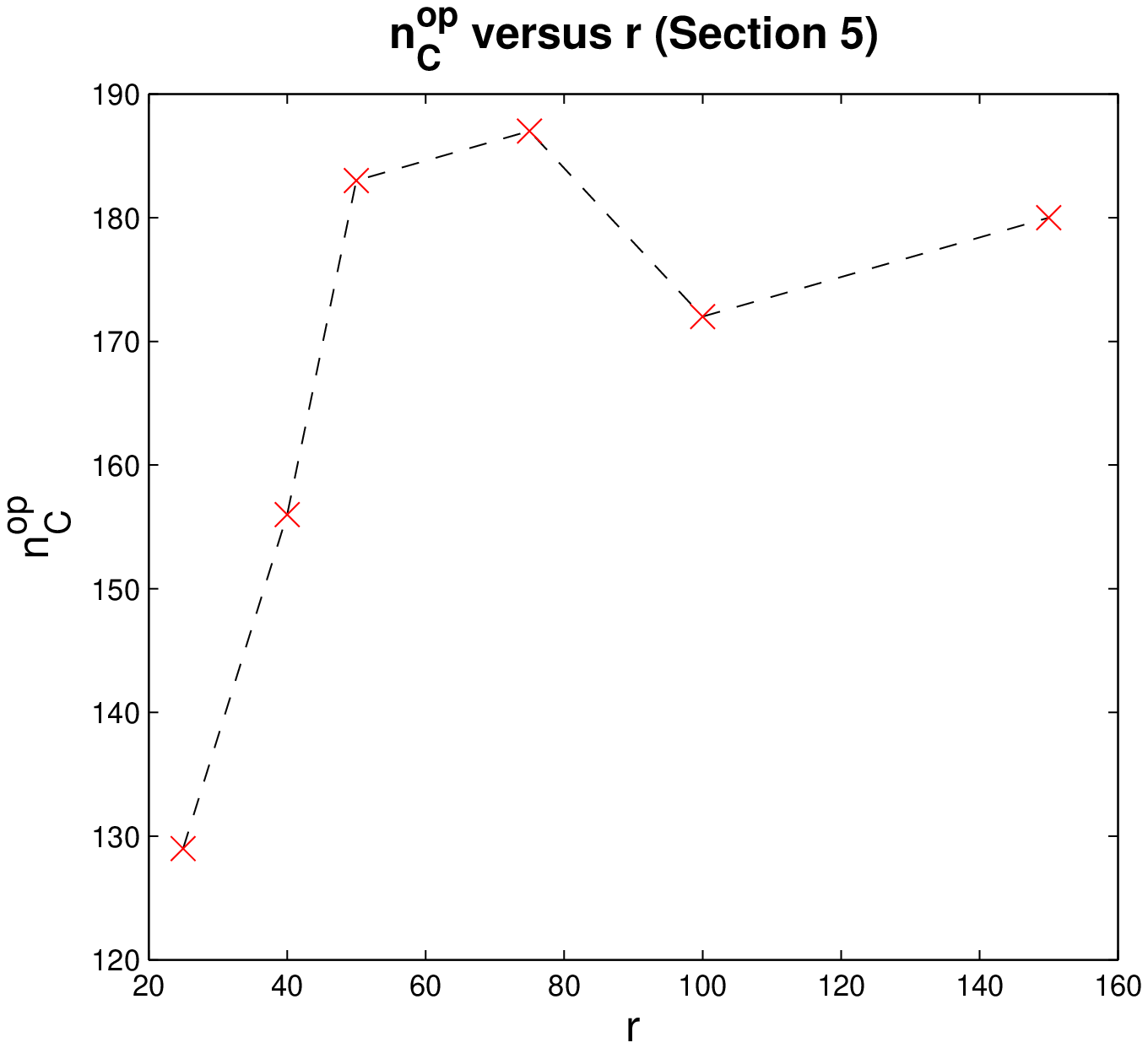}& \includegraphics[width=6.5cm,height=5cm]{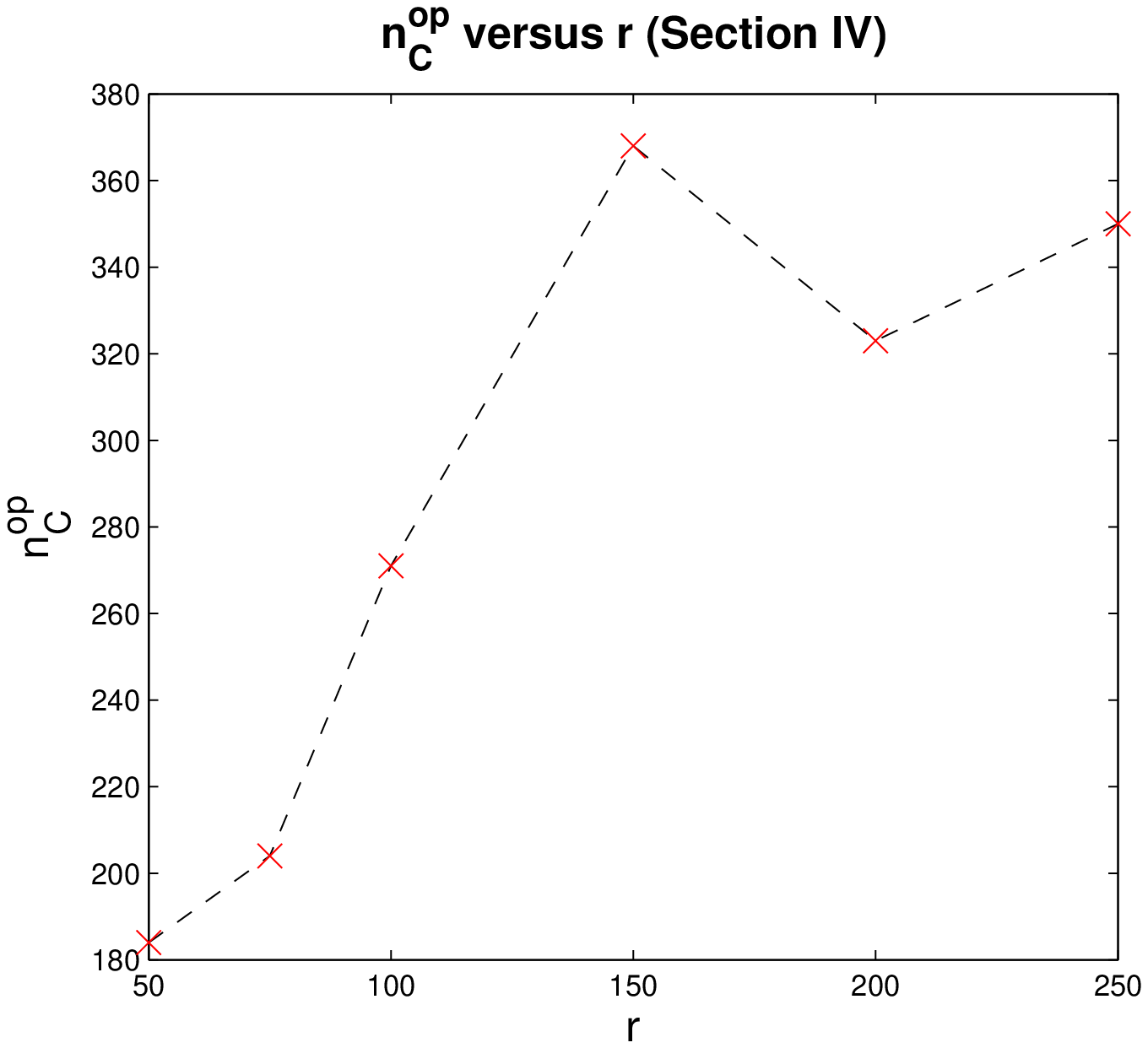}
            \end{tabular}
            \caption{The plot of $n_{C}^{op}$ versus $r$ as described in Section II.iv. In the left panel we plot $n_{C}^{op}$ versus $r$ for the ACS example presented in Section 5, and in the right panel we plot $n_{C}^{op}$ versus $r$ for the wind example in Section IV. The values of $r$ considered in the ACS example in Section 5 were 25, 40, 50, 75, 100, and 150. The values of $r$ considered in the wind example in Section IV were 50, 75, 100, 150, 200, and 250.}
            \end{center}
            \end{figure}
\subsection*{III.iv Technical Clarifications: Positive Definiteness of the Multiscale K-L Expansion} A covariance function $\mathrm{cov}\left\lbrace Y_{s}(\bs), Y_{s}(\bu)\right\rbrace$ is positive definite if \citep[][p. 68]{cressie},
\begin{equation}\label{pd}
\sum_{i = 1}^{m}\sum_{j = 1}^{m}b_{i}b_{j}\mathrm{cov}\left\lbrace Y_{s}(\bs_{i}), Y_{s}(\bs_{j}) \right\rbrace \ge 0
\end{equation}
\noindent
for \textit{any} finite number of spatial locations $\{\bs_{i}: i = 1,...,m\}$ and \textit{any} set of real numbers $\{b_{i}: i = 1,...,m\}$. That is, the covariance function, associated with the spatial random process $Y_{s}$, is positive definite if a weighted average of covariances implied by \textit{any} set $\{Y_{s}(\bs_{i}): i = 1,...,m\}$ has non-negative variance, where $\{b_{i}: i = 1,...,m\}$ are the generic weights. The validity of the covariance of $Y_{s}$ in (11) follows immediately from the definition of positive definiteness, and the quadratic form of
\begin{equation*}
\mathrm{cov}\left( \by^{(m)}\right) = \bm{\Psi}^{(m)}\bm{\Lambda}\bm{\Psi}^{(m)\prime},
\end{equation*}
\noindent
where $\bm{\Lambda}$ is defined below Equation (15) of the main text, \\
\begin{equation*}
\by^{(m)} \equiv \left\lbrace Y_{s}(\bs_{1}; \hspace{1pt} \bm{\phi}_{s}),...,Y_{s}( \bs_{m}; \hspace{1pt} \bm{\phi}_{s})\right\rbrace^{\prime},
\end{equation*}
\noindent
and
\begin{equation*}
\bm{\Psi}^{(m)} \equiv \left\lbrace\bm{\phi}_{s}(\bs_{1}),...,\bm{\phi}_{s}(\bs_{m})\right\rbrace^{\prime}.
\end{equation*}
\noindent
That is, let $\textbf{b} = (b_{1},...,b_{m})^{\prime}$, and notice that
\begin{equation*}
\sum_{i = 1}^{m}\sum_{j = 1}^{m}b_{i}b_{j}\mathrm{cov}\left\lbrace Y_{s}(\bs_{i}), Y_{s}(\bs_{j})\right\rbrace = \mathrm{cov}\left( \textbf{b}^{\prime}\by^{(m)}\right) = \textbf{b}^{\prime}\bm{\Psi}^{(m)}\bm{\Lambda}\bm{\Psi}^{(m)\prime}\textbf{b}\ge 0,
\end{equation*}
\noindent
and hence, (\ref{pd}) holds for the covariance associated with $Y_{s}$ in (11). In a similar manner, one can prove the validity of the covariance function of $Y$ in (1) using Proposition $1.ii$.

\section*{IV Application: Mediterranean Surface Winds} 
\renewcommand{\theequation}{4.\arabic{equation}}
\setcounter{equation}{0}
A critical component of the interface between the atmosphere and the upper ocean occurs due to the transfer of momentum and the exchange of heat and fresh water, which is manifested through surface winds from the atmosphere. Due to a lack of direct measurements of surface wind over the ocean, wind field estimates over such regions were historically based on a blend between mechanistic models of the atmosphere and a relatively sparse global network of wind observations from buoys and ships of opportunity.  The practical spatial resolution of these so-called ``analysis'' winds is limited to fairly large spatial and temporal scales of variability, yet they are reported on fairly high-resolution grids. The advent of space-borne scatterometer instruments in the 1990s provided the first high-volume, high-resolution in space, wind estimates over the oceans. Although these scatterometer winds have higher spatial resolution (effectively ``point'' scale), they are incomplete in space and time, necessitating an optimal blending approach (e.g., \citet{wikle2001spatiotemporal}). \citet{millif}, and \citet{wikle2013} give reviews of recent statistical approaches to generate spatially and temporally complete ocean wind fields.  

As mentioned above, the weather center analysis winds do not contain spatial information commensurate with the spatial support in which they are estimated (e.g., see \citet{millif} for discussion).  That is, the kinetic energy spectrum of the winds does not contain realistic variation at small spatial scales.  The support given by the additional (and incomplete) scatterometer wind estimates is relatively much smaller.  To date, there have been no attempts to consider an optimal spatial support for statistical wind predictions given these types of data.

\begin{figure}[t!]
      \begin{center}
      \begin{tabular}{cc}
        \includegraphics[width=8cm,height=8cm]{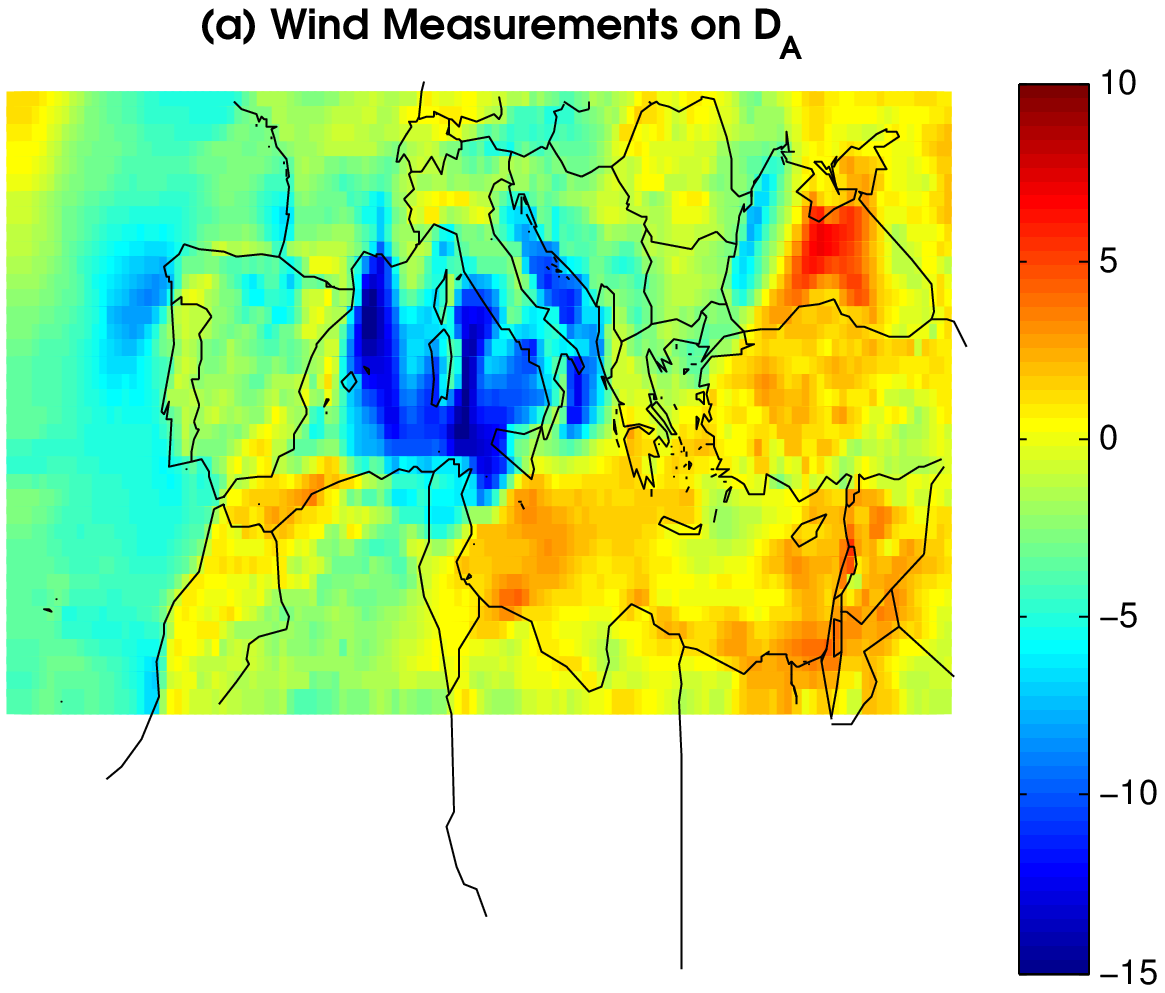}& \includegraphics[width=8cm,height=8cm]{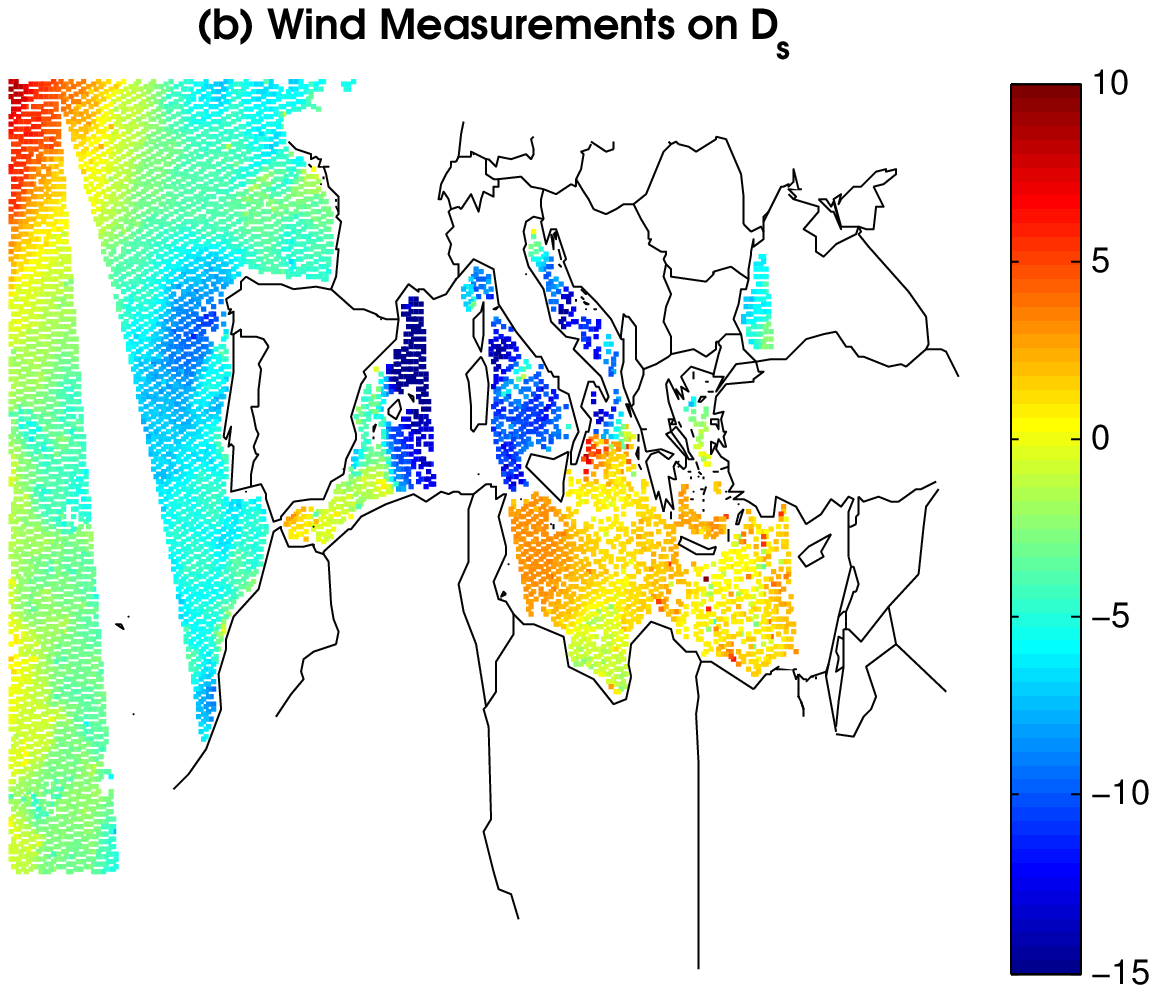}
      \end{tabular}
      \caption{Wind observations from 2 February 2005 at 12:00 UTC (Universal Coordinated Time). (a) North-south (v) component of the wind from the ECMWF-analysis winds on a  $0.5^{\circ} \times 0.5^{\circ}$ grid. (b) North-south wind component from the high resolution (25km), but spatially intermittent, QuickSCAT scatterometer wind retrievals.}
         \label{fig:v_data}
      \end{center}
      \end{figure}

In the example presented here, we consider ocean surface wind data from two sources over the Mediterranean Sea. In particular, we consider the north-south wind component for analysis winds  from the European Center for Medium range Weather Forecasting (ECMWF) and satellite wind observations from the QuikSCAT scatterometer; this is a subset of the data used in the study by \citet{millif}.  We assume that the high resolution (25-km) scatterometer wind observations are effectively ``point'' support (relative to the analysis winds). Thus, these data are recorded on both $D_{s}\subset \mathbb{R}^{2}$ and $D_{A}$. Here, $D_{s}$ ranges from 30$^{\circ}$ to 48$^{\circ}$ north latitude, and -19$^{\circ}$ to 42$^{\circ}$ east longitude, and $D_{A}$ consists of a $0.5^{\circ} \times 0.5^{\circ}$ resolution grid on $D_{s}$. In total, $D_{A}$ consists of 4,551 areal units and $D_{s}$ consists of 6,916 observations for the time of interest, resulting in a dataset of 11,467 spatial observations. Figure~\ref{fig:v_data} shows these data for a 6-hour window centered on 12:00 UTC (Universal Coordinate Time) for 2 February, 2005.  

\begin{figure}[htp]
               \begin{center}
               \begin{tabular}{ccc}
               	\includegraphics[width=4.5cm,height=4cm]{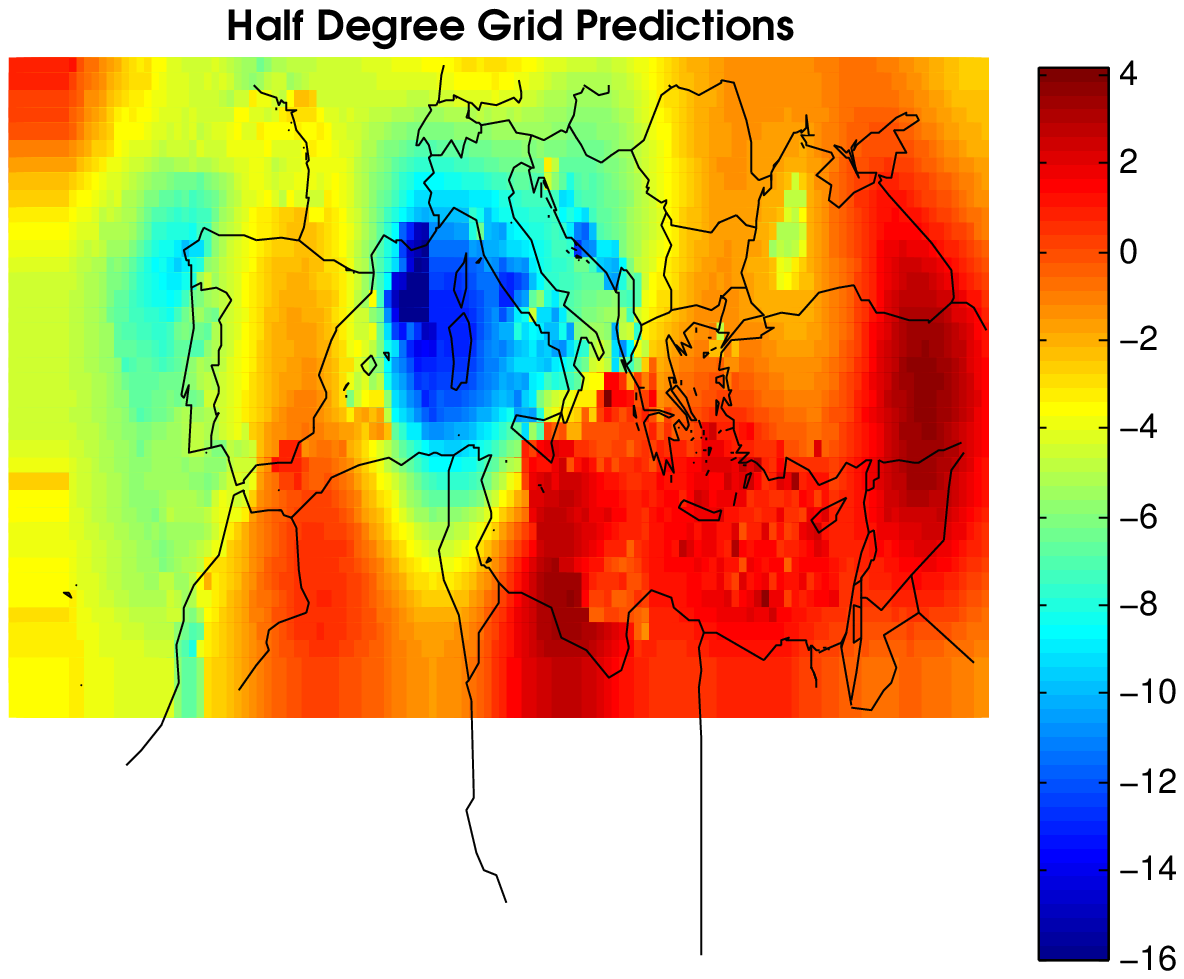}&
                   \includegraphics[width=4.5cm,height=4cm]{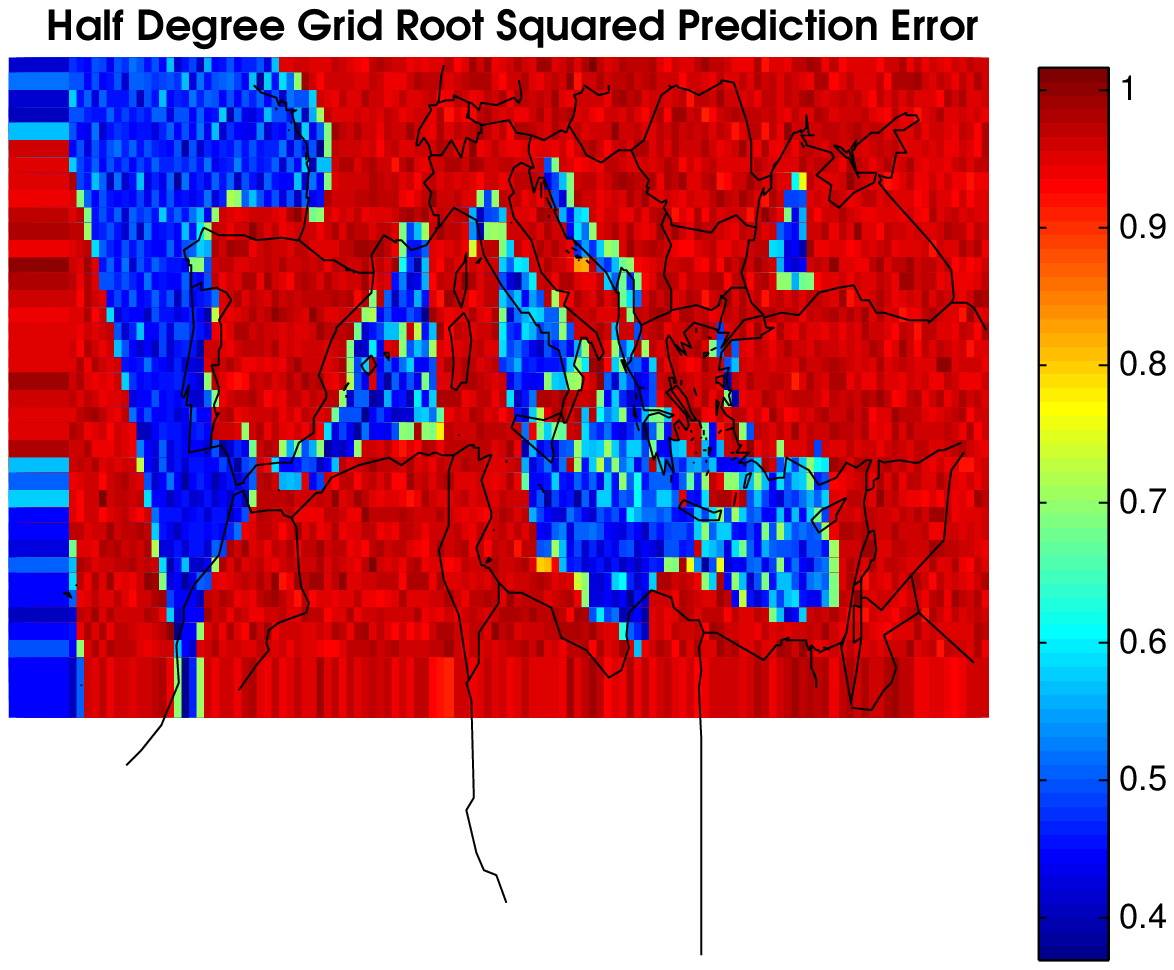}&     \includegraphics[width=4.5cm,height=4cm]{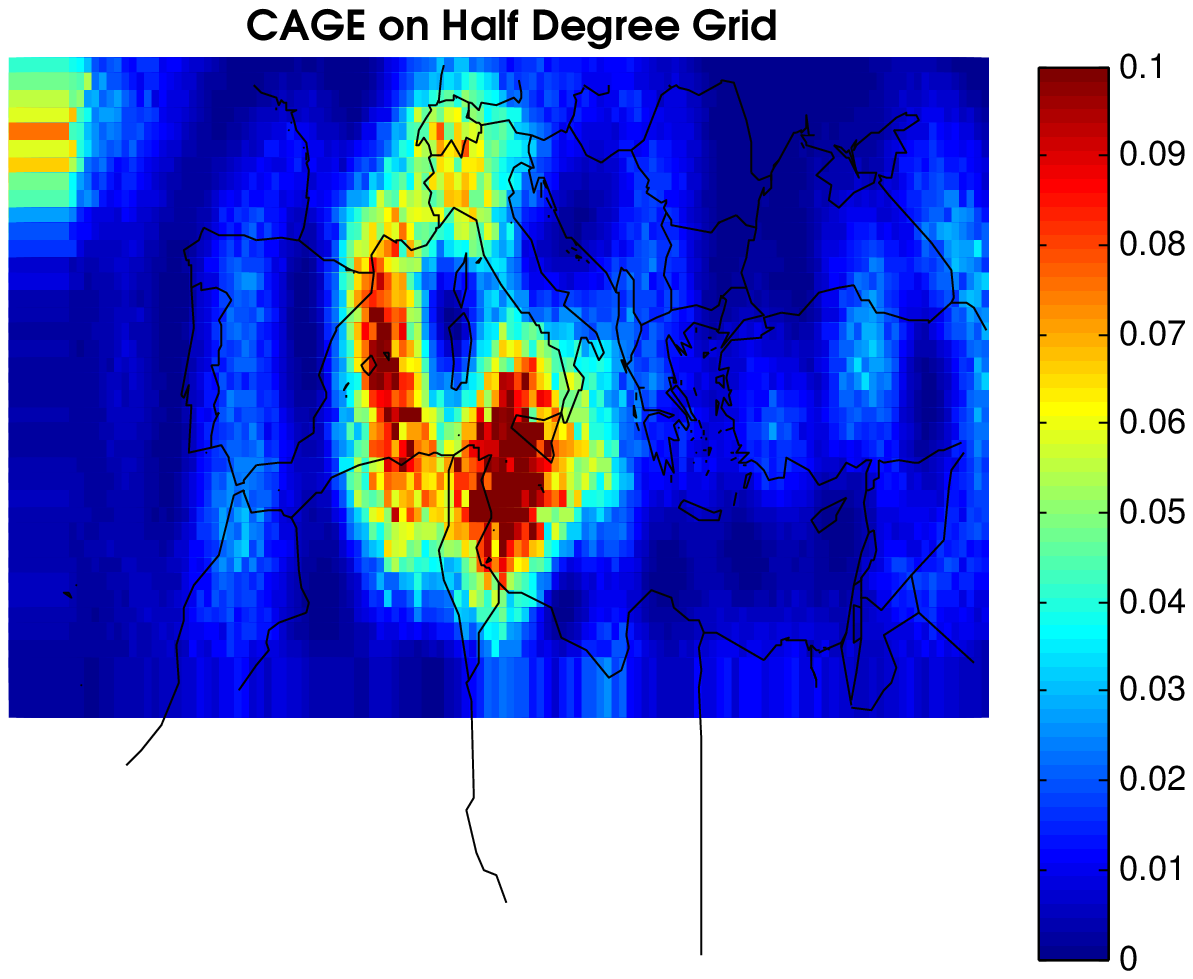}\\
                  \includegraphics[width=4.5cm,height=4cm]{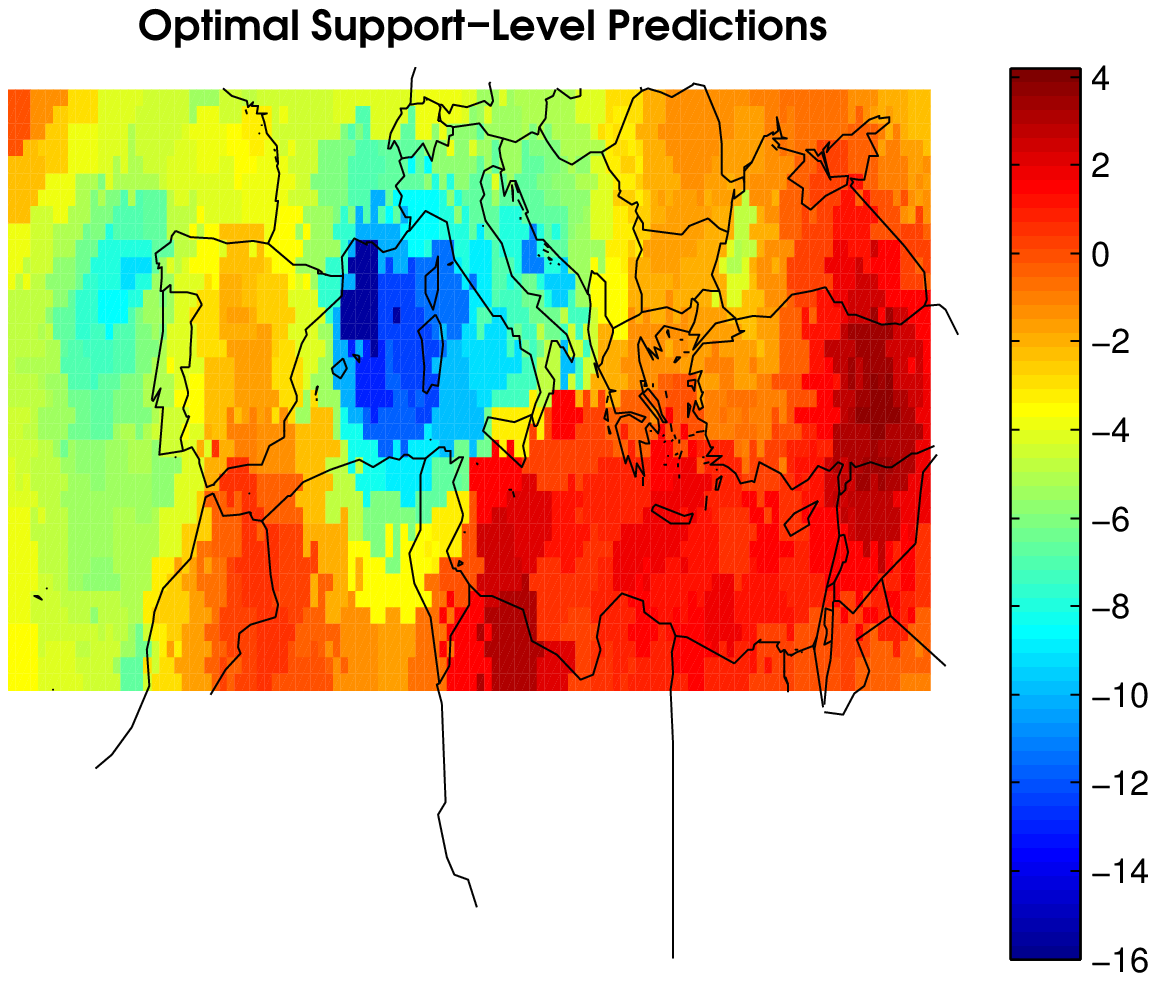}&
                  \includegraphics[width=4.5cm,height=4cm]{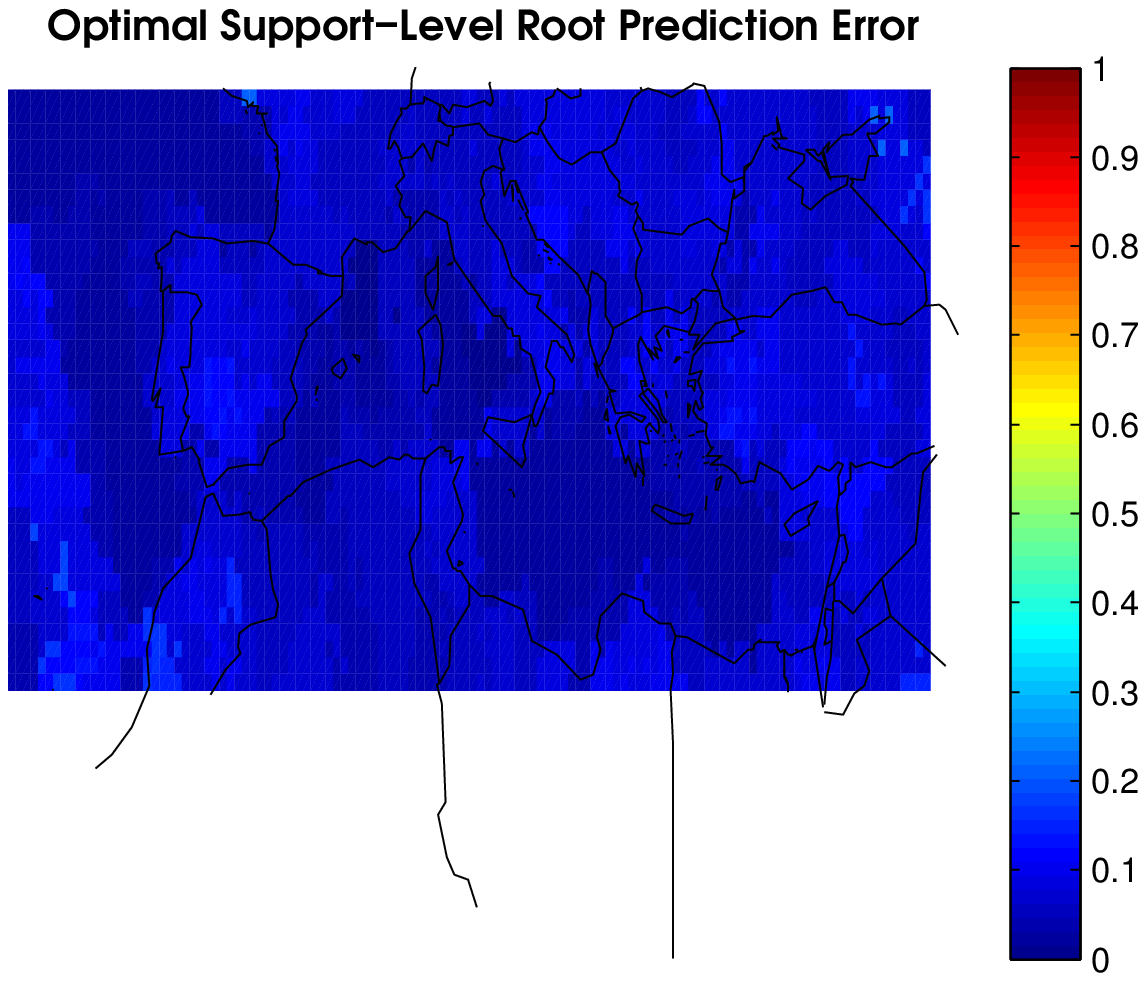}&     \includegraphics[width=4.5cm,height=4cm]{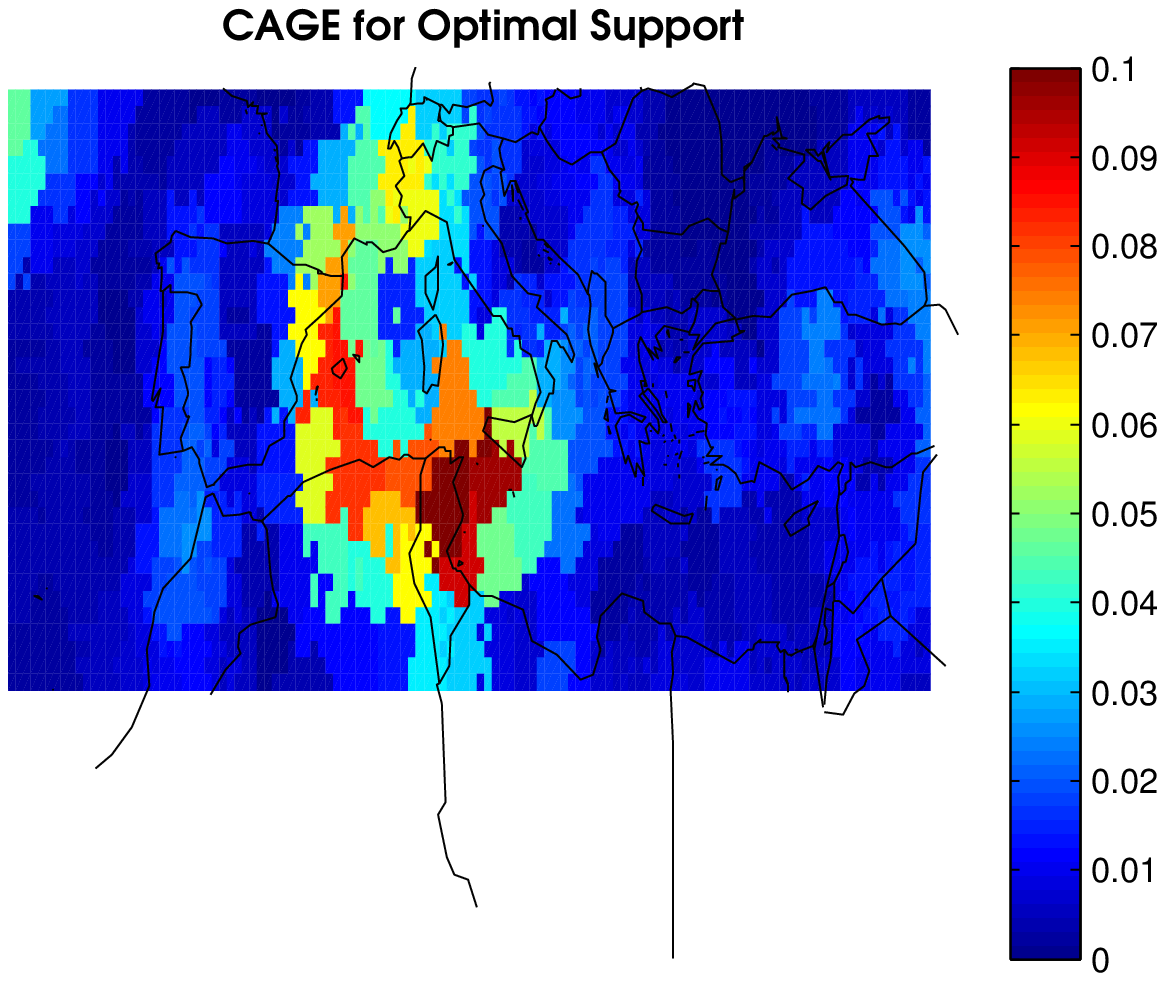}\\
                  \includegraphics[width=4.5cm,height=4cm]{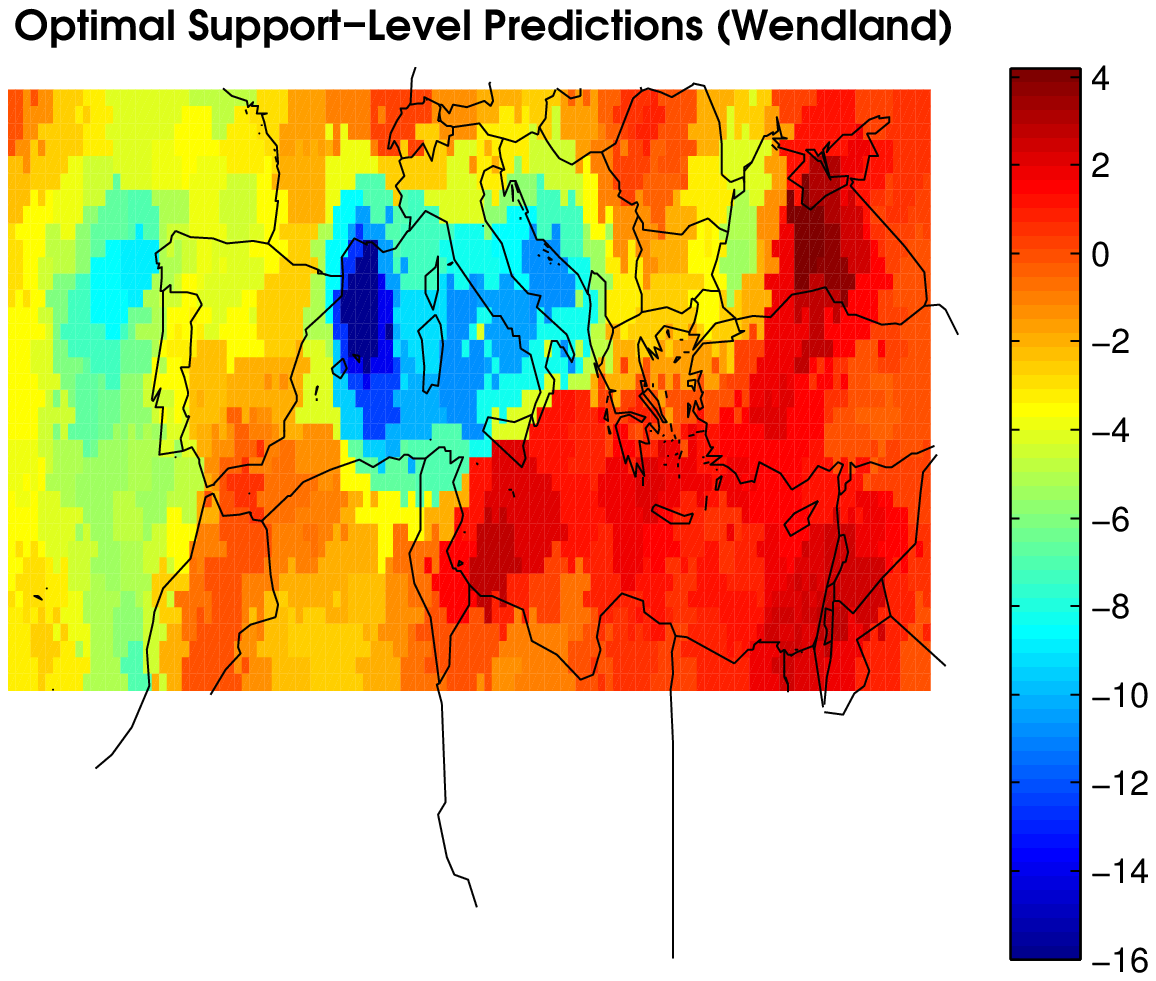}&
                  \includegraphics[width=4.5cm,height=4cm]{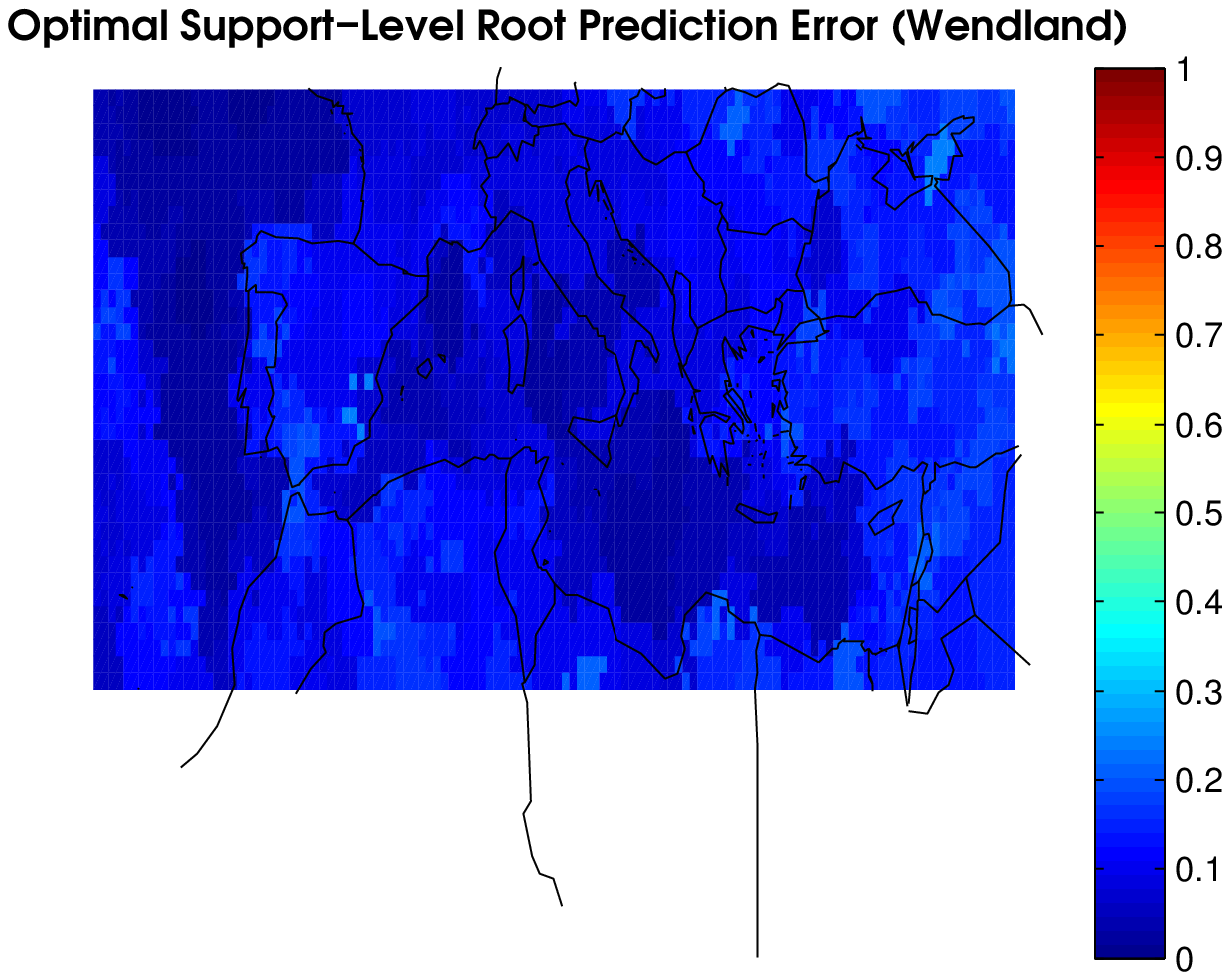}&    \includegraphics[width=4.5cm,height=4cm]{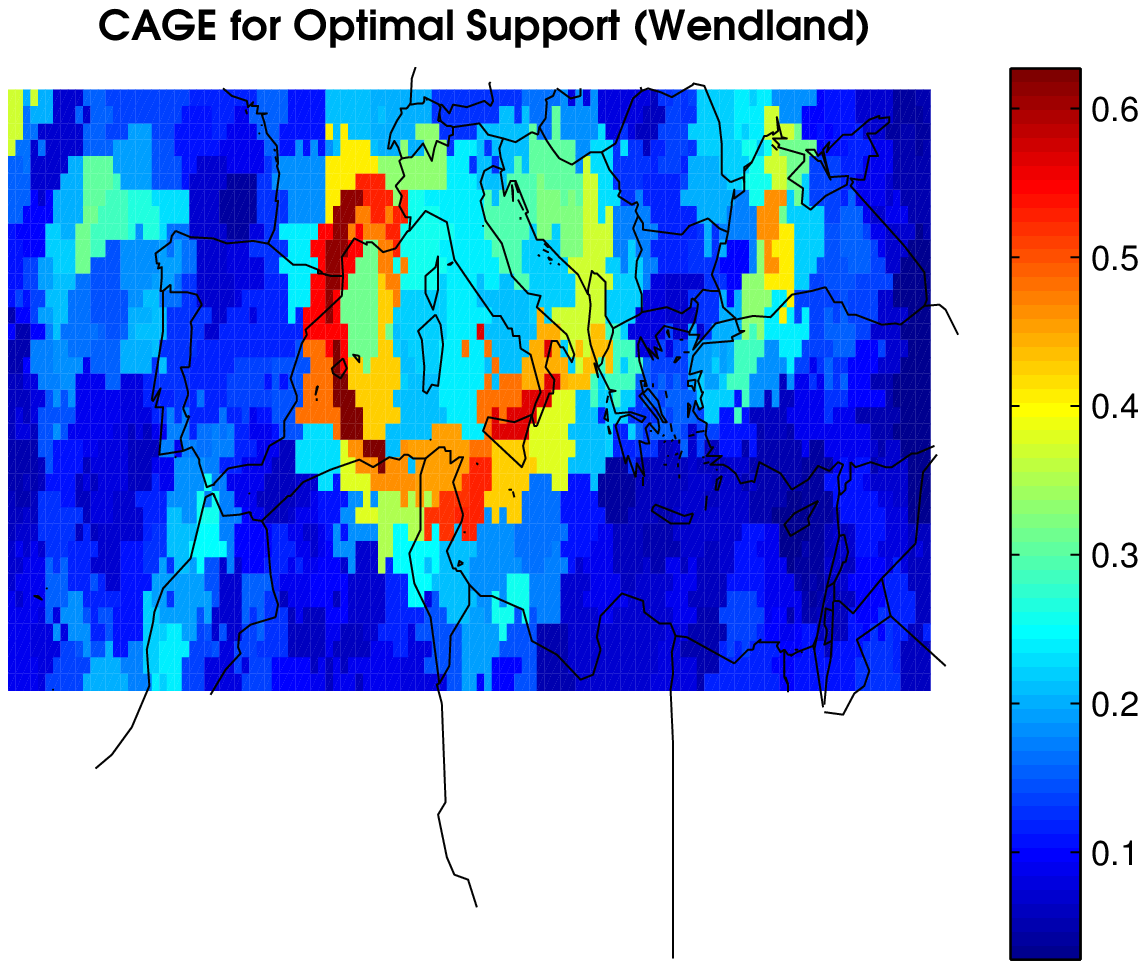}\\
                  \includegraphics[width=4.5cm,height=4cm]{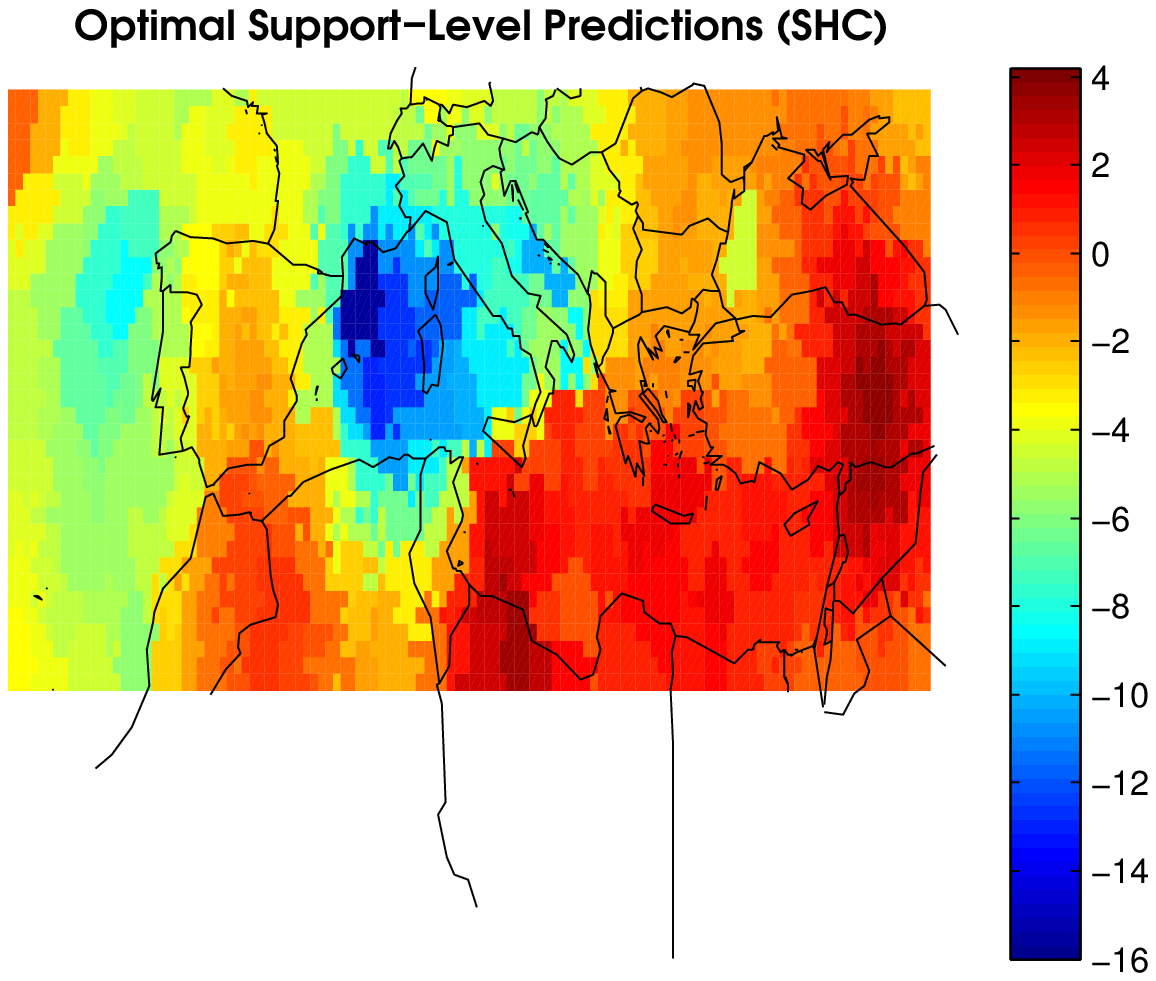}&
                  \includegraphics[width=4.5cm,height=4cm]{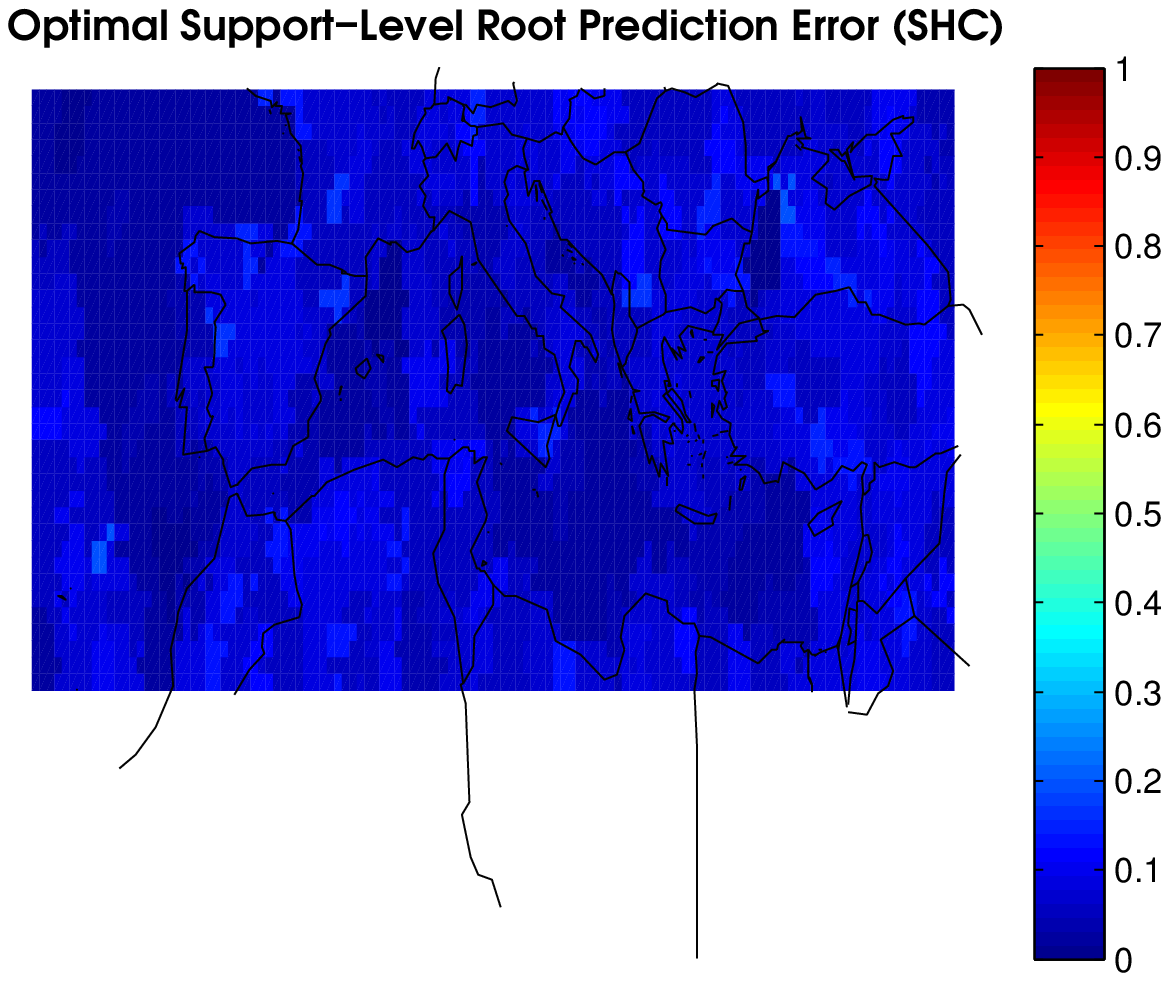}&    \includegraphics[width=4.5cm,height=4cm]{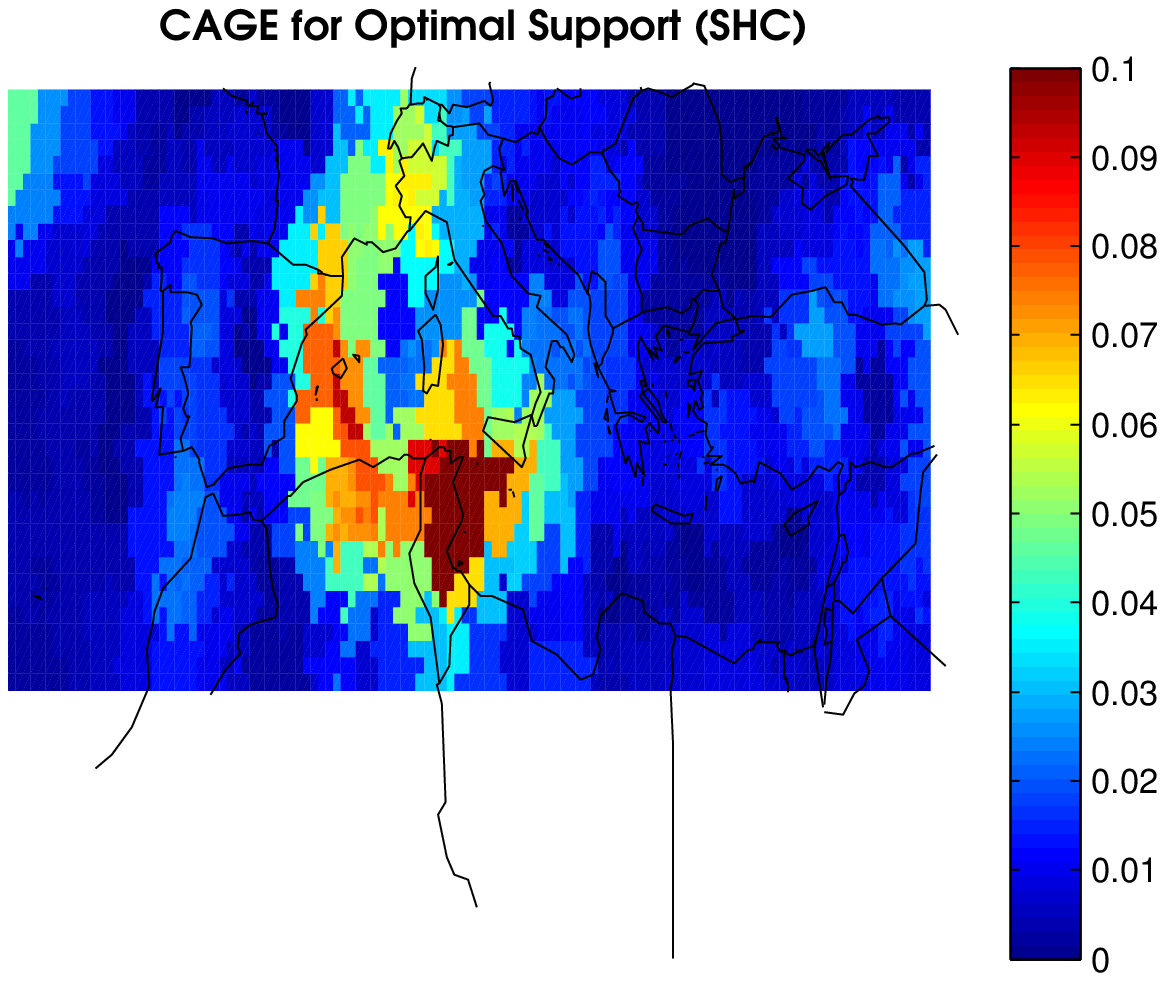}
               \end{tabular}
       
               \caption{\baselineskip=10pt CAGE-based posterior summaries of the predicted north-south wind components based on the analysis and scatterometer observations from 2 February 2005 at 12:00 UTC.  The first column displays the posterior mean; the second column displays the posterior standard deviations; and the third column contains the calculated CAGE. In the first row the values (i.e., posterior mean, posterior root prediction error, and CAGE) are all defined on a half degree grid. In the second row values are defined on the optimal spatial support found using $k$-means and the bisquare GBFs. In the third row values are defined on the optimal spatial support found using $k$-means and the Wendland GBFs. In the fourth row values are defined on the optimal spatial support using structural hierarchical clustering (SHC) and bisquare GBFs. Note that the colorbar for the predictions differ from the colorbar used in Figure 6.}
                       \label{fig:v_results1}
               \end{center}
               \end{figure}

In this application, we let $D_{B}$ be a half-degree grid. {We consider the model in Algorithm~1}, where $\bm{\psi}$ is a multiresolution bisquare basis vector consisting of local bisquare functions in (\ref{cht3.bi}). 
We chose {$r=200$} knots using a space-filling design {and the plot in Figure~5 (see Section~II.iii)}. We consider both structural hierarchical clustering and $k$-means to define $\mathcal{C}$ in (19) with $g_{L} = 280$ and $g_{U} = 380$; note that we these choices of $g_{L} = 280$ and $g_{U} = 380$ were guided {by the approach discussed in Section~II.iii using the $k$-means algorithm with initial choices of $g_L =2 $ and $g_U = 600$.} We also considered an equivalent analysis using the Wendland GBFs with $k$-means clustering. Here, the Wendland basis functions \citep{wendland} are defined as
\begin{equation}\label{wl}
\psi_{j}^{\mathrm{WL}}(\bs) =
\left\{
	\begin{array}{ll}
		(1-d_{j}(\bs))^{6}(35d_{j}(\bs)^{2} + 18d_{j}(\bs) + 3)/3  & \mbox{if } 0\le d_{j} \le 1 \\
		0  & \mathrm{otherwise}; \hspace{5pt} \bs \in D_{s}
	\end{array}
\right.
\end{equation}
\noindent
where $j = 1,...,200$, $d_{j}(\bs) = ||\bs - \textbf{c}_{j}^{*}||/w$, we choose $w = 1.5$ times the smallest distance between two different knots, and $\{\textbf{c}_{j}\}$ consists of the same 200 knot specifications used in the bisquare basis functions.
Additionally, since the latent field is interpretable on $D_{s}$, we use CAGE within the expression of $D_{C}^{op}$ in (20). Following \citet{millif}, the variances of $\epsilon(\bu)$ are set equal to 1 when $\bu \in D_{s}$, and set equal to 10 when $\bu \in D_{A}$.\\ 
\indent The results of the CAGE analysis of the posterior wind predictions is given in Figure~\ref{fig:v_results1}.  The top row of this figure shows that when using the standard 0.5$^{\circ}$ resolution support, there is a noticeable high CAGE ``crescent'' in the south central portion of the region.  This would suggest that one should be concerned about assuming that statistics on the wind field over this region can be interpreted at the point level. Note that the optimal support regions with $k$-means and bisquare GBFs (the second row of \ref{fig:v_results1}) are much larger than the $D_B$ level shown in the first row, but the predictions look qualitatively similar to the half-degree predictions, although with more smoothing and the corresponding reduction in root prediction error associated with the relatively large optimal aggregation regions. The optimal aggregation seems to pick up realistic meteorological features. For example, notice the homogeneous region centered on Corsica and Sardina, which corresponds to a region of more intense southerly winds off of the mainland (so-called ``Mistral winds'') that are important in forcing the ocean circulation (e.g., see \citet{millif}). Perhaps more importantly, although the higher CAGE crescent is still present, it is noticeably reduced in intensity relative to the $D_B$ support.  The Wendlend GBF predictions (third row) are similar to the bisquare predictions, but with generally larger regions and with higher CAGE values that are shifted northward. Finally, the last row of  Figure~\ref{fig:v_results1} shows the bisquare results with the structural hierarchical clustering method.  These are similar to the bisquare $k$-means results, but one notices more spatial detail in the predictions.\\ 
\indent There is a striking amount of dimension reduction that results from the CAGE analysis. That is, values of $n_{C}^{op}$ are considerably smaller than the number of observations, 11,467. We have that $n_{C}^{op} = 323$ when using the bisquare GBFs and $k$-means, $n_{C}^{op} = 315$ when using the Wendland GBFs and $k$-means, and $n_{C}^{op} = 327$ when using the bisquare GBFs and SHC. This suggests that optimal aggregation, such as the results presented in Figure 7, may be a viable alternative approach for dimension reduction.\\
\indent We note that there is quite a large amount of shrinkage in these wind predictions relative to the data, which is not surprising given the uncertainty in the winds and the fact that no temporal information is being considered here.  As discussed in \cite{wikle2013}, one can gain significant prediction efficiencies if temporal dynamic information is included in the model for winds.  Such an analysis is beyond the scope of this simple illustration, but the CAGE-based selection of prediction support could, in principle, be utilized in that framework.  

\section*{V Technical Proofs} 
\renewcommand{\theequation}{5.\arabic{equation}}
\setcounter{equation}{0}
In Section~V.i, we provide the proofs to Propositions 1$\--$6. In addition to these proofs, we also provide results alluded to, but not explicitly stated in the main text (Section~V.ii).

\subsection*{V.i Proof of Propositions 1$\--$6} 

\noindent
\paragraph{\large{Proof of Proposition 1:}}
\normalsize
The assumptions of Proposition 1 allow us to apply the K-L decomposition of $\{Y(\bs): \bs \in D_{s}\}$ from \citet{Karhunen}. That is, from \citet{Karhunen} we have that for $\bs \in D_{s}$
\begin{equation}\label{kl}
Y_{A}(B_{h}) = \sum_{j=1}^{\infty} \phi_{j}(\bs)\alpha_{j},
\end{equation}
\noindent
where the eigenfunctions $ \{\phi_{j}(\bs) : j = 1, 2, ...\}$ have domain $D_{s}$ and satisfies,
\begin{equation}\label{orthog}
\int_{D_{s}} \phi_{j}(\bs) \phi_{k}(\bs) d\bs = \delta_{jk},
\end{equation} 
\noindent
where $\delta_{jk}$ is the Kronecker delta function. Additionally, the random variables in the set \(\{\alpha_{j}: j = 1, 2, ...\}\) are uncorrelated with variances \(\{\lambda_{j}: j = 1,2, ...\}\), and the coefficients \(\{\alpha_{j}: j = 1, 2, ...\}\) can be found by projecting $Y_{s}(\cdot)$ onto the eigenfunctions. That is, 
\begin{equation}\label{projection}
\alpha_{j} = \int_{D_{s}} Y_{s}(\bs) \phi_{j}(\bs) d\bs,
\end{equation}
 for each $j$. Also, these eigenfunctions are solutions to the Fredholm integral equation (e.g.,\citet{Papoulis}),
\begin{equation}\label{fi}
\int_{D_{s}} C(\bs, \bu) \phi_{j}(\bs) d\bs = \lambda_{j} \phi_{j}(\bu);\hspace{5pt} \bu \in D_{s}, j = 1,2,...,
\end{equation} 
\noindent
where, from the statement of Proposition 1, \(C(\bs, \bu)\) is a valid covariance function for each $\bs, \bu \in D_{s}$. \\
\indent The statement that
\begin{equation}\label{mkl}
Y_{A}(A) = \sum_{i = 1}^{\infty} \phi_{A,j}(A) \alpha_{j},
\end{equation}
 in $L^{2}(\Omega)$ for $A \subset D_{s}$, is equivalent to saying that
\begin{equation}\label{zet}
\zeta_{n}(A) \equiv E\left\lbrace \left(Y_{A}(A) - \sum_{i = 1}^{n}\phi_{A,j}(A)\alpha_{j}\right)^{2}\right\rbrace
\end{equation} 
\noindent
converges to zero as $n$ goes to infinity. Note that in (\ref{zet}), the expectation is taken with respect to $(\Omega,\mathcal{F},\mathcal{P})$. Expanding (\ref{zet}) we have,
\begin{equation}\label{zet2}
\zeta_{n}(A) = E\left\lbrace Y_{A}(A)^{2}\right\rbrace + E\left\lbrace\left(\sum_{i = 1}^{n}\phi_{A,j}(A)\alpha_{j}\right)^{2}\right\rbrace - 2\hspace{2pt}E\left\lbrace Y_{A}(A)\left(\sum_{i = 1}^{n}\phi_{A,j}(A)\alpha_{j}\right)\right\rbrace.
\end{equation} 
The first term of the right-hand side of (\ref{zet2}) can be written as
\begin{align}
\nonumber
E\left\lbrace Y_{A}(A)^{2}\right\rbrace &= E\left\lbrace \frac{1}{|A|^{2}}\int_{A}\int_{A}Y_{s}(\bs)Y_{s}(\bu)d\bs d\bu\right\rbrace \\
\nonumber
&= \frac{1}{|A|^{2}}\int_{A}\int_{A}E(Y_{s}(\bs)Y_{s}(\bu))d\bs d\bu\\
&= \frac{1}{|A|^{2}}\int_{A}\int_{A}C(\bs,\bu)d\bs d\bu.\label{term1}
\end{align}
The second term of the right-hand side of (\ref{zet2}) can be written as\\
\begin{align}
\nonumber
E\left\lbrace \left(\sum_{i = 1}^{n}\phi_{A,j}(A)\alpha_{j}\right)^{2}\right\rbrace &= E\left\lbrace \left(\sum_{i = 1}^{n}\phi_{A,i}(A)\alpha_{i}\right) \left(\sum_{j = 1}^{n}\phi_{A,j}(A)\alpha_{j}\right)\right\rbrace \\
\nonumber
&= E\left\lbrace \sum_{i = 1}^{n}\sum_{j = 1}^{n}\phi_{A,i}(A)\phi_{A,j}(A)\alpha_{i}\alpha_{j}\right\rbrace \\
\nonumber
&= E\left\lbrace \frac{1}{|A|^{2}}\sum_{i = 1}^{n}\sum_{j = 1}^{n}\int_{A}\int_{A}\phi_{s,i}(\bs)\phi_{s,j}(\bu) \alpha_{i}\alpha_{j}d\bs d\bu \right\rbrace \\
\nonumber
&= \frac{1}{|A|^{2}}\sum_{i = 1}^{n}\sum_{j = 1}^{n}\int_{A}\int_{A}\phi_{s,i}(\bs)\phi_{s,j}(\bu)  E(\alpha_{i}\alpha_{j})d\bs d\bu  \\
&= \frac{1}{|A|^{2}}\int_{A}\int_{A}\sum_{j = 1}^{n}\phi_{s,j}(\bs)\phi_{s,j}(\bu)  \lambda_{j} d\bs d\bu,\label{term2}
\end{align}
\noindent
since recall from the K-L decomposition that $\alpha_{i}$ and $\alpha_{j}$ are uncorrelated with variances $\lambda_{i}$ and $\lambda_{j}$, respectively. Finally, the third term of the right-hand side of (\ref{zet2}) can be written as
\begin{align*}
E\left\lbrace Y_{A}(A)\left(\sum_{i = 1}^{n}\phi_{A,j}(A)\alpha_{j}\right)\right\rbrace &= E\left\lbrace \frac{1}{|A|^{2}}\int_{A}\int_{A}\sum_{i = 1}^{n}\phi_{s,i}(\bs)Y_{s}(\bu)\alpha_{i} d\bs d\bu\right\rbrace,
\end{align*}
\noindent
Since $\alpha_{i}$ is found by projecting $Y_{s}$ onto the eigenfunctions. From (\ref{projection}) we have that
\begin{align*}
E\left\lbrace Y_{A}(A)\left(\sum_{i = 1}^{n}\phi_{A,j}(A)\alpha_{j}\right)\right\rbrace &= E\left\lbrace \frac{1}{|A|^{2}}\int_{A}\int_{A}\sum_{i = 1}^{n}\phi_{s,i}(\bs)Y_{s}(\bu)\int_{D}Y_{s}(\textbf{w})\phi_{i}(\textbf{w})d\textbf{w} d\bs d\bu\right\rbrace\\
&= E\left\lbrace \frac{1}{|A|^{2}}\int_{A}\int_{A}\sum_{i = 1}^{n}\phi_{s,i}(\bs)\int_{D}Y_{s}(\bu)Y_{s}(\textbf{w})\phi_{i}(\textbf{w})d\textbf{w} d\bs d\bu\right\rbrace \\
&= \frac{1}{|A|^{2}} \int_{A}\int_{A}\sum_{i = 1}^{n}\phi_{s,i}(\bs)\int_{D}E\left\lbrace Y_{s}(\bu)Y_{s}(\textbf{w})\right\rbrace\phi_{i}(\textbf{w})d\textbf{w} d\bs d\bu \\
&=  \frac{1}{|A|^{2}} \int_{A}\int_{A}\sum_{i = 1}^{n}\phi_{s,i}(\bs)\int_{D}C(\bu, \textbf{w}) \phi_{i}(\textbf{w})d\textbf{w} d\bs d\bu.
\end{align*}
\noindent
From the Fredholm integral equation in (\ref{fi}) we have
\begin{align}
\nonumber
E\left\lbrace Y_{A}(A)\left(\sum_{i = 1}^{n}\phi_{A,j}(A)\alpha_{j}\right)\right\rbrace &= \frac{1}{|A|^{2}} \int_{A}\int_{A}\sum_{i = 1}^{n}\phi_{s,i}(\bs)\int_{D}C(\bu, \textbf{w})\phi_{i}(\textbf{w})d\textbf{w} d\bs d\bu\\
&= \frac{1}{|A|^{2}} \int_{A}\int_{A}\sum_{i = 1}^{n}\phi_{s,i}(\bs)\phi_{s,i}(\bu)\lambda_{i} d\bs d\bu.\label{term3}
\end{align}
\noindent
Substituting (\ref{term1}), (\ref{term2}), and (\ref{term3}) into (\ref{zet2}) gives
\begin{equation}\label{lastequation}
\zeta_{n}(A) = \frac{1}{|A|^{2}}\int_{A}\int_{A}C(\bs,\bu) -\sum_{i = 1}^{n}\phi_{s,i}(\bs)\phi_{s,i}(\bu)\lambda_{i} d\bs d\bu.
\end{equation}
Upon taking the limit as $n$ goes to infinity on both sides of (\ref{lastequation}), it follows from Mercer's theorem \citep{mercer} that
\begin{equation}\label{mercer2}
\underset{n\rightarrow \infty}{\lim} \zeta_{n}(A) = 0,
\end{equation}
\noindent
for each $A \subset D_{s}$; note that Mercer's theorem shows \textit{uniform} convergence at the point-level, allowing one to pass the limit through the integral. This proves the result. 

The proof of $1.ii$ follows a similar logic to (\ref{mercer2}). That is, note that
\begin{align}\label{lastequation2}
\nonumber
&\sum_{i = 1}^{n}\phi_{A,i}(A)\phi_{A,i}(B)\lambda_{i}-\mathrm{cov}\left\lbrace Y_{A}(A), Y_{A}(B)\right\rbrace \\
& = \frac{1}{|A||B|}\int_{A}\int_{B}C(\bs,\bu) -\sum_{i = 1}^{n}\phi_{s,i}(\bs)\phi_{s,i}(\bu)\lambda_{i} d\bs d\bu.
\end{align}
\noindent
Upon taking the limit as $n$ goes to infinity on both sides of (\ref{lastequation2}), it follows from Mercer's theorem \citep{mercer} that Proposition $1.ii$ holds.

\noindent
\paragraph{\large{Proof of Proposition 2:}}
\normalsize
First, we prove the following statement: If ${\phi}_{k}(\textbf{x}_{j})= {\phi}_{A,k}(A_{j})$ for $j = 1,...,n_{A}$ and for any positive integer $k$, then $\by_{s}^{(A)}$ = $\by_{A}$ almost surely. Then the continuous mapping theorem is applied to get $f(\by_{s}^{(A)})$ = $f(\by_{A})$ almost surely. \\
\indent We proceed using a proof by contradiction. Assume that $\by_{s}^{(A)}$ is not almost surely equal to $\by_{A}$. Then, for at least one $\textbf{x}_{i}$ and $A_{i}$, there exists a $\gamma>0$ such that
\begin{equation}\label{contradict}
P(|Y_{s}(\textbf{x}_{i}) - Y_{A}(A_{i})|\ge\gamma)>0.
\end{equation}
\noindent
However, we have from Chebychev's inequality
\begin{equation}\label{take2}
P(|Y_{s}(\textbf{x}_{i}) - Y_{A}(A_{i})|\ge \gamma) \le \frac{E\left[\left\lbrace Y_{s}(\textbf{x}_{i}) - Y_{A}(A_{i})\right\rbrace^{2}\right]}{\gamma^{2}}.
\end{equation}
Assume that ${\phi}_{k}(\textbf{x}_{j})= {\phi}_{A,k}(A_{j})$ for $j = 1,...,n_{A}$ and every positive integer $k$. Then, upon adding and subtracting $\sum_{k = 1}^{n}\phi_{k}(\textbf{x}_{i})$ within (\ref{take2}) we have:
\begin{align}
\nonumber
& P(|Y_{s}(\textbf{x}_{i}) - Y_{A}(A_{i})|\ge \gamma) \le \frac{1}{\gamma^{2}}E\left\lbrace Y_{s}(\textbf{x}_{i})-\sum_{k = 1}^{n}\phi_{k}(\textbf{x}_{i})\alpha_{k} + \sum_{k = 1}^{n}\phi_{A,k}(A_{i})\alpha_{k} - Y_{A}(A_{i})\right\rbrace^{2}\\
\label{line1}
&=\frac{1}{\gamma^{2}}E\left\lbrace Y_{s}(\textbf{x}_{i})-\sum_{k = 1}^{n}\phi_{k}(\textbf{x}_{i})\alpha_{k}\right\rbrace^{2} + \frac{1}{\gamma^{2}}E\left\lbrace\sum_{k = 1}^{n}\phi_{A,k}(A_{i})\alpha_{k} - Y_{A}(A_{i})\right\rbrace^{2}\\
\label{line2}
& + \frac{2}{\gamma^{2}} E\left[\left\lbrace Y_{s}(\textbf{x}_{i})-\sum_{k = 1}^{n}\phi_{k}(\textbf{x}_{i})\alpha_{k}\right\rbrace \left\lbrace \sum_{k = 1}^{n}\phi_{A,k}(A_{i})\alpha_{k} - Y_{A}(A_{i})\right\rbrace\right].
\end{align}
It follows from \citet{Karhunen} that the first term on the right-hand-side of (\ref{line1}) converges to zero. Likewise, from Proposition 1 the second term on the right-hand-side of (\ref{line2}) converges to zero as $n$ goes to infinity. Note that since $P(|Y_{s}(\textbf{x}_{i}) - Y_{A}(A_{i})|\ge \gamma)$ does not depend on $n$ we have that,
\begin{align}\label{reduce}
\nonumber
& P(|Y_{s}(\textbf{x}_{i}) - Y_{A}(A_{i})|\ge \gamma)\\
& \le \underset{n \rightarrow \infty}{\mathrm{lim}}\frac{2}{\gamma^{2}} E\left[\left\lbrace Y_{s}(\textbf{x}_{i})-\sum_{k = 1}^{n}\phi_{k}(\textbf{x}_{i})\alpha_{k}\right\rbrace \left\lbrace \sum_{k = 1}^{n}\phi_{A,k}(A_{i})\alpha_{k} - Y_{A}(A_{i})\right\rbrace\right].
\end{align}
\noindent
Thus, we are left to find the expression of the limit in (\ref{reduce}). Note,
\begin{align}
\nonumber
& E\left[\left\lbrace Y_{s}(\textbf{x}_{i})-\sum_{k = 1}^{n}\phi_{k}(\textbf{x}_{i})\alpha_{k}\right\rbrace \left\lbrace \sum_{k = 1}^{n}\phi_{A,k}(A_{i})\alpha_{k} - Y_{A}(A_{i})\right\rbrace\right] \\
\label{line11}
&=\frac{1}{|A|}E\left\lbrace\int_{A_{i}} \sum_{k = 1}^{n}Y_{s}(\textbf{x}_{i})\alpha_{k}\phi_{k}(\bs)d\bs\right\rbrace\\
\label{line12}
&-\frac{1}{|A|}E\left\lbrace\int_{A_{i}}Y_{s}(\textbf{x}_{i})Y_{s}(\bu)d\bu\right\rbrace\\
\label{line13}
&- \frac{1}{|A|}E\left\lbrace\sum_{k = 1}^{n} \sum_{j = 1}^{n} \int_{A_{i}}\phi_{k}(\textbf{x}_{i})\phi_{j}(\bs)\alpha_{j}\alpha_{k}d\bs\right\rbrace\\
\label{line14}
&+ \frac{1}{|A|}E\left\lbrace\sum_{k = 1}^{n}\int_{A_{i}}\phi_{k}(\textbf{x}_{i})\alpha_{k}Y_{s}(\bs)d\bs\right\rbrace.
\end{align}
For the term in (\ref{line11}) notice from (\ref{projection}) and (\ref{fi}) we have that
\begin{align*}
\frac{1}{|A|}E\left\lbrace\int_{A_{i}} \sum_{k = 1}^{n}Y_{s}(\textbf{x}_{i})\alpha_{k}\phi_{k}(\bs)d\bs\right\rbrace &= \frac{1}{|A|}E\left\lbrace\int_{A_{i}} \sum_{k = 1}^{n}Y_{s}(\textbf{x}_{i})\int_{D_{s}}Y_{s}(\bu)\phi_{k}(\bu)d\bu\hspace{5pt}\phi_{k}(\bs)d\bs\right\rbrace\\
&= \frac{1}{|A|}\int_{A_{i}} \sum_{k = 1}^{n}\int_{D_{s}}E\left\lbrace Y_{s}(\textbf{x}_{i})Y_{s}(\bu)\right\rbrace\phi_{j}(\bu)d\bu\hspace{5pt}\phi_{k}(\bs)d\bs\\
&= \frac{1}{|A|}\int_{A_{i}} \sum_{k = 1}^{n}\int_{D_{s}}C(\textbf{x}_{i},\bu)\phi_{k}(\bu)d\bu\hspace{5pt}\phi_{k}(\bs)d\bs\\
&= \frac{1}{|A|}\int_{A_{i}} \sum_{k = 1}^{n}\phi_{k}(\bs)\phi_{k}(\textbf{x}_{i})\lambda_{k}d\bs.
\end{align*}
The terms in (\ref{line12}) and (\ref{line13}) can be written as
\begin{align*}
&-\frac{1}{|A|}E\left\lbrace\int_{A_{i}}Y_{s}(\textbf{x}_{i})Y_{s}(\bu)d\bu\right\rbrace = -\frac{1}{|A|}E\left\lbrace\int_{A_{i}}C(\textbf{x}_{i},\bu)d\bu\right\rbrace,\\
&- \frac{1}{|A|}E\left\lbrace\sum_{k = 1}^{n} \sum_{j = 1}^{n} \int_{A_{i}}\phi_{k}(\textbf{x}_{i})\phi_{j}(\bs)\alpha_{j}\alpha_{k}d\bs\right\rbrace = -\frac{1}{|A|}\int_{A_{i}} \sum_{k = 1}^{n}\phi_{k}(\bs)\phi_{k}(\textbf{x}_{i})\lambda_{k}d\bs.
\end{align*}
For the term in (\ref{line14}) notice from (\ref{projection}) and (\ref{fi}) we have that
\begin{align*}
\nonumber
\frac{1}{|A|}E\left\lbrace\sum_{k = 1}^{n}\int_{A_{i}}\phi_{k}(\textbf{x}_{i})\alpha_{k}Y_{s}(\bs)d\bs\right\rbrace
=\frac{1}{|A|}\int_{A_{i}} \sum_{k = 1}^{n}\phi_{k}(\bs)\phi_{k}(\textbf{x}_{i})\lambda_{k}d\bs.
\end{align*}
\noindent
Thus, it follows that
\begin{align}
\nonumber
& E\left[\left\lbrace Y_{s}(\textbf{x}_{i})-\sum_{k = 1}^{n}\phi_{k}(\textbf{x}_{i})\alpha_{k}\right\rbrace \left\lbrace \sum_{k = 1}^{n}\phi_{A,k}(A_{i})\alpha_{k} - Y_{A}(A_{i})\right\rbrace\right]\\
&=\frac{2}{|A|}\int_{A_{i}}\sum_{k = 1}^{n}\phi_{k}(\textbf{x}_{i})\phi_{k}(\bs)\lambda_{k} - C(\textbf{x}_{i},\textbf{s}) d\bs,
\end{align}
which, again by Mercer's theorem, converges to 0 as $n$ goes to infinity. Thus, from (\ref{reduce}) we have that
\begin{equation*}
P(|Y_{s}(\textbf{x}_{i}) - Y_{A}(A_{i})|\ge \gamma)=0,
\end{equation*}
which contradicts (\ref{contradict}). One can prove forward implication of Proposition $2.ii$ in a similar manner. \\
\indent To prove the reverse statement of Proposition $2.i$, suppose that $f(\by_{s}^{(A)}) = f(\by_{A})$ almost surely for any measurable real-valued function $f$. Thus, the functions $f_{i}(\textbf{b}) =b_{i}$ for $i = 1,...,n_{A}$ and $\textbf{b} = (b_{i}: i = 1,...,n_{A})^{\prime}\in \mathbb{R}^{n_{A}}$, imply that
\begin{equation} \label{almostsure}
Y_{s}(\textbf{x}_{i}) = Y_{A}(A_{i}),
\end{equation}
\noindent
almost surely. Multiplying both sides by $\alpha_{j}$ we have 
\begin{equation*}
Y_{s}(\textbf{x}_{i})\alpha_{j} = Y_{A}(A_{i})\alpha_{j}
\end{equation*}
\noindent
almost surely. Substituting (\ref{projection}) into the equation above gives,
\begin{equation*}
Y_{s}(\textbf{x}_{i})\int_{D_{s}}Y_{s}(\bs)\phi_{j}(\bs)d\bs = \frac{1}{|A_{i}|}\int_{A_{i}}\int_{D_{s}}Y_{s}(\bu)Y_{s}(\bs)\phi_{j}(\bs)d\bs d\bu.
\end{equation*}
Taking the expectation on both sides we have
\begin{equation*}
\int_{D_{s}}C(\textbf{x}_{i},\bs)\phi_{j}(\bs)d\bs = \frac{1}{|A_{i}|}\int_{A_{i}}\int_{D_{s}}C(\bu,\bs)\phi_{j}(\bs)d\bs d\bu,
\end{equation*}
and then from (\ref{fi}) we have
\begin{equation*}
\phi_{j}(\textbf{x}_{i})\lambda_{j} = \frac{1}{|A_{i}|}\int_{A_{i}}\phi_{j}(\bu)d\bu \lambda_{j}.
\end{equation*}
Dividing by $\lambda_{j}$
\begin{equation*}
\phi_{j}(\textbf{x}_{i}) = \frac{1}{|A_{i}|}\int_{A_{i}}\phi_{j}(\bu)d\bu.
\end{equation*}
This proves the result. One can prove the reverse statement of Proposition $2.ii$ in a similar manner.

By the condition in Proposition $2.iii$, we have that for a given ${\phi}_{k}$,
\begin{align}\label{assumpProp1}
\phi_{k}(B_{j}) = \frac{1}{|B_{j}|}\int_{B_{j}}{\phi}_{k}(\bs)d\bs &= \frac{1}{|B_{j}|}\int_{B_{j}}{\phi}_{k}(A_{j})d\bs\\
& = {\phi}_{k}(A_{j})\frac{1}{|B_{j}|}\int_{B_{j}}1d\bs = {\phi}_{k}(A_{j}).
\end{align}
It follows from Proposition $2.ii$ that Proposition $2.iii$ holds.

\noindent
\paragraph{\large{Proof of Proposition 3:}}
\normalsize
We now prove the equalities listed in {Equations (8), (9), and (10)} of Proposition 3. We start with Equation (8). Notice that for a given $\bs \in D_{s}$, $A \in D_{A}$, $\{{\phi}_{k}(\cdot)\}$, and $\{\lambda_{k}\}$,
\begin{align}
\nonumber
& E\left[ \left\lbrace Y_{s}(\bs) - Y_{A}(A)\right\rbrace^{2}|\{\phi_{k}\},\{\lambda_{k}\}\right]\\
\label{prop31}
&= E\left[ \left\lbrace Y_{s}(\bs) - \sum_{k = 1}^{n}\phi_{k}(\bs)\alpha_{k}\right\rbrace^{2}\vert\{\phi_{k}\},\{\lambda_{k}\}\right]\\
\label{prop32}
& + E\left[ \left\lbrace \sum_{k = 1}^{n}\phi_{k}(\bs)\alpha_{k}- \sum_{k = 1}^{n}\phi_{A,k}(A)\alpha_{k}\right\rbrace^{2}\vert\{\phi_{k}\},\{\lambda_{k}\}\right] \\
\label{prop33}
&+ \frac{1}{|A|}E\left[ \left\lbrace \sum_{k = 1}^{n}\phi_{A,k}(A)\alpha_{k} - Y_{A}(A)\right\rbrace^{2}\vert\{\phi_{k}\},\{\lambda_{k}\}\right]
\\
\label{prop34}
&+ 2 E\left[\left\lbrace Y_{s}(\bs) - \sum_{k = 1}^{n}\phi_{k}(\bs)\alpha_{k}\right\rbrace \left\lbrace\sum_{k = 1}^{n}\phi_{k}(\bs)\alpha_{k}- \sum_{k = 1}^{n}\phi_{A,k}(A)\alpha_{k}\right\rbrace\vert\{\phi_{k}\},\{\lambda_{k}\}\right]\\
\label{prop35}
&+ 2 E\left[\left\lbrace Y_{s}(\bs) - \sum_{k = 1}^{n}\phi_{k}(\bs)\alpha_{k}\right\rbrace \left\lbrace\sum_{k = 1}^{n}\phi_{A,k}(A)\alpha_{k} - Y_{A}(A)\right\rbrace\vert\{\phi_{k}\},\{\lambda_{k}\}\right]
\\
\label{prop36}
&+ 2 E\left[\left\lbrace \sum_{k = 1}^{n}\phi_{k}(\bs)\alpha_{k}- \sum_{k = 1}^{n}\phi_{A,k}(A)\alpha_{k}\right\rbrace \left\lbrace\sum_{k = 1}^{n}\phi_{A,k}(A)\alpha_{k} - Y_{A}(A)\right\rbrace\vert\{\phi_{k}\},\{\lambda_{k}\}\right]
.
\end{align}
\noindent
Through an application of Mercer's theorem we have that the sum of the cross-product terms in (\ref{prop34}), (\ref{prop35}), and (\ref{prop36}) converge to zero as $n$ goes to infinity. Similarly, it follows from \citet{Karhunen} that (\ref{prop31}) goes to zero as $n$ goes to infinity, and from Proposition 1 that (\ref{prop33}) goes to zero as $n$ goes to infinity. Thus,
\begin{equation}
E\left[ \left\lbrace Y_{s}(\bs) - Y_{A}(A)\right\rbrace^{2}|\{\phi_{k}\},\{\lambda_{k}\}\right]= \sum_{j = 1}^{\infty} \left(\phi_{j}(\bs) - \phi_{A,j}(A)\right)^{2} \lambda_{j},
\end{equation}
\noindent
Then, upon taking the expectation with respect to $\{\phi_{k}\},\{\lambda_{k}\}\vert \bz$ we have the desired result.\\
\indent To prove Equation (9) recall from Mercer's theorem and Proposition $1.ii$ that,
\begin{align}\label{covarrelation2}
\nonumber
\mathrm{var}\left\lbrace Y_{s}(\bs)\right\rbrace &= \sum_{k = 1}^{\infty}{\phi}_{k}(\bs)^{2}\lambda_{j}\\
\mathrm{var}\left\lbrace Y_{A}(A)\right\rbrace &= \sum_{k = 1}^{\infty}{\phi}_{A,k}(A)^{2}\lambda_{j}.
\end{align}
\noindent
Expanding (9) and substituting (\ref{covarrelation2}) we have
\begin{align}\label{cagecovarrelation3}
\nonumber
\mathrm{CAGE}(A) &=  E\left[\int_{A}\frac{\sum_{j = 1}^{\infty}\left\lbrace\phi_{j}(\bs) - \phi_{A,j}(A)\right\rbrace^{2}\lambda_{j}}{|A|} d\bs \vert \bz\right]\\
\nonumber
&=  E\left\lbrace\int_{A}\frac{\sum_{j = 1}^{\infty}\phi_{j}(\bs)^{2}\lambda_{j} -2 \sum_{j = 1}^{\infty}\phi_{j}(\bs)\phi_{A,j}(A)\lambda_{j}}{|A|} d\bs  + \sum_{k = 1}^{\infty}{\phi}_{A,k}(A)^{2}\lambda_{j}\vert \bz\right\rbrace\\
\nonumber
&=  E\left\lbrace\int_{A}\frac{\sum_{j = 1}^{\infty}\phi_{j}(\bs)^{2}\lambda_{j}}{|A|} d\bs  -2\sum_{k = 1}^{\infty}{\phi}_{A,k}(A)^{2}\lambda_{j}+ \sum_{k = 1}^{\infty}{\phi}_{A,k}(A)^{2}\lambda_{j}\vert \bz\right\rbrace \\
\nonumber
&=  E\left\lbrace\int_{A}\frac{\sum_{j = 1}^{\infty}\phi_{j}(\bs)^{2}\lambda_{j}}{|A|} d\bs  -\sum_{k = 1}^{\infty}{\phi}_{A,k}(A)^{2}\lambda_{j}\vert \bz\right\rbrace\\
\nonumber
&= \hspace{5pt}E\left[\int_{A}\frac{\mathrm{var}\left\lbrace Y_{s}(\bs)\right\rbrace}{|A|} d\bs - \mathrm{var}\left\lbrace Y_{A}(A)\right\rbrace\vert \bz\right];\hspace{5pt} A \subset D_{s}.
\end{align}
\noindent
This proves (9).\\
\indent We now prove Equation (10). From (8) we have for any $A \subset D_{s}$, 
\begin{equation}
\mathrm{CAGE}(A) = \hspace{5pt}E\left[ \int_{A}\frac{\left\lbrace Y_{s}(\bs) - Y_{A}(A)\right\rbrace^{2}}{|A|} d\bs \vert \bz\right].
\end{equation}
\noindent
Adding and subtracting $\widehat{Y}_{A}$,
\begin{align}
\nonumber
\mathrm{CAGE}(A) &= \hspace{5pt}E\left[ \int_{A}\frac{\left\lbrace Y_{s}(\bs) -\widehat{Y}_{A}(A) + \widehat{Y}_{A}(A) - Y_{A}(A)\right\rbrace^{2}}{|A|} d\bs \vert \bz\right]\\
\nonumber
&= \hspace{5pt}E\left[ \int_{A}\frac{\left\lbrace Y_{s}(\bs) -\widehat{Y}_{A}(A)\right\rbrace^{2}}{|A|} d\bs \vert \bz\right] + E\left[ \int_{A}\frac{\left\lbrace\widehat{Y}_{A}(A) - Y_{A}(A)\right\rbrace^{2}}{|A|} d\bs \vert \bz\right]\\
\nonumber
&+2E\left[ \int_{A}\frac{\left\lbrace Y_{s}(\bs) -\widehat{Y}_{A}(A)\right\rbrace \left\lbrace \widehat{Y}_{A}(A) - Y_{A}(A)\right\rbrace}{|A|} d\bs \vert \bz\right]\\
\nonumber
&= \hspace{5pt}E\left[ \int_{A}\frac{\left\lbrace Y_{s}(\bs) -\widehat{Y}_{A}(A)\right\rbrace^{2}}{|A|} d\bs \vert \bz\right] + E\left[ \left\lbrace \widehat{Y}_{A}(A) - Y_{A}(A)\right\rbrace^{2}d\bs \vert \bz\right]\\
\nonumber
&-2E\left[ \left\lbrace\widehat{Y}_{A}(A) - Y_{A}(A)\right\rbrace^{2}\vert \bz\right]\\
\nonumber
&= \hspace{5pt}E\left[ \int_{A}\frac{\left\lbrace Y_{s}(\bs) -\widehat{Y}_{A}(A)\right\rbrace^{2}}{|A|} d\bs \vert \bz\right] - E\left[ \left\lbrace\widehat{Y}_{A}(A) - Y_{A}(A)\right\rbrace^{2} \vert \bz\right].
\end{align}
\noindent
This proves Equation (11).

\noindent
\paragraph{\large{Proof of Proposition 4:}}
\normalsize
The fine-scale variation term $\delta$ in (16) can be written as
\begin{equation*}
\delta(\bu;\bm{\xi})= \textbf{h}(\bu)^{\prime}\bm{\xi};\hspace{5pt}\bu \in D_{s}\cup D_{A},
\end{equation*}
where
\begin{equation*}
\textbf{h}(\bu)\equiv \left\{
	\begin{array}{ll}
		\left(I(\bu \in B): B \in D_{B}\right)^{\prime}& \mbox{if } \bu \in D_{s} \\
		\left(\frac{|\bu \cap B|}{|B|}: B \in D_{B}\right)^{\prime}  & \mbox{if }  \bu \in D_{A},
	\end{array}
	\right.
\end{equation*}
\noindent
and $I(\cdot)$ is the indicator function. Then, from Equation (15) we have that for a given $\bm{\phi}_{s}$ and $\bm{\alpha}$,
\begin{align}\label{byassump2}
\nonumber
\textbf{Y}_{s}^{(C)} &= \mu\bm{1}_{n_{C}} + \bm{\Phi}_{s}^{(C)}\bm{\alpha}+\textbf{H}_{s}^{(C)}\bm{\xi}\\
\textbf{Y}_{C} &=\mu\bm{1}_{n_{C}}+\bm{\Phi}_{C}\bm{\alpha}+\textbf{H}_{C}\bm{\xi},
\end{align}
where the $n_{C}\times r$ matrices $\bm{\Phi}_{s}^{(C)}\equiv (\bm{\phi}_{s}(\textbf{x}_{j})^{\prime}:j = 1,...,n_{C})^{\prime}$ and $\bm{\Phi}_{C}\equiv (\bm{\phi}(C_{j}; \hspace{5pt}\bm{\phi}_{s})^{\prime}:j = 1,...,n_{C})^{\prime}$, and the $n_{C}\times n_{B}$ matrices $\textbf{H}_{s}^{(C)}\equiv (\textbf{h}(\textbf{x}_{j})^{\prime}:j = 1,...,n_{C})^{\prime}$ and $\textbf{H}_{C}\equiv (\textbf{h}(C_{j})^{\prime}:j = 1,...,n_{C})^{\prime}$. Notice that for the values of $\{\textbf{x}_{j}\}$ and $\{\textbf{C}_{j}\}$ given in the statement of Proposition 5, we have $\textbf{H}_{s}^{(C)} = \textbf{H}_{C} = \textbf{I}_{n_{C}}$ (the $n_{C}\times n_{C}$ identity matrix), and thus,
\begin{align}\label{byassump33}
\nonumber
\textbf{Y}_{s}^{(C)} &= \mu\bm{1}_{n_{C}} + \bm{\Phi}_{s}^{(C)}\bm{\alpha}+\bm{\xi}\\
\textbf{Y}_{C} &=\mu\bm{1}_{n_{C}}+\bm{\Phi}_{C}\bm{\alpha}+\bm{\xi}.
\end{align}
The condition for the forward implication of Proposition $4.i$ is that $\bm{\Phi}_{s}^{(C)} = \bm{\Phi}_{C}$; thus, from (\ref{byassump33}) we have that
\begin{equation}\label{byassump3}
\textbf{Y}_{s}^{(C)} = \mu\bm{1}_{n_{C}} + \bm{\Phi}_{s}^{(C)}\bm{\alpha}+\bm{\xi}=\textbf{Y}_{C}.
\end{equation}
When applying any real-valued measurable $f$ to both sides of (\ref{byassump3}), we obtain that $f(\textbf{Y}_{s}^{(C)})$ = $f(\textbf{Y}_{C})$ almost surely. One can prove forward implication of Proposition $4.ii$ in a similar manner. 

To prove the reverse statement of Proposition $4.i$, suppose that $f(\textbf{Y}_{s}^{(C)})$ = $f(\textbf{Y}_{C})$ almost surely for any real-valued function $f$. Thus, the functions $f_{i}(\textbf{b}) = b_{i}$ for $i = 1,...,n_{A}$ and $\textbf{b} = (b_{j}: j = 1,...,n_{A})^{\prime}\in \mathbb{R}^{n_{A}}$, imply that
\begin{equation} \label{almostsure5}
\textbf{Y}_{s}^{(C)} = \textbf{Y}_{C},
\end{equation}
\noindent
almost surely. From (\ref{byassump33}) and (\ref{almostsure5}) we see that
\begin{equation}\label{almostsure6}
\bm{\Phi}_{s}^{(C)}\bm{\alpha}=\bm{\Phi}_{C}\bm{\alpha},
\end{equation}
almost surely. Multiply both sides of (\ref{almostsure6}) by $\bm{\alpha}^{\prime}$, and take the expectation with respect to $Y\vert \bm{\phi}_{s}, \bm{\Lambda}$ to obtain
\begin{equation}\label{almostsure4}
\bm{\Phi}_{s}^{(C)}\bm{\Lambda}=\bm{\Phi}_{C}\bm{\Lambda}.
\end{equation}
\noindent
Provided that $\lambda_{j}>0$ for all $j$, we can take the inverse of $\bm{\Lambda}$ on both sides of (\ref{almostsure4}) so that $\bm{\Phi}_{s}^{(C)}=\bm{\Phi}_{C}$, which is the desired result. One can prove the reverse statement of Proposition $4.ii$ in a similar manner.

By the condition in Proposition $4.iii$, we have that for a given $\bm{\phi}_{s}$ and $\bm{\alpha}$,
\begin{equation}\label{assumpProp6}
\bm{\phi}_{s}(\textbf{x}_{j})^{\prime}\bm{\alpha} = \bm{\phi}(C_{j}; \hspace{5pt}\bm{\phi}_{s})^{\prime}\bm{\alpha}; \hspace{5pt} j = 1,...,n_{C}.
\end{equation}
Integrating (\ref{assumpProp6}) with respect to $\textbf{x}_{j}$ we have
\begin{equation*}
\bm{\phi}(B_{j}; \hspace{5pt}\bm{\phi}_{s})^{\prime}\bm{\alpha} = \bm{\phi}(C_{j}; \hspace{5pt}\bm{\phi}_{s})^{\prime}\bm{\alpha}; \hspace{5pt} j = 1,...,n_{C}.
\end{equation*}
\noindent
Since $\lambda_{j}>0$ for all $j$, this leads to the condition for the forward implication of Proposition $4.ii$, and thus, it follows that Proposition $4.iii$ holds.

\noindent
\paragraph{\large{Proof of Proposition 5:}}
\normalsize
From Equation (1) we see that for $Y(\cdot;\bm{\phi}_{s}^{\mathrm{OC}})$ to be a multiscale truncated K-L expansion, we only need to show that $Y_{s}(\cdot;\bm{\phi}_{s}^{\mathrm{OC}})$ is a truncated K-L expansion. Many of the following equations can be found in {\citet{obled-creutin}.}

To show that $Y_{s}(\cdot;\bm{\phi}_{s}^{\mathrm{OC}})$ is a truncated K-L expansion, we need to establish three items: the eigenvalues must be nonnegative with at least one eigenvalue strictly positive; the Fredholm integral equations must hold; and the eigenvectors must be orthonormal. Notice that
\begin{align}\label{covapprox}
\nonumber
&\mathrm{cov}\left[Y_{s}\left\lbrace \bs;\bm{\phi}_{s}^{\mathrm{OC}}(\cdot;\hspace{5pt}\textbf{F})\right\rbrace,Y_{s}\left\lbrace \bu;\bm{\phi}_{s}^{\mathrm{OC}}(\cdot;\hspace{5pt}\textbf{F})\right\rbrace\right]\\
\nonumber
 & = E\left[ \left\lbrace \sum_{k = 1}^{r}\sum_{i = 1}^{r} \psi_{i}(\bs)F_{ik}\alpha_{k} \right\rbrace \left\lbrace\sum_{q= 1}^{r} \sum_{p = 1}^{r} \psi_{q}(\bu)F_{qp}\alpha_{p} \right\rbrace \right] \\
&= \sum_{k = 1}^{r} \lambda_{k} \left\lbrace\sum_{i = 1}^{r}\psi_{i}(\bs)F_{ik}\right\rbrace\left\lbrace\sum_{q = 1}^{r}\psi_{q}(\bu)F_{qk}\right\rbrace.
\end{align}
\noindent
 Substituting (\ref{covapprox}) into the Fredholm integral equation we have, for $k=1,...,r$,
\begin{equation}\label{fredholm}
\int_{D_{s}} \left\lbrace\sum_{i = 1}^{r} \sum_{k = 1}^{r}\sum_{q = 1}^{r} F_{qk}\lambda_{k}F_{ik}\psi_{i}(\bs)\psi_{q}(\bu)\right\rbrace \left\lbrace\sum_{m = 1}^{r}\psi_{m}(\bs)F_{mp}\right\rbrace d\bs = \omega_{p}\left\lbrace\sum_{q = 1} \psi_{q}(\bu)F_{qp}\right\rbrace,
\end{equation}
\noindent
where $\{\omega_{k}\}$ represents the eigenvalues of $Y_{s}(\cdot;\bm{\phi}_{s}^{\mathrm{OC}})$. Distributing the sums and integral through (\ref{fredholm}), we obtain
\begin{equation}\label{fredholm2}
\sum_{q = 1}^{r} \psi_{q}(\bu)  \left\lbrace\sum_{i = 1}^{r} \sum_{k = 1}^{r} \sum_{m = 1}^{r}F_{qk} \lambda_{k}F_{ik}\right\rbrace \int_{D_{s}}\psi_{i}(\bs)\psi_{m}(\bs)F_{mp}d\bs = \omega_{p}\left\lbrace\sum_{q = 1} \psi_{q}(\bu)F_{qp}\right\rbrace.
\end{equation}
\noindent
Matching terms in (\ref{fredholm2}), we have

\begin{equation}\label{termbyterm}
\sum_{i = 1}^{r} \sum_{k = 1}^{r} \sum_{m = 1}^{r}F_{qk}\lambda_{k}F_{ik} W_{im}F_{mp} = \omega_{p}F_{qp}; q = 1, ..., r.
\end{equation}
\noindent
In matrix form, (\ref{termbyterm}) becomes,

\begin{equation}\label{fiapprox}
\textbf{F}\bm{\Lambda}\textbf{F}^{\prime}\textbf{W}\textbf{F} = \textbf{F}\bm{\Omega},
\end{equation}

\noindent
where \(\bm{\Lambda} \equiv\) diag\((\lambda_{k})\) and \(\bm{\Omega} \equiv\) diag\((\omega_{k})\). The assumption that $\textbf{F}^{\prime}\textbf{W}\textbf{F} = \textbf{I}$ and (\ref{fiapprox}) implies that the Fredholm-integral equation holds provided that 
\begin{equation}\label{fireduced}
\textbf{F}\bm{\Lambda} = \textbf{F}\bm{\Omega}.
\end{equation}
\noindent
Since, $\textbf{F}$ is invertible we have that (\ref{fireduced}) verifies that the eigenvalues of $Y_{s}(\cdot;\bm{\phi}_{s}^{\mathrm{OC}})$ are nonnegative with $\bm{\Lambda} = \bm{\Omega}$ (and at least one eigenvalue is strictly positive), and that the Fredholm integral equations for $Y_{s}(\cdot;\bm{\phi}_{s}^{\mathrm{OC}})$ hold. The orthogonality of $\bm{\phi}_{s}^{\mathrm{OC}}$ holds by assumption since
\begin{align}
\nonumber
&\int \phi_{i}^{\mathrm{OC}}(\bs;\hspace{5pt}\textbf{F}) \phi_{j}^{\mathrm{OC}}(\bs;\hspace{5pt}\textbf{F}) d\bs = \sum_{k = 1}^{r} \sum_{ p = 1}^{r}F_{ki}F_{pj}\int \psi_{k}(\bs) \psi_{p}(\bs) d\bs \\
\nonumber
&= \sum_{k = 1}^{r} \sum_{ p = 1}^{r}F_{ki}W_{kp}F_{pj} = I(i=j),
\end{align}
which results in the relation,
\begin{equation}
\nonumber
\textbf{F}^{\prime}\textbf{W}\textbf{F} = \textbf{I}.
\end{equation}
\noindent
This completes the proof.

\noindent \paragraph{\large{Proof of Proposition 6:}}
\normalsize
Let $\textbf{W} = \textbf{P}_{\mathrm{W}}\bm{\Lambda}_{\mathrm{W}}\textbf{P}_{\mathrm{W}}^{\prime}$ be the spectral decomposition of $\textbf{W}$. It follows that the Cholesky square root of $\textbf{W}$ and $\textbf{W}^{-1}$ is given by $\textbf{P}_{\mathrm{W}}\bm{\Lambda}_{\mathrm{W}}^{1/2}$ and $\textbf{P}_{\mathrm{W}}\bm{\Lambda}_{\mathrm{W}}^{-1/2}$, respectively. It follows immediately that $\textbf{G}^{\prime}(\textbf{P}_{\mathrm{W}}\bm{\Lambda}_{\mathrm{W}}^{-1/2})^{\prime}\textbf{W}\textbf{P}_{\mathrm{W}}\bm{\Lambda}_{\mathrm{W}}^{-1/2}\textbf{G} = \textbf{I}$.\\

\subsection*{V.ii Additional Results} In the main-text, three results were discussed, but not formally stated. Thus, in this section we state and prove these results. In particular, at the end of Remark 6, we mentioned that the CAGE identities in Proposition 3 also hold for DCAGE; this extension of Proposition 3 is referred to as \textit{Result 1}. Also, at the end of Section~3.2 we mention that a version of Proposition 3 exists for CAGE in (17) and DCAGE in (18); these two extensions are referred to as \textit{Result 2} and \textit{Result 3}, respectively.\\

\noindent
\textit{Result 1: Assume that the conditions of Proposition 1 hold. Assume that the stochastic process $Z: D_{s}\times \Omega \rightarrow \mathbb{R}$ is generated based on any generic probability space $(\Omega,\mathcal{F},\mathcal{P})$ such that the conditional probability density function of $Y(\bu)\vert \bz$ exists for each $\bu \in D_{s}\cup D_{A}$, where $Z$ is defined in Remark 2. Then, DCAGE in (7) has the following alternative expressions:
\begin{align}
\label{intuitive102}
DCAGE(C) &= \hspace{5pt}E\left[ \underset{h \in H}{\sum}\frac{\left\lbrace Y_{A}(B_{h}) - Y_{A}(C)\right\rbrace^{2}}{|C|}  \vert \bz\right]\\
\label{intuitive202}
DCAGE(C) &= \hspace{5pt}E\left[ \underset{h \in H}{\sum}\frac{\mathrm{var}\left\lbrace Y_{A}(B_{h})\right\rbrace}{|C|} - \mathrm{var}\left\lbrace Y_{A}(C)\right\rbrace\vert \bz\right]\\
\label{intuitive302}
DCAGE(C) &=\hspace{5pt}E\left[ \underset{h \in H}{\sum}\frac{\left\lbrace Y_{A}(B_{h}) -\widehat{Y}_{A}(C)\right\rbrace^{2}}{|C|} \vert \bz\right] - E\left[ \left\lbrace\widehat{Y}_{A}(C) - Y_{A}(C)\right\rbrace^{2} \vert \bz\right],
\end{align}
\noindent
where $C = \cup_{h \in H} B_{h}$, $H \subset \{1,...,n_{B}\}$, and $B_{h} \in D_{B}$ for each $h\in H$.}\\ 

\noindent
\paragraph{\large{Proof of Result 1:}}
\normalsize
We now prove the equalities listed in {Equations (\ref{intuitive102}), (\ref{intuitive202}), and (\ref{intuitive302})} of Proposition 3. We start with Equation (\ref{intuitive102}). Notice that for a given $B_{h} \in D_{B}$, $C = \cup_{h \in H} B_{h}$, $H \subset \{1,...,n_{B}\}$, $\{{\phi}_{k}(\cdot)\}$, and $\{\lambda_{k}\}$,
\begin{align}
\nonumber
& E\left[ \left\lbrace Y_{A}(B_{h}) - Y_{A}(C)\right\rbrace^{2}|\{\phi_{k}\},\{\lambda_{k}\}\right]\\
\label{prop312}
&= E\left[ \left\lbrace Y_{A}(B_{h}) - \sum_{k = 1}^{n}\phi_{A,k}(B_{h})\alpha_{k}\right\rbrace^{2}\vert\{\phi_{k}\},\{\lambda_{k}\}\right]\\
\label{prop322}
& + E\left[ \left\lbrace \sum_{k = 1}^{n}\phi_{A,k}(B_{h})\alpha_{k}- \sum_{k = 1}^{n}\phi_{A,k}(C)\alpha_{k}\right\rbrace^{2}\vert\{\phi_{k}\},\{\lambda_{k}\}\right] \\
\label{prop332}
&+ \frac{1}{|C|}E\left[ \left\lbrace \sum_{k = 1}^{n}\phi_{A,k}(C)\alpha_{k} - Y_{A}(C)\right\rbrace^{2}\vert\{\phi_{k}\},\{\lambda_{k}\}\right]
\\
\label{prop342}
&+ 2 E\left[\left\lbrace Y_{A}(B_{h}) - \sum_{k = 1}^{n}\phi_{A,k}(B_{h})\alpha_{k}\right\rbrace \left\lbrace\sum_{k = 1}^{n}\phi_{A,k}(B_{h})\alpha_{k}- \sum_{k = 1}^{n}\phi_{A,k}(C)\alpha_{k}\right\rbrace\vert\{\phi_{k}\},\{\lambda_{k}\}\right]\\
\label{prop352}
&+ 2 E\left[\left\lbrace Y_{A}(B_{h}) - \sum_{k = 1}^{n}\phi_{A,k}(B_{h})\alpha_{k}\right\rbrace \left\lbrace\sum_{k = 1}^{n}\phi_{A,k}(C)\alpha_{k} - Y_{A}(C)\right\rbrace\vert\{\phi_{k}\},\{\lambda_{k}\}\right]
\\
\label{prop362}
&+ 2 E\left[\left\lbrace \sum_{k = 1}^{n}\phi_{A,k}(B_{h})\alpha_{k}- \sum_{k = 1}^{n}\phi_{A,k}(C)\alpha_{k}\right\rbrace \left\lbrace\sum_{k = 1}^{n}\phi_{A,k}(C)\alpha_{k} - Y_{A}(C)\right\rbrace\vert\{\phi_{k}\},\{\lambda_{k}\}\right]
.
\end{align}
\noindent
Through an application of Mercer's theorem we have that the sum of the cross-product terms in (\ref{prop342}), (\ref{prop352}), and (\ref{prop362}) converge to zero as $n$ goes to infinity. Similarly, it follows from Proposition 1 that (\ref{prop312}) and (\ref{prop332}) go to zero as $n$ goes to infinity. Thus,
\begin{equation}
E\left[ \left\lbrace Y_{A}(B_{h}) - Y_{A}(C)\right\rbrace^{2}|\{\phi_{k}\},\{\lambda_{k}\}\right]= \sum_{j = 1}^{\infty} \left(\phi_{A,j}(B_{h}) - \phi_{A,j}(C)\right)^{2} \lambda_{j},
\end{equation}
\noindent
Then, upon taking the expectation with respect to $\{\phi_{k}\},\{\lambda_{k}\}\vert \bz$ we have the desired result.\\
\indent To prove Equation (\ref{intuitive202}) recall from Proposition $1.ii$ that,
\begin{align}\label{covarrelation22}
\nonumber
\mathrm{var}\left\lbrace Y_{A}(B_{h})\right\rbrace &= \sum_{k = 1}^{\infty}{\phi}_{A,k}(B_{h})^{2}\lambda_{j}\\
\mathrm{var}\left\lbrace Y_{A}(C)\right\rbrace &= \sum_{k = 1}^{\infty}{\phi}_{A,k}(C)^{2}\lambda_{j}.
\end{align}
\noindent
Expanding (\ref{intuitive202}) and substituting (\ref{covarrelation2}) we have
\begin{align}\label{cagecovarrelation32}
\nonumber
\mathrm{CAGE}(C) &=  E\left[\underset{h \in H}{\sum}\frac{\sum_{j = 1}^{\infty}\left\lbrace\phi_{A,j}(B_{h}) - \phi_{A,j}(C)\right\rbrace^{2}\lambda_{j}}{|C|}  \vert \bz\right]\\
\nonumber
&=  E\left\lbrace\underset{h \in H}{\sum}\frac{\sum_{j = 1}^{\infty}\phi_{A,j}(B_{h})^{2}\lambda_{j} -2 \sum_{j = 1}^{\infty}\phi_{A,j}(B_{h})\phi_{A,j}(C)\lambda_{j}}{|C|}  + \sum_{k = 1}^{\infty}{\phi}_{A,k}(\bs)^{2}\lambda_{j}\vert \bz\right\rbrace\\
\nonumber
&=  E\left\lbrace\underset{h \in H}{\sum}\frac{\sum_{j = 1}^{\infty}\phi_{A,j}(B_{h})^{2}\lambda_{j}}{|C|}  -2\sum_{k = 1}^{\infty}{\phi}_{A,k}(C)^{2}\lambda_{j}+ \sum_{k = 1}^{\infty}{\phi}_{A,k}(C)^{2}\lambda_{j}\vert \bz\right\rbrace \\
\nonumber
&=  E\left\lbrace\underset{h \in H}{\sum}\frac{\sum_{j = 1}^{\infty}\phi_{A,j}(B_{h})^{2}\lambda_{j}}{|C|}  -\sum_{k = 1}^{\infty}{\phi}_{A,k}(C)^{2}\lambda_{j}\vert \bz\right\rbrace\\
\nonumber
&= \hspace{5pt}E\left[\underset{h \in H}{\sum}\frac{\mathrm{var}\left\lbrace Y_{A}(B_{h})\right\rbrace}{|C|} - \mathrm{var}\left\lbrace Y_{A}(C)\right\rbrace\vert \bz\right];\hspace{5pt} A \subset D_{s}.
\end{align}
\noindent
This proves (\ref{intuitive202}).\\
\indent We now prove Equation (\ref{intuitive302}). From (\ref{intuitive102}) we have, 
\begin{equation}
\mathrm{CAGE}(C) = \hspace{5pt}E\left[ \underset{h \in H}{\sum}\frac{\left\lbrace Y_{A}(B_{h}) - Y_{A}(C)\right\rbrace^{2}}{|C|}  \vert \bz\right].
\end{equation}
\noindent
Adding and subtracting $\widehat{Y}_{A}$,
\begin{align}
\nonumber
\mathrm{CAGE}(C) &= \hspace{5pt}E\left[ \underset{h \in H}{\sum}\frac{\left\lbrace Y_{A}(B_{h}) -\widehat{Y}_{A}(C) + \widehat{Y}_{A}(C) - Y_{A}(C)\right\rbrace^{2}}{|C|}  \vert \bz\right]\\
\nonumber
&= \hspace{5pt}E\left[ \underset{h \in H}{\sum}\frac{\left\lbrace Y_{A}(B_{h}) -\widehat{Y}_{A}(C)\right\rbrace^{2}}{|C|}  \vert \bz\right] + E\left[ \underset{h \in H}{\sum}\frac{\left\lbrace\widehat{Y}_{A}(C) - Y_{A}(C)\right\rbrace^{2}}{|C|}  \vert \bz\right]\\
\nonumber
&+2E\left[ \underset{h \in H}{\sum}\frac{\left\lbrace Y_{A}(B_{h}) -\widehat{Y}_{A}(C)\right\rbrace \left\lbrace \widehat{Y}_{A}(C) - Y_{A}(C)\right\rbrace}{|C|}  \vert \bz\right]\\
\nonumber
&= \hspace{5pt}E\left[ \underset{h \in H}{\sum}\frac{\left\lbrace Y_{A}(B_{h}) -\widehat{Y}_{A}(C)\right\rbrace^{2}}{|C|}  \vert \bz\right] + E\left[ \left\lbrace \widehat{Y}_{A}(C) - Y_{A}(C)\right\rbrace^{2} \vert \bz\right]\\
\nonumber
&-2E\left[ \left\lbrace\widehat{Y}_{A}(C) - Y_{A}(C)\right\rbrace^{2}\vert \bz\right]\\
\nonumber
&= \hspace{5pt}E\left[ \underset{h \in H}{\sum}\frac{\left\lbrace Y_{A}(B_{h}) -\widehat{Y}_{A}(C)\right\rbrace^{2}}{|C|}  \vert \bz\right] - E\left[ \left\lbrace\widehat{Y}_{A}(C) - Y_{A}(C)\right\rbrace^{2} \vert \bz\right].
\end{align}
\noindent
This proves Equation (\ref{intuitive302}).\\

\noindent
\textit{Result 2: For $Z$ defined in (14) and $Y(\cdot;\hspace{5pt}\bm{\phi}_{s})$ defined in (13), we have that CAGE in (17) has the following alternative expressions:
\begin{align}
\label{intuitive1}
CAGE(A) &= \hspace{5pt}E\left[\int_{A}\frac{\left\lbrace Y_{s}(\bs;\hspace{5pt} \bm{\phi}_{s}) - Y_{A}(A;\hspace{5pt} \bm{\phi}_{s})\right\rbrace^{2}}{|A|} d\bs \vert \bz\right]\\
\label{intuitive2}
CAGE(A) &= \hspace{5pt}E\left[\int_{A}\frac{\mathrm{var}\left\lbrace Y_{s}(\bs;\hspace{5pt} \bm{\phi}_{s})\right\rbrace}{|A|} d\bs - \mathrm{var}\left\lbrace Y_{A}(A;\hspace{5pt} \bm{\phi}_{s})\right\rbrace\vert \bz\right]\\
\label{intuitive3}
CAGE(A) &=\hspace{5pt}E\left[ \int_{A}\frac{\left\lbrace Y_{s}(\bs;\hspace{5pt} \bm{\phi}_{s}) -\widehat{Y}_{A}(A)\right\rbrace^{2}}{|A|} d\bs \vert \bz\right] - E\left[ \left\lbrace\widehat{Y}_{A}(A) - Y_{A}(A;\hspace{5pt} \bm{\phi}_{s})\right\rbrace^{2} \vert \bz\right], 
\end{align}
\noindent
where $A$ is a generic areal unit (i.e., $A \subset D_{s}$), and $\widehat{Y}_{A}(A) \equiv E\left\lbrace Y_{A}(A)|\bz\right\rbrace$.}

\noindent
\paragraph{\large{Proof of Result 2:}}
\normalsize
We now prove the equalities listed in {Equations (\ref{intuitive1}), (\ref{intuitive2}), and (\ref{intuitive3})} of Proposition 5. We start with Equation (\ref{intuitive1}). Notice that for a given $\bs \in D_{s}$, $A \in D_{A}$, $\bm{\alpha}$, $\bm{\phi}_{s}$, and $\bm{\Lambda}$,
\begin{equation*}
\frac{1}{|A|}\left\lbrace Y_{s}(\bs;\hspace{5pt}\bm{\phi}_{s}) - Y_{A}(A;\hspace{5pt}\bm{\phi}_{s})\right\rbrace^{2} = \frac{1}{|A|}\left\lbrace\bm{\phi}_{s}(\bs) - \bm{\phi}(A;\hspace{5pt}\bm{\phi}_{s})\right\rbrace^{\prime}\bm{\alpha}\bm{\alpha}^{\prime}\left\lbrace\bm{\phi}_{s}(\bs) - \bm{\phi}(A;\hspace{5pt}\bm{\phi}_{s})\right\rbrace.
\end{equation*}
\noindent
Taking the expectation with respect to $\bm{\alpha}|\bm{\phi}_{s},\bm{\Lambda}$ we have\\
\begin{equation}\label{expect}
\frac{1}{|A|}E\left[ \{Y_{s}(\bs;\hspace{5pt}\bm{\phi}_{s}) - Y_{A}(A;\hspace{5pt}\bm{\phi}_{s})\}^{2}|\bm{\phi}_{s},\bm{\Lambda}\right] = \frac{1}{|A|} \{\bm{\phi}_{s}(\bs) - \bm{\phi}(A;\hspace{5pt}\bm{\phi}_{s})\}^{\prime}\bm{\Lambda}\{\bm{\phi}_{s}(\bs) - \bm{\phi}(A;\hspace{5pt}\bm{\phi}_{s})\}.
\end{equation}
\noindent
Then, upon taking the expectation of (\ref{expect}) with respect to $\bm{\phi}_{s},\bm{\Lambda}|\bz$ and integrating $\bs$ over $A$, we obtain Equation (\ref{intuitive1}).\\
\indent To prove Equation (\ref{intuitive2}) notice that
\begin{align}\label{covarrelation}
\nonumber
\mathrm{var}\{Y_{s}(\bs;\hspace{5pt}\bm{\phi}_{s})\} &= \bm{\phi}_{s}(\bs)^{\prime}\Lambda\bm{\phi}_{s}(\bs)\\
\mathrm{var}\{Y_{A}(A;\hspace{5pt}\bm{\phi}_{s})\} &= \bm{\phi}(A;\hspace{5pt}\bm{\phi}_{s})^{\prime}\Lambda\bm{\phi}(A;\hspace{5pt}\bm{\phi}_{s}).
\end{align}
\noindent
Expanding (\ref{intuitive2}) and substituting (\ref{covarrelation}) we have
\begin{align}\label{cagecovarrelation}
\nonumber
\mathrm{CAGE}(A) &=  E\left[\int_{A}\frac{\left\lbrace\bm{\phi}_{s}(\bs) - \bm{\phi}(A;\hspace{5pt}\bm{\phi}_{s})\right\rbrace^{\prime}\bm{\Lambda}\left\lbrace\bm{\phi}_{s}(\bs) - \bm{\phi}(A;\hspace{5pt}\bm{\phi}_{s})\right\rbrace}{|A|} d\bs \vert \bz\right]\\
\nonumber
&=  E\left\lbrace\int_{A}\frac{\bm{\phi}_{s}(\bs)^{\prime}\Lambda\bm{\phi}_{s}(\bs) -2 \bm{\phi}_{s}(\bs)^{\prime}\Lambda\bm{\phi}(A;\hspace{5pt}\bm{\phi}_{s})}{|A|} d\bs  + \bm{\phi}(A;\hspace{5pt}\bm{\phi}_{s})^{\prime}\Lambda\bm{\phi}(A;\hspace{5pt}\bm{\phi}_{s})\vert \bz\right\rbrace\\
\nonumber
&=  E\left\lbrace\int_{A}\frac{\bm{\phi}_{s}(\bs)^{\prime}\Lambda\bm{\phi}_{s}(\bs)}{|A|} d\bs  -2\bm{\phi}(A;\hspace{5pt}\bm{\phi}_{s})^{\prime}\Lambda\bm{\phi}(A;\hspace{5pt}\bm{\phi}_{s})+ \bm{\phi}(A;\hspace{5pt}\bm{\phi}_{s})^{\prime}\Lambda\bm{\phi}(A;\hspace{5pt}\bm{\phi}_{s})\vert \bz\right\rbrace \\
\nonumber
&=  E\left\lbrace\int_{A}\frac{\bm{\phi}_{s}(\bs)^{\prime}\Lambda\bm{\phi}_{s}(\bs)}{|A|} d\bs  -\bm{\phi}(A;\hspace{5pt}\bm{\phi}_{s})^{\prime}\Lambda\bm{\phi}(A;\hspace{5pt}\bm{\phi}_{s})\vert \bz\right\rbrace\\
\nonumber
&= \hspace{5pt}E\left[\int_{A}\frac{\mathrm{var}\left\lbrace Y_{s}(\bs;\hspace{5pt} \bm{\phi}_{s})\right\rbrace}{|A|} d\bs - \mathrm{var}\left\lbrace Y_{A}(A;\hspace{5pt} \bm{\phi}_{s})\right\rbrace\vert \bz\right];\hspace{5pt} A \subset D_{s}.
\end{align}
\noindent
This proves (\ref{intuitive2}).\\
\indent We now prove Equation (\ref{intuitive3}). From (\ref{intuitive1}) we have for any $A \subset D_{s}$, 
\begin{equation}
\mathrm{CAGE}(A) = \hspace{5pt}E\left[ \int_{A}\frac{\left\lbrace Y_{s}(\bs;\hspace{5pt} \bm{\phi}_{s}) - Y_{A}(A;\hspace{5pt} \bm{\phi}_{s})\right\rbrace^{2}}{|A|} d\bs \vert \bz\right].
\end{equation}
\noindent
Adding and subtracting $\widehat{Y}_{A}$,
\begin{align}
\nonumber
\mathrm{CAGE}(A) &= \hspace{5pt}E\left[ \int_{A}\frac{\left\lbrace Y_{s}(\bs;\hspace{5pt} \bm{\phi}_{s}) -\widehat{Y}_{A}(A) + \widehat{Y}_{A}(A) - Y_{A}(A;\hspace{5pt} \bm{\phi}_{s})\right\rbrace^{2}}{|A|} d\bs \vert \bz\right]\\
\nonumber
&= \hspace{5pt}E\left[ \int_{A}\frac{\left\lbrace Y_{s}(\bs;\hspace{5pt} \bm{\phi}_{s}) -\widehat{Y}_{A}(A)\right\rbrace^{2}}{|A|} d\bs \vert \bz\right] + E\left[ \int_{A}\frac{\left\lbrace\widehat{Y}_{A}(A) - Y_{A}(A;\hspace{5pt} \bm{\phi}_{s})\right\rbrace^{2}}{|A|} d\bs \vert \bz\right]\\
\nonumber
&+2E\left[ \int_{A}\frac{\left\lbrace Y_{s}(\bs;\hspace{5pt} \bm{\phi}_{s}) -\widehat{Y}_{A}(A)\right\rbrace \left\lbrace \widehat{Y}_{A}(A) - Y_{A}(A;\hspace{5pt} \bm{\phi}_{s})\right\rbrace}{|A|} d\bs \vert \bz\right]\\
\nonumber
&= \hspace{5pt}E\left[ \int_{A}\frac{\left\lbrace Y_{s}(\bs;\hspace{5pt} \bm{\phi}_{s}) -\widehat{Y}_{A}(A)\right\rbrace^{2}}{|A|} d\bs \vert \bz\right] + E\left[ \left\lbrace \widehat{Y}_{A}(A) - Y_{A}(A;\hspace{5pt} \bm{\phi}_{s})\right\rbrace^{2}d\bs \vert \bz\right]\\
\nonumber
&-2E\left[ \left\lbrace\widehat{Y}_{A}(A) - Y_{A}(A;\hspace{5pt} \bm{\phi}_{s})\right\rbrace^{2}\vert \bz\right]\\
\nonumber
&= \hspace{5pt}E\left[ \int_{A}\frac{\left\lbrace Y_{s}(\bs;\hspace{5pt} \bm{\phi}_{s}) -\widehat{Y}_{A}(A)\right\rbrace^{2}}{|A|} d\bs \vert \bz\right] - E\left[ \left\lbrace\widehat{Y}_{A}(A) - Y_{A}(A;\hspace{5pt} \bm{\phi}_{s})\right\rbrace^{2} \vert \bz\right].
\end{align}
\noindent
This proves Equation (\ref{intuitive3}).\\

\noindent
\textit{Result 3: For $Z$ defined in (14) and $Y(\cdot;\hspace{5pt}\bm{\phi}_{s})$ defined in (13), we have that DCAGE in (18) has the following alternative expressions:
\begin{align}
\label{intuitive100}
DCAGE(C) &= \hspace{5pt}E\left\lbrace \underset{h \in H}{\sum}\frac{(Y_{A}(B_{h};\hspace{5pt} \bm{\phi}_{s}) - Y_{A}(C;\hspace{5pt} \bm{\phi}_{s}))^{2}}{|C|}  \vert \bz\right\rbrace\\
\label{intuitive200}
DCAGE(C) &= \hspace{5pt}E(\underset{h \in H}{\sum}\frac{\mathrm{var}(Y_{A}(B_{h};\hspace{5pt} \bm{\phi}_{s}))}{|C|} - \mathrm{var}(Y_{A}(C;\hspace{5pt} \bm{\phi}_{s}))\vert \bz)\\
\label{intuitive300}
DCAGE(C) &=\hspace{5pt}E\left\lbrace \underset{h \in H}{\sum}\frac{(Y_{A}(B_{h};\hspace{5pt} \bm{\phi}_{s}) -\widehat{Y}_{A}(C))^{2}}{|C|} \vert \bz\right\rbrace - E\left\lbrace (\widehat{Y}_{A}(C) - Y_{A}(C;\hspace{5pt} \bm{\phi}_{s}))^{2} \vert \bz\right\rbrace,
\end{align}
\noindent
where $C = \cup_{h \in H} B_{h}$, $H \subset \{1,...,n_{B}\}$, and $B_{h} \in D_{B}$ for each $h\in H$.}\\ 

\noindent
\textit{Proof of Result 3:} In the proof of Result 2, replace the integral with sums, and replace $\bm{\phi}_{s}(\bs)$ and $Y_{s}(\bs; \bm{\phi}_{s})$ with $\bm{\phi}_{A}(B_{h}; \bm{\phi}_{s})$ and $Y_{A}(B_{h}; \bm{\phi}_{s})$, respectively.

\bibliographystyle{jasa}  
\bibliography{myref}
\end{document}